\documentclass[usenatbib,useAMS]{mn2e}

\usepackage{times} 
\usepackage{amssymb}
\usepackage[dvips]{graphicx} 

\newcommand{\etal}{{\it et al.\ }}
\newcommand{\gsim}{\gtrsim} 
\newcommand{\lsim}{\lesssim} 
\newcommand{\Msol}{M_\odot}
\newcommand{\Lsol}{L_\odot}

\newcommand{\m}{{\rm m}}
\newcommand{\mum}{\mu{\rm m}}
\newcommand{\pc}{{\rm pc}}

\newcommand{\Mpc}{{\rm Mpc}}
\newcommand{\kms}{{\rm km\,s^{-1}}}

\newcommand{\Myr}{{\rm Myr}}
\newcommand{\Gyr}{{\rm Gyr}}
\newcommand{\Jy}{{\rm mJy}}
\newcommand{\mJy}{{\rm mJy}}
\newcommand{\muJy}{\mu{\rm Jy}}
\newcommand{\GALFORM}{{\tt GALFORM}} 
\newcommand{\GRASIL}{{\tt GRASIL}}
\newcommand{\SPITZER}{{\it Spitzer}}
\newcommand{\IRAS}{{\it IRAS}}
\newcommand{\COBE}{{\it COBE}}
\newcommand{\ISO}{{\it ISO}}
\newcommand{\SCUBA}{{\it SCUBA}}
\newcommand{\HERSCHEL}{{\it Herschel}}

\newcommand{\Vc}{V_{\rm c}}
\newcommand{\tesc}{t_{\rm esc}}

\title[Galaxy evolution in the IR]
{Galaxy evolution in the infra-red: comparison of a hierarchical galaxy
formation model with SPITZER data}

\author[Lacey  et al.]{C. G. Lacey
\thanks{E-mail: Cedric.Lacey@durham.ac.uk (CGL)},$^1$
C. M. Baugh,$^1$ C.S. Frenk,$^1$ L. Silva,$^2$ G.L. Granato,$^3$  and A. Bressan,$^3$\\
$^{1}$Institute for Computational Cosmology, Department of Physics,
University of Durham, South Road, Durham, DH1 3LE, UK\\
$^2$INAF, Osservatorio Astronomico di Trieste, Via Tiepolo 11, I-34131
Trieste, Italy\\
$^3$INAF, Osservatorio Astronomico di Padova, Vicolo dell'Osservatorio
2, I-35122 Padova, Italy.
}

\begin{document}

\maketitle
  
\begin{abstract}
We present predictions for the evolution of the galaxy luminosity
function, number counts and redshift distributions in the IR based on
the $\Lambda$CDM cosmological model. We use the combined \GALFORM\
semi-analytical galaxy formation model and \GRASIL\ spectrophotometric
code to compute galaxy SEDs including the reprocessing of radiation by
dust. The model, which is the same as that in \cite{Baugh05}, assumes
two different IMFs: a normal solar neighbourhood IMF for quiescent
star formation in disks, and a very top-heavy IMF in starbursts
triggered by galaxy mergers. We have shown previously that the
top-heavy IMF seems to be necessary to explain the number counts of
faint sub-mm galaxies.  We compare the model with observational data
from the \SPITZER\ Space Telescope, with the model parameters fixed at
values chosen before \SPITZER\ data became available.  We find that
the model matches the observed evolution in the IR remarkably well
over the whole range of wavelengths probed by \SPITZER. In particular,
the \SPITZER\ data show that there is strong evolution in the mid-IR
galaxy luminosity function over the redshift range $z \sim 0-2$, and
this is reproduced by our model without requiring any adjustment of
parameters. On the other hand, a model with a normal IMF in starbursts
predicts far too little evolution in the mid-IR luminosity function,
and is therefore excluded.
\end{abstract}

\begin{keywords}
galaxies: evolution -- galaxies: formation -- galaxies: high-redshift
-- infrared: galaxies -- ISM: dust, extinction
\end{keywords}

\section{Introduction}

In recent years, the evolution of galaxies at mid- and far-infrared
wavelengths has been opened up for direct observational study by
infrared telescopes in space. Already in the 1980s, the \IRAS\
satellite surveyed the local universe in the IR, showing that much of
present-day star formation is optically obscured, revealing a
population of luminous and ultra-luminous infrared galaxies (LIRGs
with total IR luminosities $L_{IR} \sim 10^{11}-10^{12}\Lsol$ and
ULIRGs with $L_{IR} \gsim 10^{12}\Lsol$), and providing the first
hints of strong evolution in the number density of ULIRGs at recent
cosmic epochs \citep[e.g.][]{Wright84,Soifer87a,Sanders96}.  The next major
advance came with the discovery by \COBE\ of the cosmic far-IR
background which has an energy density comparable to that in the
optical/near-IR background \citep{Puget96,Hauser98}. This implies
that, over the history of the universe, as much energy has been
emitted by dust in galaxies as reaches us directly in starlight, after
dust extinction is taken into account. This discovery made apparent
the need to understand the IR as much as the optical emission from
galaxies in order to have a complete picture of galaxy evolution. In
particular, it is essential to understand IR emission from dust in
order to understand the cosmic history of star formation, since most
of the radiation from young stars must have been absorbed by dust over
the history of the universe, in order to account for the far-IR
background \citep[e.g.][]{Hauser98}.

Following these early discoveries, the \ISO\ satellite enabled the
first deep surveys of galaxies in the mid- and far-IR. The deepest of
these surveys were in the mid-IR at 15$\mum$, and probed the evolution
of LIRGs and ULIRGs out to $z\sim 1$, showing strong evolution in
these populations, and directly resolving most of the cosmic infrared
background at that wavelength
\citep{Elbaz99,Elbaz02,Gruppioni02}. Deep \ISO\ surveys in the far-IR
at 170$\mum$ \citep{Dole01,Patris03} probed lower redshifts, $z \sim
0.5$.  Around the same time, sub-mm observations using the \SCUBA\
instrument on the JCMT revealed a huge population of high-z ULIRGs
\citep{Smail97,Hughes98} which were subsequently found to have a
redshift distribution peaking at $z\sim 2$ \citep{Chapman05},
confirming the dramatic evolution in number density for this
population seen at shorter wavelengths and lower redshifts. The sub-mm
galaxies have been studied in more detail in subsequent \SCUBA\
surveys \citep[e.g. SHADES,][]{Mortier05}.

Now observations using the \SPITZER\ satellite \citep{Werner04}, with
its hugely increased sensitivity and mapping speed are revolutionizing
our knowledge of galaxy evolution at IR wavelengths from 3.6 to 160
$\mum$. \SPITZER\ surveys have allowed direct determinations of the
evolution of the galaxy luminosity function out to $z\sim 1$ in the
rest-frame near-IR and to $z\sim 2$ in the mid-IR
\citep{LeFloch05,Perez05,Babbedge06,Franceschini06}. Individual
galaxies have been detected by \SPITZER\ out to $z\sim 6$
\citep{Eyles05}. In the near future, the \HERSCHEL\ satellite
\citep{Pilbratt03} should make it possible to measure the far-IR
luminosity function out to $z\sim 2$, and thus directly measure the
total IR luminosities of galaxies over most of the history of the
universe.

Accompanying these observational advances, various types of
theoretical models have been developed to interpret or explain the
observational data on galaxy evolution in the IR. We can distinguish
three main classes of model: 

{\em (a) Purely phenomenological models:} In these models, the galaxy
luminosity function and its evolution are described by a purely
empirical expression, and this is combined with observationally-based
templates for the IR spectral energy distribution (SED). The free
parameters in the expression for the luminosity function are then
chosen to obtain the best match to some set of observational data,
such as number counts and redshift distributions in different IR
bands. These parameters are purely descriptive and provide little
insight into the physical processes which control galaxy
evolution. Examples of these models are
\citet{Pearson96,Xu98,Blain99,Franceschini01,Chary01,Rowan01,Lagache03,Gruppioni05}.

{\em (b) Hierarchical galaxy formation models with phenomenological
SEDs:} In these models, the evolution of the luminosity functions of
the stellar and total dust emission are calculated from a detailed
model of galaxy formation based on the cold dark matter (CDM) model of
structure formation, including physical modelling of processes such as
gas cooling and galaxy mergers.  The stellar luminosity of a model
galaxy is computed from its star formation history, and the stellar
luminosity absorbed by dust, which equals the total IR luminosity
emitted by dust, is calculated from this based on some treatment of
dust extinction. However, the SED shapes required to calculate the
distribution of the dust emission over wavelength from the total IR
dust emission are either observationally-based templates
\citep[e.g.][]{Guiderdoni98,Devriendt00} or are purely
phenomenological, e.g. a modified Planck function with an empirically
chosen dust temperature \citep[e.g.][]{Kaviani03}. In this approach,
the shape of the IR SED assumed for a model galaxy may be incompatible
with its other predicted properties, such as its dust mass and radius.

{\em (c) Hierarchical galaxy formation models with theoretical SEDs:}
These models are similar to those of type (b), in that the evolution
of the galaxy population is calculated from a detailed physical model
of galaxy formation based on CDM, but instead of using
phenomenological SEDs for the dust emission, the complete SED of each
model galaxy, from the far-UV to the radio, is calculated by combining
a theoretical stellar population synthesis model for the stellar
emission with a theoretical radiative transfer and dust heating model
to predict both the extinction of starlight by dust and the IR/sub-mm
SED of the dust emission. The advantages of this type of model are
that it is completely {\em ab initio}, with the maximum possible
theoretical self-consistency, and all of the model parameters relate
directly to physical processes. For example, the typical dust
temperature and the shape of the SED of dust emission depend on the
stellar luminosity and the dust mass, and evolution in all of these
quantities is computed self-consistently in this type of model.
Following this modelling approach thus allows more rigorous testing of
the predictions of physical models for galaxy formation against
observational data at IR wavelengths, as well as shrinking the
parameter space of the predictions. Examples of such models are
\citet{Granato00} and \citet{Baugh05}. (An alternative modelling
approach also based on theoretical IR SEDs but with a simplified
treatment of the assembly of galaxies and halos has been presented by
\citet{Granato04} and \citet{Silva05}.)

In this paper, we follow the third approach, with physical modelling
both of galaxy formation and of the galaxy SEDs, including the effects
of dust. This paper is the third in a series, where we combine the
\GALFORM\ semi-analytical model of galaxy formation \citep{Cole00}
with the \GRASIL\ model for stellar and dust emission from galaxies
\citep{Silva98}. The \GALFORM\ model computes the evolution of
galaxies in the framework of the $\Lambda$CDM model for structure formation,
based on physical treatments of the assembly of dark matter halos by
merging, gas cooling in halos, star formation and supernova feedback,
galaxy mergers, and chemical enrichment. The \GRASIL\ model computes
the SED of a model galaxy from the far-UV to the radio, based on
theoretical models of stellar evolution and stellar atmospheres,
radiative transfer through a two-phase dust medium to calculate both
the dust extinction and dust heating, and a distribution of dust
temperatures in each galaxy calculated from a detailed dust grain
model. In the first paper in the series \citep{Granato00}, we modelled
the IR properties of galaxies in the local universe. While this model
was very successful in explaining observations of the local universe,
we found subsequently that it failed when confronted with observations
of star-forming galaxies at high redshifts, predicting far too few
sub-mm galaxies (SMGs) at $z\sim 2$ and Lyman-break galaxies (LBGs) at
$z\sim 3$. Therefore, in the second paper \citep{Baugh05}, we proposed
a new version of the model which assumes a top-heavy IMF in starbursts
(with slope $x=0$, compared to Salpeter slope $x=1.35$), but a normal
solar neighbourhood IMF for quiescent star formation. In this new
model, the star formation parameters were also changed to force more
star formation to happen in bursts. This revised model agreed well
with both the number counts and redshift distributions of SMGs
detected at 850$\mum$, and with the rest-frame far-UV luminosity
function of LBGs at $z\sim 3$, while still maintaining consistency
with galaxy properties in the local universe such as the optical,
near-IR and far-IR luminosity functions, and gas fractions,
metallicities, morphologies and sizes.

This same model of \citet{Baugh05} was found by \citet{LeD05a,LeD05b}
to provide a good match to the observed evolution of the population of
Ly$\alpha$-emitting galaxies over the redshift range $z\sim
3-6$. Support for the controversial assumption of a top-heavy IMF in
bursts came from the studies of chemical enrichment in this model by
\citet{Nagashima05a,Nagashima05b}, who found that the metallicities of
both the intergalactic gas in galaxy clusters and the stars in
elliptical galaxies were predicted to be significantly lower than
observed values if a normal IMF was assumed for all star formation,
but agreed much better if a top-heavy IMF in bursts was assumed, as in
\citeauthor{Baugh05}. In this third paper in the series, we extend the
\citet{Baugh05} model to make predictions for galaxy evolution in the
IR, and compare these predictions with observational data from
\SPITZER. We emphasize that all of the model parameters for the
predictions presented in this paper were fixed by \citeauthor{Baugh05}
prior to the publication of any results from \SPITZER, and we have not
tried to obtain a better fit to any of the \SPITZER\ data by adjusting
these parameters\footnote 
{A closely related model of galaxy formation
obtained by applying \GALFORM\ principles to the Millennium simulation
of \citet{Springel05} has recently been published by \citet{Bower06}.
This model differs from the current one primarily in that it includes
feedback from AGN activity, but does not have a top-heavy IMF in
bursts. We plan to investigate the IR predictions of this alternative
model in a subsequent paper.}.

Our goals in this paper are to test our model of galaxy evolution with
a top-heavy IMF in starbursts against observations of dust-obscured
star-forming galaxies over the redshift range $z\sim 0 - 2$, and also
to test our predictions for the evolution of the stellar populations
of galaxies against observational data in the rest-frame near- and
mid-IR. The plan of the paper is as follows: In
Section~\ref{sec:model}, we give an overview of the \GALFORM\ and
\GRASIL\ models, focusing on how the predictions we present later on
are calculated. In Section~\ref {sec:ncts}, we compare the galaxy
number counts predicted by our model with observational data in all 7
\SPITZER\ bands, from 3.6 to 160 $\mum$. In
Section~\ref{sec:lf-evoln}, we investigate galaxy evolution in the IR
in more detail, by comparing model predictions directly with galaxy
luminosity functions constructed from \SPITZER\ data.  In
Section~\ref{sec:mstar-sfr}, we present the predictions of our model
for the evolution of the galaxy stellar mass function and star
formation rate distribution, and investigate the insight our model
offers on how well stellar masses and star formation rates can be
estimated from \SPITZER\ data. We present our conclusions in
Section~\ref{sec:conc}. In the Appendix, we present model predictions
for galaxy redshift distributions in the different \SPITZER\ bands, to
assist in interpreting data from different surveys.


\section{Model}
\label{sec:model}

In this paper use the \GALFORM\ semi-analytical model to predict the
physical properties of the galaxy population at different redshifts,
and combine it with the \GRASIL\ spectrophotometric model to predict
the detailed SEDs of model galaxies. Both \GALFORM\ and \GRASIL\ have
been described in detail in various previous papers, but since the
descriptions of the different model components, as well as of our
particular choice of parameters, are spread among different papers, we
give an overview of both of these here. \GALFORM\ is described in
\S\ref{sec:GALFORM}, and \GRASIL\ in \S\ref{sec:GRASIL}.  Particularly
important features of our model are the triggering of starbursts by
mergers (discussed in \S\ref{sec:mergers}) and the assumption of a
top-heavy IMF in starbursts (discussed in \S\ref{sec:IMF}). We further
discuss the choice of model parameters in
\S\ref{sec:parameters}. Readers who are already familiar with the
\citet{Baugh05} model can skip straight to the results, starting in
\S\ref{sec:ncts}.

\subsection{\GALFORM\ galaxy formation model}
\label{sec:GALFORM}

We compute the formation and evolution of galaxies within the
framework of the $\Lambda$CDM model of structure formation using the
semi-analytical galaxy formation model \GALFORM. The general
methodology and approximations behind the \GALFORM\ model are set out
in detail in \citet{Cole00} (see also the review by
\citet{Baugh06}). In summary, the \GALFORM\ model follows the main
processes which shape the formation and evolution of galaxies. These
include: (i) the collapse and merging of dark matter halos; (ii) the
shock-heating and radiative cooling of gas inside dark halos, leading
to the formation of galaxy disks; (iii) quiescent star formation in
galaxy disks; (iv) feedback both from supernova explosions and from
photo-ionization of the IGM; (v) chemical enrichment of the stars and
gas; (vi) galaxy mergers driven by dynamical friction within common
dark matter halos, leading to the formation of stellar spheroids, and
also triggering bursts of star formation.  The end product of the
calculations is a prediction of the numbers and properties of galaxies
that reside within dark matter haloes of different masses. The model
predicts the stellar and cold gas masses of the galaxies, along with
their star formation and merger histories, their sizes and
metallicities.

The prescriptions and parameters for the different processes which we
use in this paper are identical to those adopted by \citet{Baugh05},
but differ in several important respects from \citet{Cole00}. All of
these parameters were chosen by comparison with pre-\SPITZER\
observational data. The background cosmology is a spatially flat CDM
universe with a cosmological constant, with ``concordance'' parameters
$\Omega_{m}=0.3$, $\Omega_{\Lambda}=0.7$, $\Omega_{b}=0.04$, and $h
\equiv H_0/(100\kms\Mpc^{-1})=0.7$. The amplitude of the initial
spectrum of density fluctuations is set by the r.m.s. linear
fluctuation in a sphere of radius 8$h^{-1}\Mpc$, $\sigma_8=0.93$. For
completeness, we now summarize the prescriptions and parameters used,
but give details mainly where they differ from those in
\citet{Cole00}, or where they are particularly relevant to predicting
IR emission from dust.

\subsubsection{Halo assembly histories}
As in \citet{Cole00}, we describe the assembly histories of dark
matter halos through halo merger trees which are calculated using a
Monte Carlo method based on the extended Press-Schechter approach
\citep[e.g.][]{LC93}. The process of galaxy formation is then
calculated separately for each halo merger tree, following the
baryonic physics in all of the separate branches of the tree. As has
been shown by \citet{Helly03}, the statistical properties of galaxies
calculated in semi-analytical models using these Monte Carlo merger
trees are very similar to those computed using merger trees extracted
directly from N-body simulations.

\subsubsection{Gas cooling in halos}
The cooling of gas in halos is calculated using the same simple
spherical model as in \citet{Cole00}. The diffuse gas in halos
(consisting of all of the gas which has not previously condensed into
galaxies) is assumed to be shock-heated to the halo virial temperature
when the halo is assembled, and then to cool radiatively by atomic
processes. The cooling time depends on radius through the gas density
profile, which is assumed to have a core radius which grows as gas is
removed from the diffuse phase by condensing into galaxies. The gas at
some radius $r$ in the halo then cools and collapses to the halo
centre on a timescale which is the larger of the cooling time $t_{\rm
cool}$ and the free-fall time $t_{\rm ff}$ at that radius. Thus, for
$t_{\rm cool}(r) > t_{\rm ff}(r)$, we have {\em hot accretion}, and
for $t_{\rm cool}(r) < t_{\rm ff}(r)$, we have {\em cold accretion}
\footnote{Note that contrary to claims by \citet{Birnboim03}, 
the process of ``cold accretion'', if not the name,
has always been part of semi-analytical models (see \citet{Croton06}
for a detailed discussion)}. In our model, gas only accretes onto the
{\em central} galaxy in a halo, not onto any {\em satellite} galaxies
which share that halo. We denote all of the diffuse gas in halos as
``hot'', and all of the gas which has condensed into galaxies as
``cold''.

\subsubsection{Star formation timescale in disks}
The global rate of star formation $\psi$ in galaxy disks is assumed to
be related to the cold gas mass, $M_{\rm gas}$, by $\psi = M_{\rm
gas}/\tau_{\rm *,disk}$, where the star formation timescale is taken to be
\begin{equation}
\tau_{\rm *,disk} = \tau_{*0} \left( V_{\rm c}/200\, \kms
\right)^{\alpha_*},
\label{eq:taustar_disk}
\end{equation}
where $V_{\rm c}$ is the circular velocity of the galaxy disk (at its
half-mass radius) and $\tau_{*0}$ is a constant. We adopt values
$\tau_{*0} = 8\, \Gyr$ and $\alpha_* = -3$, chosen to reproduce the
observed relation between gas mass and B-band luminosity for
present-day disk galaxies. As discussed in \citet{Baugh05}, this
assumption means that the disk star formation timescale is independent
of redshift (at a given $V_{\rm c}$), resulting in disks at high
redshift that are much more gas-rich than at low redshift, and have
more gas available for star formation in bursts triggered by galaxy
mergers at high redshift.

\subsubsection{Galaxy mergers and triggering of starbursts}
\label{sec:mergers}
In the model, all galaxies originate as central galaxies in some halo,
but can then become satellite galaxies if their host halo merges into
another halo.  Mergers can then occur between satellite and central
galaxies within the same halo, after dynamical friction has caused the
satellite galaxy to sink to the centre of the halo. Galaxy mergers can
produce changes in galaxy morphology and trigger bursts.  We classify
galaxy mergers according to the ratio of masses (including stars and
gas) $M_2/M_1 \leq 1$ of the secondary to primary galaxy involved. We
define mergers to be {\em major} or {\em minor} according to whether
$M_2/M_1 > f_{\rm ellip}$ or $M_2/M_1 < f_{\rm ellip}$
\citep{Kauffmann93}. In major mergers, any stellar disks in either the
primary or secondary are assumed to be disrupted, and the stars
rearranged into a spheroid. In minor mergers, the stellar disk in the
primary galaxy is assumed to remain intact, while all of the stars in
the secondary are assumed to be added to the spheroid of the
primary. We adopt a threshold $f_{\rm ellip}=0.3$ for major mergers,
consistent with the results of numerical simulations
\citep[e.g.][]{Barnes98}, which reproduces the observed present-day
fraction of spheroidal galaxies.  We assume that major mergers always
trigger a starburst if any gas is present. We also assume that minor
mergers can trigger bursts, if they satisfy both $M_2/M_1 > f_{\rm
burst}$ and the gas fraction in the disk of the primary galaxy exceeds
$f_{\rm gas,crit}$. Following \citet{Baugh05}, we adopt $f_{\rm
burst}=0.05$ and $f_{\rm gas,crit}=0.75$. The parameters for bursts in
minor mergers were motivated by trying to explain the number of sub-mm
galaxies. An important consequence of assuming
eqn.(\ref{eq:taustar_disk}) for the star formation timescale in disks,
combined with the triggering of starbursts in minor mergers, is that
the global star formation rate at high redshifts is dominated by
bursts, while that at low redshifts it is dominated by quiescent disks
(see \citeauthor{Baugh05} for a detailed discussion of these points).

In either kind of starburst, we assume that the burst consumes all of
the cold gas in the two galaxies involved in the merger, and that the
stars produced are added to the spheroid of the merger remnant. During
the burst, we assume that star formation proceeds according to the
relation $\psi = M_{\rm gas}/\tau_{\rm *,burst}$. For the burst timescale,
we assume
\begin{equation}
\tau_{\rm *,burst} = \max \left[ f_{\rm dyn}\tau_{\rm dyn,sph} ;
  \tau_{\rm *,burst,min} \right],
\end{equation}
where $\tau_{\rm dyn,sph}$ is the dynamical time in the newly-formed
spheroid. We adopt $f_{\rm dyn}=50$ and $\tau_{\rm *,burst,min} = 0.2 \,
\Gyr$ (these parameters were chosen by \citet{Baugh05} to allow a
simultaneous match to the sub-mm number counts and to the local
60$\mum$ luminosity function). The star formation rate in a burst thus
decays exponentially with time after the galaxy merger. It is assumed
to be truncated after 3 e-folding times (where the e-folding time
takes account of stellar recycling and feedback - see
\citet{Granato00} for details), with the remaining gas being ejected
into the galaxy halo at that time.

\subsubsection{Feedback from photo-ionization}
After the intergalactic medium (IGM) has been reionized at redshift
$z_{\rm reion}$, the formation of low-mass galaxies is inhibited, both
by the effect of the IGM pressure inhibiting collapse of gas into
halos, and by the reduction of gas cooling in halos due to the
photo-ionizing background. We model this in a simple way, by assuming
that for $z<z_{\rm reion}$, cooling of gas is completely suppressed in
halos with circular velocities $V_{\rm c} < V_{\rm crit}$. We adopt
$V_{\rm crit} = 60\,\kms$, based on the detailed modelling by
\citet{Benson02}. We assume in this paper that reionization occurs at
$z_{\rm reion}=6$, for consistency with \citet{Baugh05}, but
increasing this to $z_{\rm reion} \sim 10$ in line with the WMAP
3-year estimate of the polarization of the microwave background
\citep{Spergel07} has no significant effect on the model results
presented in this paper.

\subsubsection{Feedback from supernovae}
Photo-ionization feedback only affects very low mass galaxies. More
important for most galaxies is feedback from supernova explosions. We
assume that energy input from supernovae causes gas to be ejected from
galaxies at a rate
\begin{equation}
{\dot M}_{\rm ej} = \beta(\Vc)\, \psi = \left[ \beta_{\rm reh}(\Vc) +  \beta_{\rm sw}(\Vc) \right]\, \psi
\label{eq:SNfeedback}
\end{equation}
The supernova feedback is assumed to operate for both quiescent star
formation in disks and for starbursts triggered by galaxy mergers,
with the only difference being that we take $\Vc$ to be the circular
velocity at the half mass radius of the disk in the former case, and
at the half-mass radius of the spheroid in the latter case. For
simplicity, we keep the same feedback parameters for starbursts as for
quiescent star formation.

The supernova feedback has two components: the {\em reheating} term
$\beta_{\rm reh}\psi$ describes gas which is reheated and ejected into
the galaxy halo, from where it is allowed to cool again after the halo
mass has doubled through hierachical mass build-up. For this, we use
the parametrization of \citet{Cole00}:
\begin{equation}
\beta_{\rm reh} = \left( \Vc/V_{\rm hot} \right)^{-\alpha_{\rm hot}},
\end{equation}
where we adopt parameter values $V_{\rm hot} = 300\,\kms$ and
$\alpha_{\rm hot}=2$. The {\em reheating} term has the largest effect
on low-mass galaxies, for which ejection of gas from galaxies flattens
the faint-end slope of the galaxy luminosity function.

The second term $\beta_{\rm sw}\psi$ in eqn.(\ref{eq:SNfeedback}) is
the {\em superwind} term, which describes ejection of gas out of the
halo rather than just the galaxy. Once ejected, this gas is assumed
never to re-accrete onto any halo. We model the superwind ejection
efficiency as
\begin{equation}
\beta_{\rm sw} = f_{\rm sw} \min\left[ 1, \left(\Vc/V_{\rm sw}\right)^{-2} \right]
\end{equation}
based on \citet{Benson03}. We adopt parameter values $f_{\rm sw}=2$
and $V_{\rm sw} = 200\,\kms$, as in \citet{Baugh05}.  The {\em
superwind} term mainly affects higher mass galaxies, where the
ejection of gas from halos causes an increase in the cooling time of
gas in halos by reducing the gas densities. This brings the predicted
break at the bright end of the local galaxy luminosity function into
agreement with observations, as discussed in \citet{Benson03}. The
various parameters for supernova feedback are thus chosen in order to
match the observed present-day optical and near-IR galaxy luminosity
functions, as well as the galaxy metallicity-luminosity relation.

We note that the galaxy formation model in this paper, unlike some
other recent semi-analytical models, does not include AGN
feedback. Instead, the role of AGN feedback in reducing the amount of
gas cooling to form massive galaxies is taken by superwinds driven by
supernova explosions.  The first semi-analytical model to include AGN
feedback was that of \citet{Granato04}, who introduced a detailed
model of feedback from QSO winds during the formation phase of
supermassive black holes (SMBHs), with the aim of explaining the
co-evolution of the spheroidal components of galaxies and their
SMBHs. The predictions of the \citeauthor{Granato04} model for number
counts and redshift distributions in the IR have been computed by
\citet{Silva05} using the \GRASIL\ spectrophotometric model, and
compared to \ISO\ and \SPITZER\ data. However, the \citet{Granato04}
model has the limitations that it does not include the merging of
galaxies or of dark halos, and does not treat the formation and
evolution of galactic disks. More recently, several semi-analytical
models have been published which propose that heating of halo gas by
relativistic jets from an AGN in an optically inconspicuous or
``radio'' mode can balance radiative cooling of gas in high-mass
halos, thus suppressing {\em hot accretion} of gas onto galaxies
\citep{Bower06, Croton06, Cattaneo06, Monaco07}. However, these AGN
feedback models differ in detail, and all are fairly schematic. None
of these models has been shown to reproduce the observed number counts
and redshifts of the faint sub-mm galaxies.

 The effects of our superwind feedback are qualitatively quite similar
to those of the radio-mode AGN feedback. Both superwind and AGN
feedback models contain free parameters, which are adjusted in order
to make the model fit the bright end of the observed present-day
galaxy luminosity function at optical and near-IR
wavelengths. However, since the physical mechanisms are different,
they make different predictions for how the galaxy luminosity function
should evolve with redshift. Current models for the radio-mode AGN
feedback are very schematic, but they have the advantage over the
superwind model that the energetic constraints are greatly relaxed,
since accretion onto black holes can convert mass into energy with a
much higher efficiency than can supernova explosions. We will
investigate the predictions of models with AGN feedback for the IR and
sub-mm evolution of galaxies in a future paper.

\subsubsection{The Stellar Initial Mass Function and Chemical
  Evolution}
\label{sec:IMF}

Stars in our model are assumed to form with different Initial Mass
Functions (IMFs), depending on whether they form in disks or in
bursts. Both IMFs are taken to be piecewise power laws, with slopes
$x$ defined by ${\rm d}N/{\rm d}\ln m \propto m^{-x}$, with $N$ the
number of stars and $m$ the stellar mass (so the Salpeter slope is
$x=1.35$), and covering a stellar mass range $0.15 < m <
120\Msol$. Quiescent star formation in galaxy disks is assumed to have
a solar neighbourhood IMF, for which we use the \citet{Kennicutt83}
paramerization, with slope $x=0.4$ for $m< \Msol$ and $x=1.5$ for $m>
\Msol$. (The \citet{Kennicutt83} IMF is similar to other popular
parametrizations of the solar neighbourhood IMF, such as that of
\citet{Kroupa01}.) Bursts of star formation triggered by galaxy
mergers are assumed to form stars with a top-heavy IMF with slope
$x=0$. As discussed in detail in \citet{Baugh05}, the top-heavy IMF in
bursts was found to be required in order to reproduce the observed
number counts and redshift distributions of the faint sub-mm
galaxies. Furthermore, as shown by \citet{Nagashima05a,Nagashima05b},
the predicted chemical abundances of the X-ray emitting gas in galaxy
clusters and of the stars in elliptical galaxies also agree better
with observational data in a model with the top-heavy IMF in bursts,
rather than a universal solar neighbourhood IMF.

A variety of other observational evidence has accumulated which
suggests that the IMF in some environments may be top-heavy compared
to the solar neighbourhood IMF. \citet{Rieke93} argued for a top-heavy
IMF in the nearby starburst M82, based on modelling its integrated
properties, while \citet{Parra07} found possible evidence for a
top-heavy IMF in the ultra-luminous starburst Arp220 from the relative
numbers of supernovae of different types observed at radio wavelengths.
Evidence has been found for a top-heavy IMF in some star
clusters in intensely star-forming regions, both in M82
\citep[e.g.][]{Mcgrady03}, and in our own Galaxy
\citep[e.g.][]{Figer99,Stolte05,Harayama07}. Observations of both the
old and young stellar populations in the central 1~pc of our Galaxy
also favour a top-heavy IMF
\citep{Paumard06,Maness07}. \citet{Fardal06} found that reconciling
measurements of the optical and IR extragalactic background with
measurements of the cosmic star formation history also seemed to
require an average IMF that was somewhat top-heavy. Finally,
\citet{Dokkum07} found that reconciling the colour and luminosity
evolution of early-type galaxies in clusters also favoured a top-heavy
IMF. \citet{Larson98} summarized other evidence for a top-heavy IMF
during the earlier phases of galaxy evolution, and argued that this
could be a natural consequence of the temperature-dependence of the
Jeans mass for gravitational instability in gas
clouds. \citet{Larson05} extended this to argue that a top-heavy IMF
might also be expected in starburst regions, where there is strong
heating of the dust by the young stars.

In our model, the fraction of star formation occuring
in the burst mode increases with redshift (see \citet{Baugh05}), so
the average IMF with which stars are being formed shifts from being
close to a solar neighbourhood IMF at the present-day to being very
top-heavy at high redshift.  In this model, 30\% of star formation
occured in the burst mode when integrated over the past history of the
universe, but only 7\% of the current stellar mass was formed in
bursts, because of the much larger fraction of mass recycled by dying
stars for the top-heavy IMF. We note that our predictions for the IR
and sub-mm luminosities of starbursts are not sensitive to the precise
form of the top-heavy IMF, but simply require a larger fraction of $m
\sim 5-20 \Msol$ stars relative to a solar neighbourhood IMF.

In this paper, we calculate chemical evolution using the instantaneous
recycling approximation, which depends on the total fraction of mass
recycled from dying stars ($R$), and the total yield of heavy elements
($p$). Both of these parameters depend on the IMF. We use the results
of stellar evolution computations to calculate values of $R$ and $p$
consistent with each IMF (see \citet{Nagashima05a} for details of the
stellar evolution data used). Thus, we use $R=0.41$ and $p=0.023$ for
the quiescent IMF, and $R=0.91$ and $p=0.15$ for the burst IMF. Our
chemical evolution model then predicts the masses and total
metallicities of the gas and stars in each galaxy as a function of
time.

\subsubsection{Galaxy sizes and dust masses}
\label{sec:dustmass}
For calculating the extinction and emission by dust, it is essential
to have an accurate calculation of the dust optical depths in the
model galaxies, which in turn depends on the mass of dust and the size
of the galaxy. The dust mass is calculated from the gas mass and
metallicity predicted by the chemical enrichment model, assuming that
the dust-to-gas ratio is proportional to metallicity, normalized to
match the local ISM value at solar metallicity.  The sizes of galaxies
are computed exactly as in \citet{Cole00}: gas which cools in a halo
is assumed to conserve its angular momentum as it collapses, forming a
rotationally-supported galaxy disk; the radius of this disk is then
calculated from its angular momentum, including the gravity of the
disk, spheroid (if any) and dark halo. Galaxy spheroids are built up
both from pre-existing stars in galaxy mergers, and from the stars
formed in bursts triggered by these mergers; the radii of spheroids
formed in mergers are computed using an energy conservation argument.
In calculating the sizes of disks and spheroids, we include the
adiabatic contraction of the dark halo due to the gravity of the
baryonic components.  This model was tested for disks by
\citet{Cole00} and for spheroids by \citet{Almeida07} (see also Coenda
\etal in preparation, and Gonzalez \etal in preparation). During a
burst, we assume that the gas and stars involved in the burst have a
distribution with the same half-mass radius as the spheroid
(i.e. $\eta=1$ in the notation of \citet{Granato00}, who used a value
$\eta=0.1$).

\subsection{\GRASIL\ model for stellar and dust emission}
\label{sec:GRASIL}

For each galaxy in our model, we compute the spectral energy
distribution using the spectrophotometric model \GRASIL\ 
\citep{Silva98,Granato00}. \GRASIL\ computes the emission
from the stellar population, the absorption and emission of radiation
by dust, and also radio emission (thermal and synchrotron) powered by
massive stars \citep{Bressan02}.

\subsubsection{SED model}
The main features of the \GRASIL\ model are as follows:\\
(i) The stars
are assumed to have an axisymmetric distribution in a disk and a
bulge. Given the distribution of stars in age and metallicity
(obtained from the star formation and chemical enrichment history),
the SED of the stellar population is calculated using a population
synthesis model based on the Padova stellar evolution tracks and
Kurucz model atmospheres \citep{Bressan98}. This is done separately
for the disk and bulge.\\
(ii) The cold gas and dust in a galaxy are
assumed to be in a 2-phase medium, consisting of dense gas in giant
molecular clouds embedded in a lower-density diffuse component. In a
quiescent galaxy, the dust and gas are assumed to be confined to the
disk, while for a galaxy undergoing a burst, the dust and gas are
confined to the spheroidal burst component. \\
(iii) Stars are assumed to
be born inside molecular clouds, and then to leak out into the diffuse
medium on a timescale $\tesc$. As a result, the youngest and most
massive stars are concentrated in the dustiest regions, so they
experience larger dust extinctions than older, typically lower-mass
stars, and dust in the clouds is also much more strongly heated than
dust in the diffuse medium. \\
(iv) The extinction of the starlight by
dust is computed using a radiative transfer code; this is used also to
compute the intensity of the stellar radiation field heating the dust
at each point in a galaxy. \\
(v) The dust is modelled as a mixture of
graphite and silicate grains with a continuous distribution of grain
sizes (varying between 8\AA\ and 0.25 $\mum$), and
also Polycyclic Aromatic Hydrocarbon (PAH) molecules with a
distribution of sizes. The equilibrium temperature in the local
interstellar radiation field is calculated for each type and size of
grain, at each point in the galaxy, and this information is then used
to calculate the emission from each grain. In the case of very small
grains and PAH molecules, temperature fluctuations are important, and
the probability distribution of the temperature is calculated. The
detailed spectrum of the PAH emission is obtained using the PAH
cross-sections from \citet{Li01}, as described in \citet{vega05}. The
grain size distribution is chosen to match the mean dust extinction
curve and emissivity in the local ISM, and is not varied, except that
the PAH abundance in molecular clouds is assumed to be $10^{-3}$ of
that in the diffuse medium \citep{vega05}. \\
(vi) Radio emission from
ionized gas in HII regions and from synchrotron radiation from
relativistic electrons accelerated in supernova remnant shocks are
calculated as described in \citet{Bressan02}. 

The output from \GRASIL\ is then the complete SED of a galaxy from the
far-UV to the radio (wavelengths $100\AA \lsim \lambda \lsim
1\m$). The SED of the dust emission is computed as a sum over the
different types of grains, having different temperatures depending on
their size and their position in the galaxy. The dust SED is thus
intrinsically multi-temperature. \GRASIL\ has been
shown to give an excellent match to the measured SEDs of both
quiescent (e.g. M51) and starburst (e.g. M82) galaxies
\citep{Silva98,Bressan02}.

The assumption of axisymmetry in \GRASIL\ is a limitation when
considering starbursts triggered by galaxy mergers. However,
observations of local ULIRGs imply that most of the star formation
happens in a single burst component after the galaxy merger is
substantially complete, so the assumption of axisymmetry for the burst
component may not be so bad.

\subsubsection{\GRASIL\ parameters}
The main parameters in the \GRASIL\ dust model are the fraction
$f_{\rm mc}$ of the cold gas which is in molecular clouds, the
timescale $\tesc$ for newly-formed stars to escape from their parent
molecular cloud, and the cloud masses $M_c$ and radii $r_c$ in the
combination $M_c/r_c^2$, which determines the dust optical depth of
the clouds. We assume $f_{\rm mc}=0.25$, $M_c = 10^6\Msol$ and $r_c =
16\pc$ as in \citet{Granato00}, and also adopt the same geometrical
parameters as in that paper. We make the following two changes in
\GRASIL\ parameters relative to \citeauthor{Granato00}, as discussed
in \citet{Baugh05}: (a) We assume $\tesc = 1\,\Myr$ in both disks and
bursts (instead of the \citeauthor{Granato00} values $\tesc = 2$ and
$10\,\Myr$ respectively). This value was chosen in order to obtain a
better match of the predicted rest-frame far-UV luminosity function of
galaxies at $z\sim 3$ to that measured for Lyman-break galaxies. (b)
The dust emissivity law in bursts at long wavelengths is modified from
$\epsilon_{\nu} \propto \nu^{-2}$ to $\epsilon_{\nu} \propto
\nu^{-1.5}$ for $\lambda > 100\mum$. This was done in order to improve
slightly the fit of the model to the observed sub-mm number counts. In
applying \GRASIL\ to model the SEDs of a sample of nearby galaxies,
\citet{Silva98} found that a similar modification (to $\epsilon_{\nu}
\propto \nu^{-1.6}$) seemed to be required in the case of Arp220 (the
only ultra-luminous starburst in their sample), in order to reproduce
the observed sub-mm data for that galaxy. This modification in fact
has little effect on the IR predictions presented in the present
paper, but we retain it for consistency with \citet{Baugh05}.

\subsubsection{Interface with \GALFORM}
For calculating the statistical properties of the galaxy population
from the combined \GALFORM+\GRASIL\ model, we follow the same strategy
as described in \citet{Granato00}. We first run the \GALFORM\ code to
generate a large catalogue of model galaxies at any redshift, and then
run the \GRASIL\ code on subsamples of these. For the quiescent
galaxies, we select a subsample which has equal numbers of galaxies in
equal logarithmic bins of stellar mass, while for the bursting
galaxies, we select a subsample with equal numbers of galaxies in
equal logarithmic bins of burst mass. For the burst sample, we compute
SEDs at several different representative stages in the burst
evolution, while for the quiescent sample, we only compute SEDs at a
single epoch. Using this sampling strategy, we obtain a good
coverage of all the different masses, types and evolutionary stages of
galaxies, while minimizing the computational cost of running the
\GRASIL\ code. The statistical properties of the galaxy population are
then obtained by assigning the model galaxies appropriate weights
depending on their predicted number density in a representative
cosmological volume. 

The outputs from the \GALFORM\ galaxy formation model required by
\GRASIL\ to calculate the galaxy SEDs are: the combined star formation
history and metallicity distribution for the disk and bulge, the radii
of both components, and the total mass of dust. The dust mass is
calculated from the mass and metallicity of the cold gas in the
galaxy, assuming that the dust-to-gas ratio is proportional to the
metallicity. Since the gas mass and metallicity both evolve, so does
the dust mass, and this evolution is fully taken into account in
\GRASIL. For simplicity, we assume that the size distribution of the
dust grains and PAH molecules does not evolve, apart from the
normalization.

Once we have calculated the SEDs for the model galaxies, we compute
luminosities in different observed bands (e.g. the optical B-band or
the \SPITZER\ 24$\mum$ band) by convolving the SED with the
filter+detector response function for that band. For computing the
predicted fluxes from galaxies in a fixed observer-frame band, we
redshift the SED before doing the convolution.

The \GRASIL\ code is quite CPU-intensive, requiring several minutes of
CPU time per galaxy. Consequently, we are limited to running samples
of a few thousand galaxies at each redshift. As a result, quantities
such as luminosity functions and redshift distributions still show
some small amount of noise, rather than being completely smooth
curves, as can be seen in many of the figures in this paper.

\subsection{Choice of parameters in the \GALFORM+\GRASIL\ model}
\label{sec:parameters}

The combined \GALFORM+\GRASIL\ model has a significant number of
parameters, but this is inevitable given the very wide range of
physical processes which are included. The parameters are constrained
by requiring the model predictions to reproduce a limited set of
observational data - once this is done, there is rather little freedom
in the choice of parameters. We have described above how the main
parameters are fixed, and more details can be found in \citet{Cole00}
and \citet{Baugh05}. For both of these papers, large grids of GALFORM
models were run with different parameters, in order to decide which
set of parameters gave the best overall fit to the set of calibrating
observational data. These papers also show the effects of varying some
of the main model parameters around their best-fit values.  The
parameters in the standard model for which we present results in this
paper were chosen to reproduce the following properties for
present-day galaxies: the luminosity functions in the B- and K-bands
and at 60$\mum$, the relations between gas mass and luminosity and
metallicity and luminosity, the size-luminosity relation for galaxy
disks, and the fraction of spheroidal galaxies. In addition, the model
was required to reproduce the observed rest-frame far-UV (1500\AA)
luminosity function at $z=3$, and the observed sub-mm number counts
and redshift distribution at 850$\mum$ \citep{Baugh05}. The sub-mm
number counts are the main factor driving the need to include a
top-heavy IMF in bursts.

The parameters for our standard model are exactly the same as in
\citet{Baugh05}, which were chosen before \SPITZER\ data became
available. Since these parameters were not adjusted to match any data
obtained with \SPITZER, the predictions of our model in the \SPITZER\
bands are genuine predictions. We could obviously have fine-tuned our
parameters in order to match better the observational data we
considered in this paper, but this would have conflicted with our main
goal, which is to present predictions for a wide set of observable
properties based on a single physical model in a series of papers.

Since our assumption of a top-heavy IMF in bursts is a controversial
one, we will also show some predictions from a variant model, which is
identical to the standard model, except that we assume the same solar
neighbourhood (Kennicutt) IMF in bursts and in disks. Comparing the
predictions for the standard and variant models then shows directly
the effects of changing the IMF in bursts. We note that the variant
model matches the present-day optical and near-IR luminosity functions
almost as well as the standard model, though it is a poorer fit to the
local 60$\mum$ luminosity function for the brightest galaxies (see
Fig.~\ref{fig:lf60}). The variant model underpredicts the 850$\mum$
counts by a factor of 10--30.


\begin{figure*}

\begin{center}

\begin{minipage}{7cm}
\includegraphics[width=7cm]{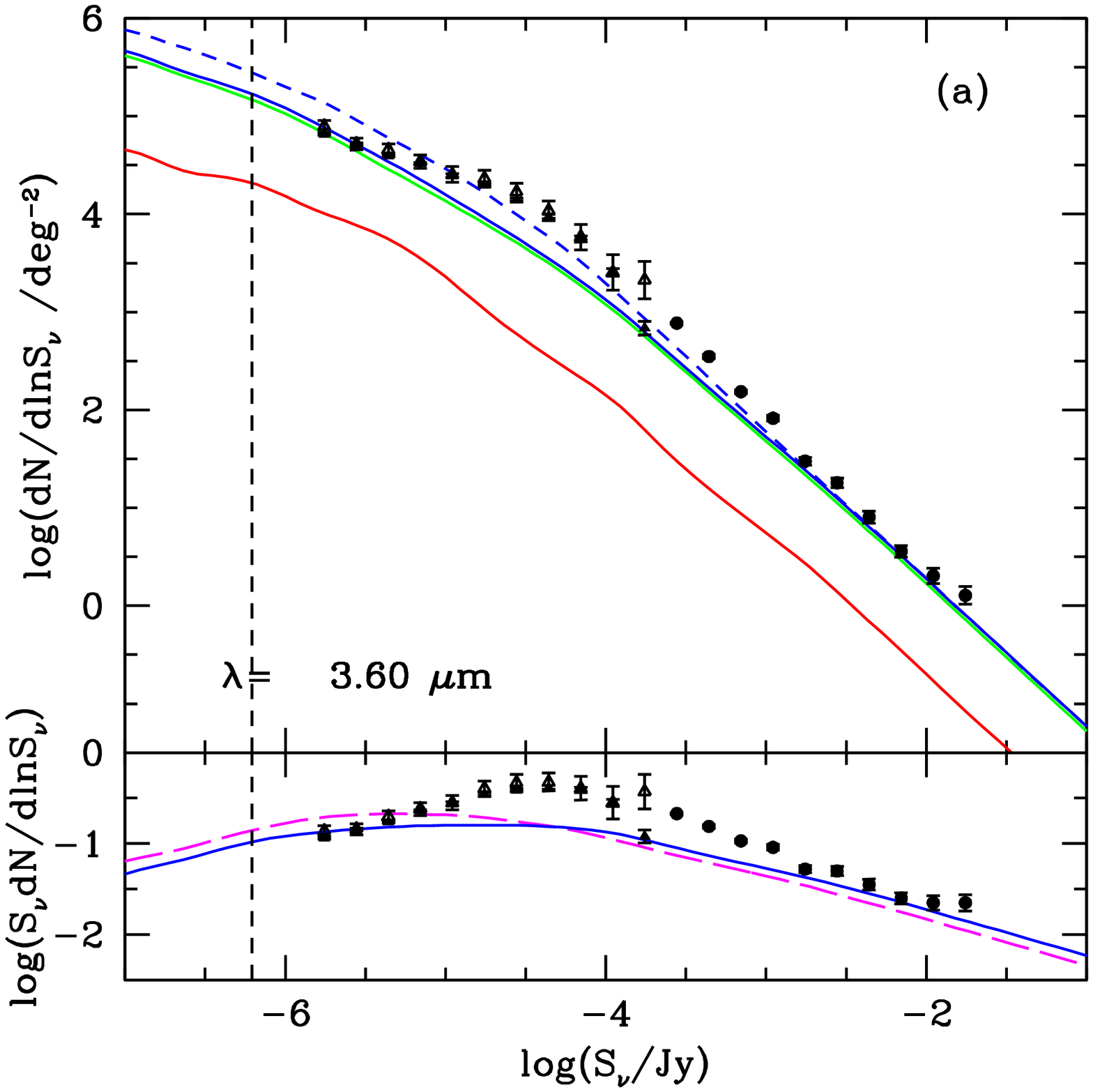}
\end{minipage}
\hspace{1cm}
\begin{minipage}{7cm}
\includegraphics[width=7cm]{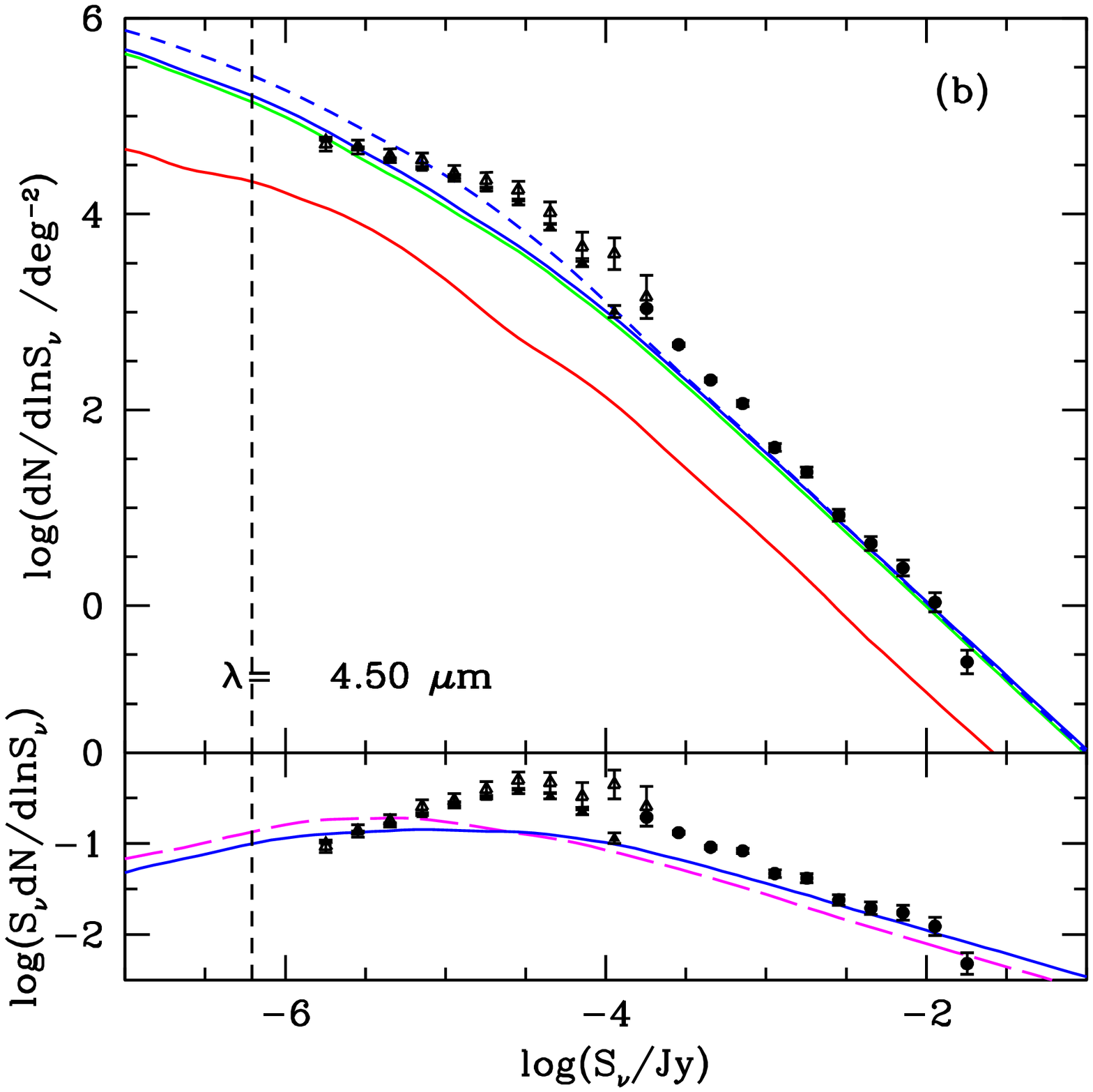}
\end{minipage}

\begin{minipage}{7cm}
\includegraphics[width=7cm]{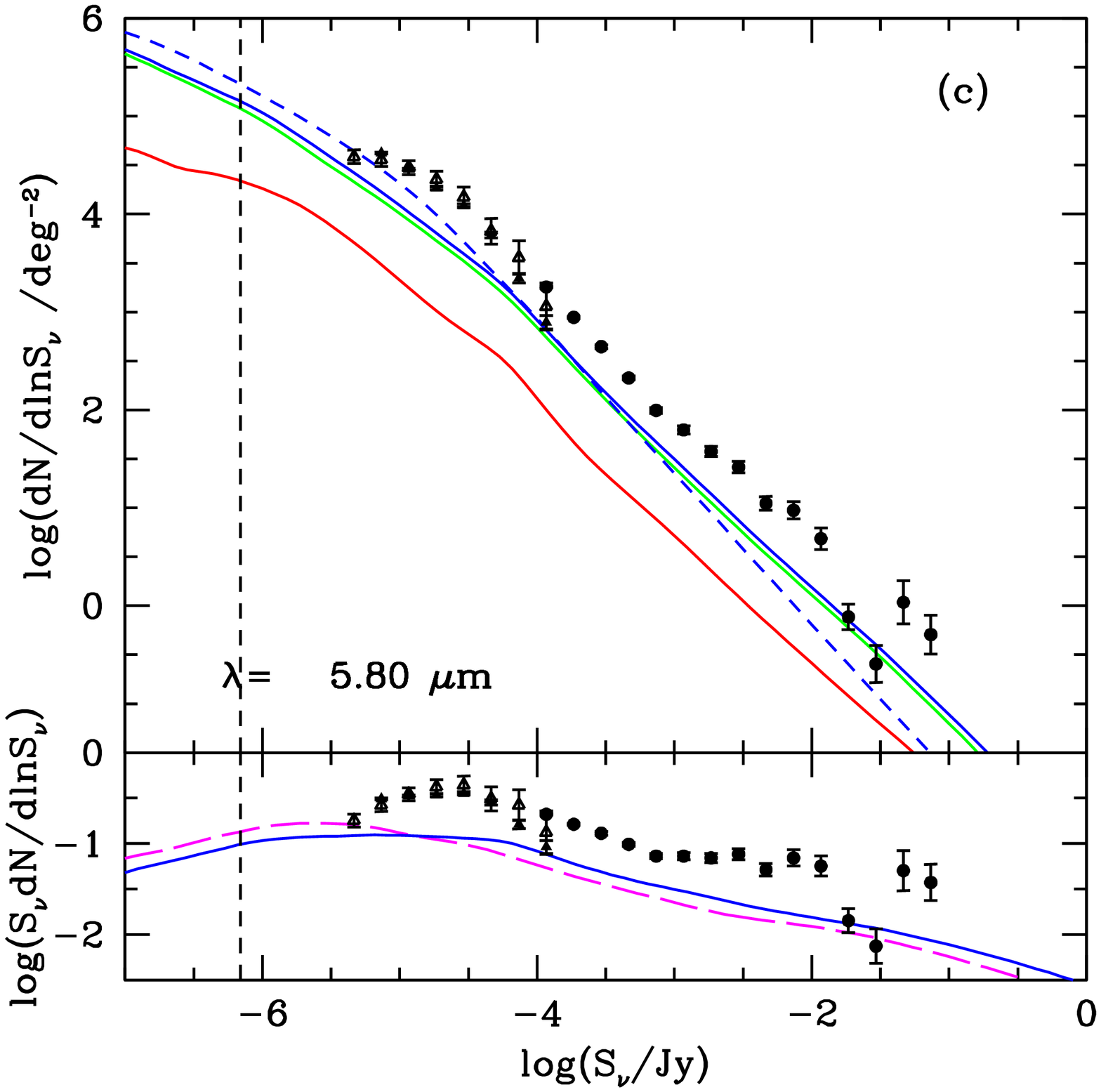}
\end{minipage}
\hspace{1cm}
\begin{minipage}{7cm}
\includegraphics[width=7cm]{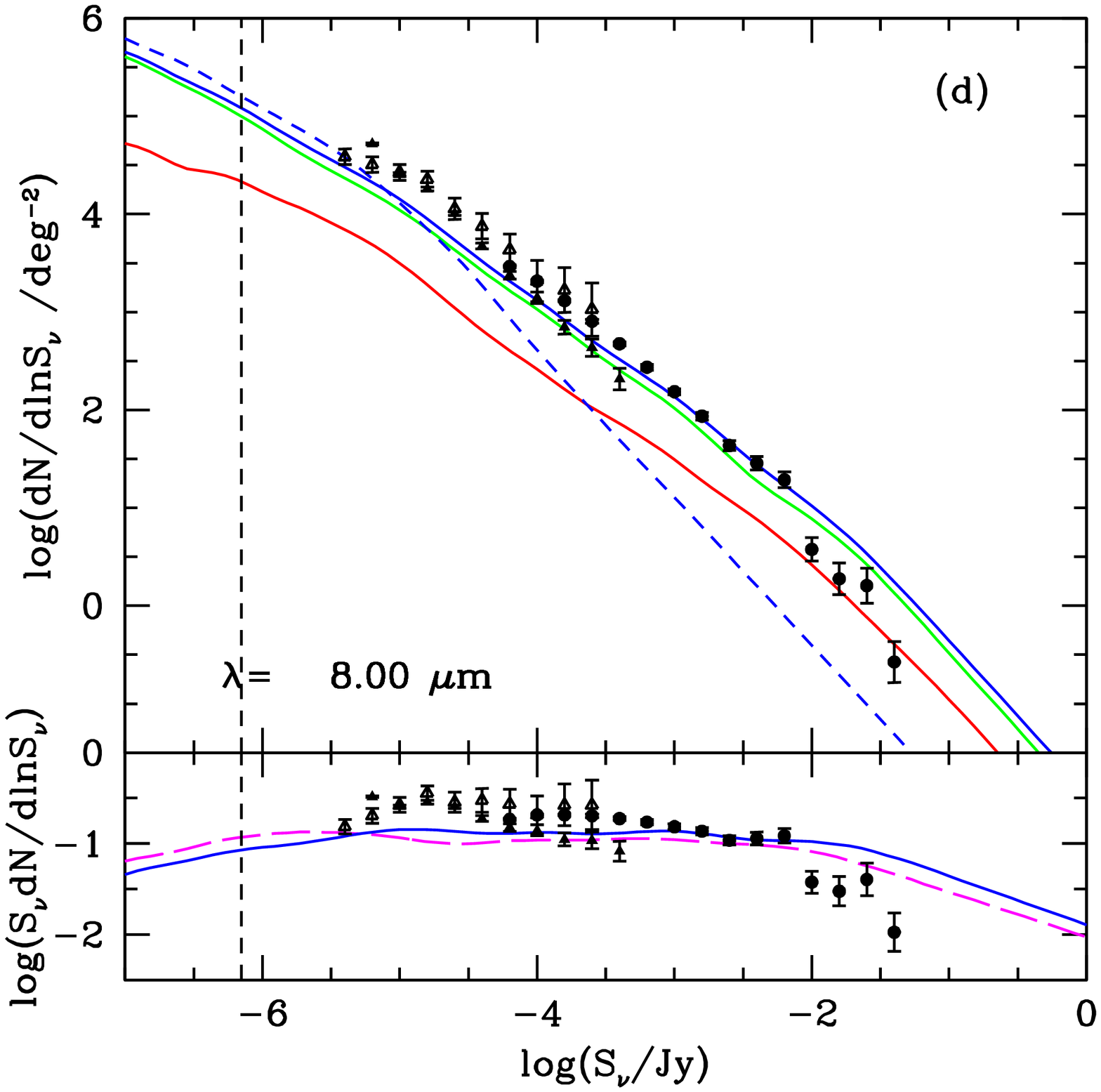}
\end{minipage}

\end{center}

\caption{Galaxy differential number counts in the four IRAC bands. The
  curves show model predictions, while the symbols with error bars
  show observational data from \citet{Fazio04} (with different symbols
  for data from different survey fields). Each panel is split in two:
  the upper sub-panel plots the counts as $dN/d\ln S_{\nu}$ vs
  $S_{\nu}$, while the lower sub-panel plots $S_{\nu} dN/d\ln S_{\nu}$
  (in units $\Jy \deg^{-2}$) on the same horizontal scale. The upper
  sub-panels show four different curves for our standard model - solid
  blue: total counts including dust extinction and emission; dashed
  blue: total counts excluding interstellar dust; solid red: ongoing
  bursts (including dust); solid green: quiescent galaxies (including
  dust). The lower sub-panels compare the total counts including dust
  for the standard model (solid blue line) with those for a variant
  model with a normal IMF for all stars (dashed magenta line). The
  vertical dashed line shows the estimated confusion limit for the
  model. (a) 3.6 $\mum$. (b) 4.5 $\mum$. (c) 5.8 $\mum$. (d) 8.0
  $\mum$. }

\label{fig:ncts-IRAC}
\end{figure*}

\section{Number counts}
\label{sec:ncts}

\begin{figure}

\begin{center}

\includegraphics[width=7cm]{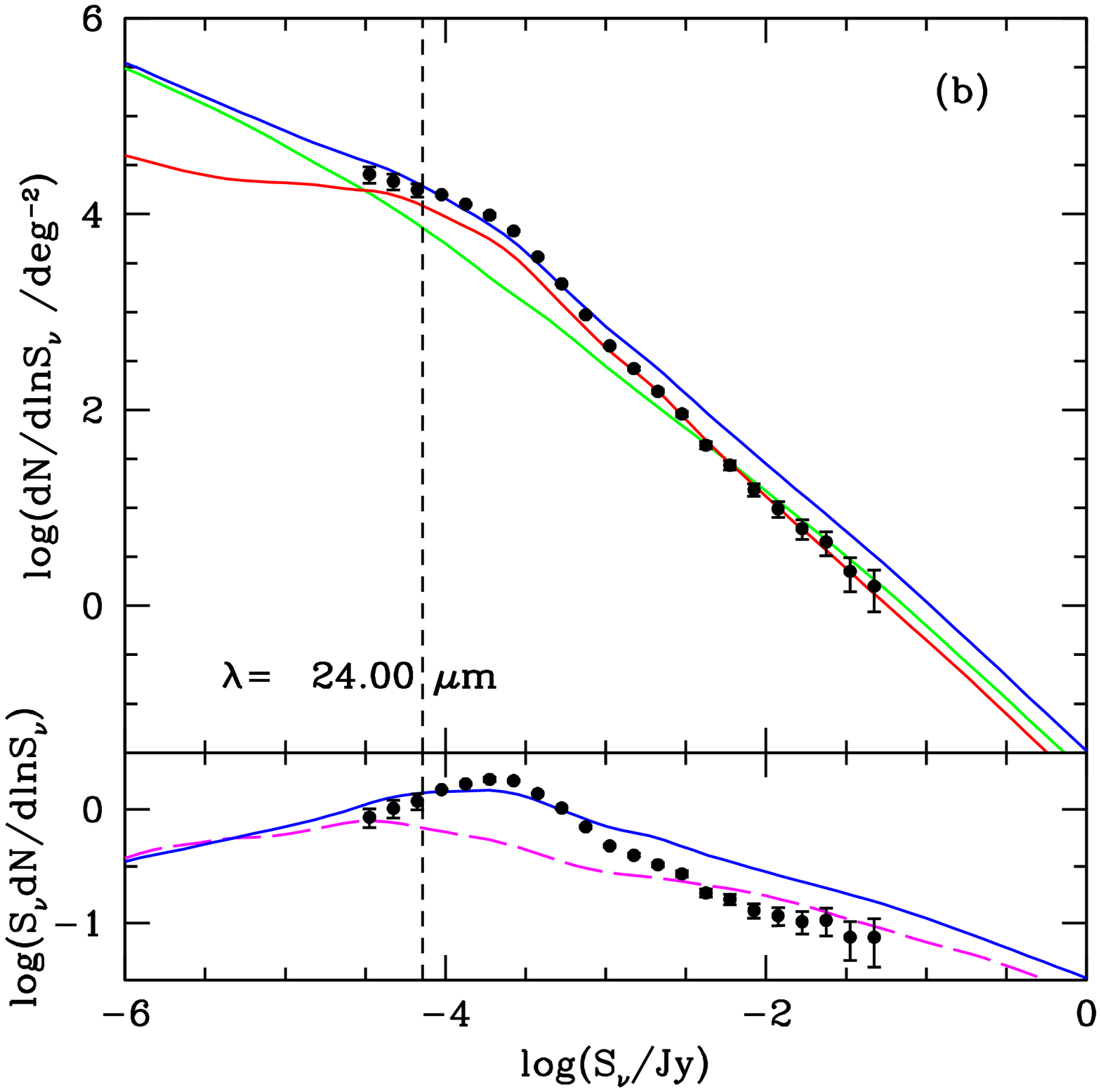}
\includegraphics[width=7cm]{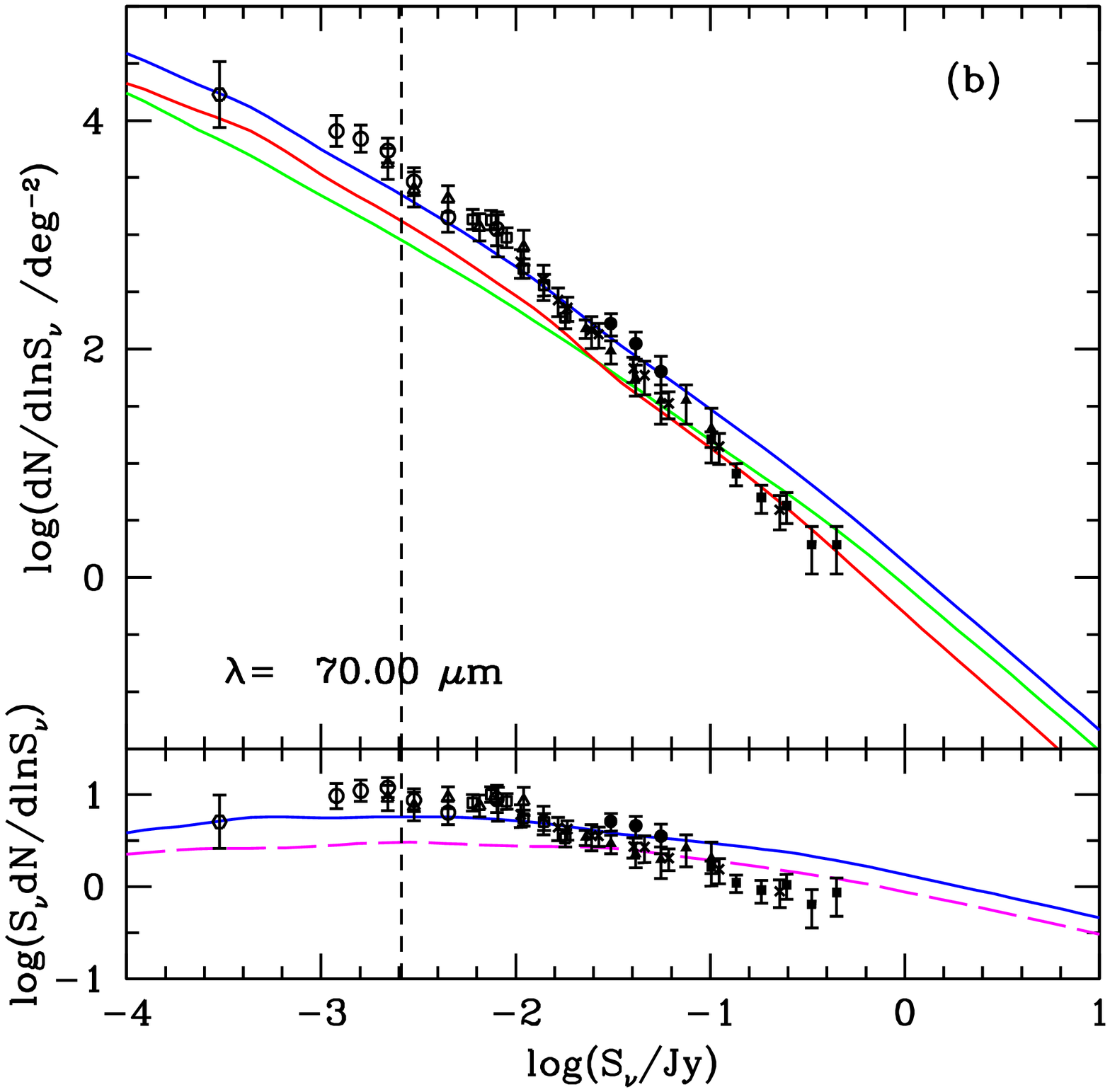}
\includegraphics[width=7cm]{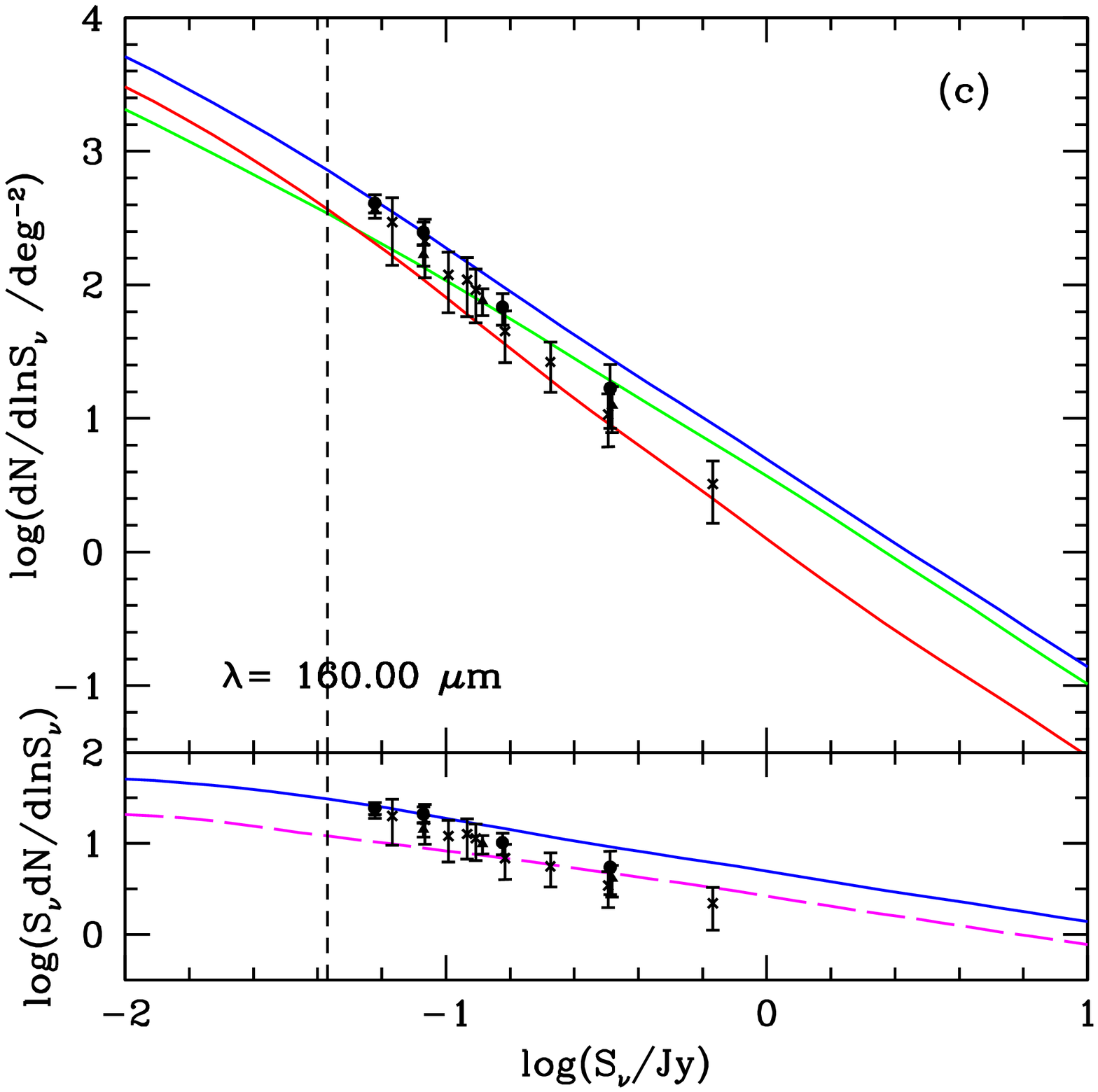}

\end{center}

\caption{Galaxy differential number counts in the three MIPS
  bands. The curves show model predictions while the symbols with
  error bars show observational data. The meaning of the different
  model lines is the same as in Fig.~\ref{fig:ncts-IRAC}. (a) 24
  $\mum$, with observational data from \citet{Papovich04}. (b) 70
  $\mum$, with observational data from \citet{Dole04} (filled
  symbols), \citet{Frayer06a} (crosses), and \citet{Frayer06b} (open
  symbols). (c) 160 $\mum$ (bottom panel), with observational data
  from \citet{Dole04} (filled symbols) and \citet{Frayer06a}
  (crosses).}

\label{fig:ncts-MIPS}
\end{figure}

We begin our comparison of the predictions of our galaxy formation
model against \SPITZER\ data with the galaxy number
counts. Fig.~\ref{fig:ncts-IRAC} shows number counts in the four IRAC
bands (3.6, 4.5, 5.8 and 8.0 $\mum$), and Fig.~\ref{fig:ncts-MIPS}
does the same for the three MIPS bands (24, 70 and 160 $\mum$). Each
panel is split in two: the upper sub-panel plots the counts per
logarithmic flux interval, $dN/d\ln S_{\nu}$, while the lower
sub-panel instead plots $S_{\nu} dN/d\ln S_{\nu}$. The latter is
designed to take out much of the trend with flux, in order to show
more clearly the differences between the model and the onservational
data.  In each case we plot three curves for our standard model: the
solid blue line shows the total number counts including both
extinction and emission by dust, the solid red line shows the
contribution to this from galaxies currently forming stars in a burst,
and the solid green line shows the contribution from all other
galaxies (star-forming or not), which we denote as ``quiescent''. In
Fig.~\ref{fig:ncts-IRAC}, we also plot a dashed blue line which shows
the predicted total counts if we ignore absorption and emission from
interstellar dust (emission from dust in the envelopes of AGB stars is
still included in the stellar contribution, however). In the MIPS
bands, the predicted counts are negligible in the absence of
interstellar dust, so we do not plot them in
Fig.~\ref{fig:ncts-MIPS}. In the lower sub-panels, we also show by a
dashed magenta line the prediction from a variant model which assumes
a normal (Kennicutt) IMF for all star formation, but is otherwise
identical to our standard model (which has a top-heavy IMF in
bursts). This variant model fits the local B- and K-band and 60 $\mum$
luminosity functions about as well as our standard model, but
dramatically underpredicts the 850 $\mum$ number counts.  The observed
number counts are shown by black symbols with error bars.

Overall, the agreement between the predictions of our standard model
and the observed counts is remarkably good, when one takes account of
the fact that no parameters of the model were adjusted to improve the
fit to any data from \SPITZER. Consider first the results for the IRAC
bands, shown in Fig.~\ref{fig:ncts-IRAC}.  Here, the agreement of the
model with observations seems best at 3.6 and 8.0 $\mum$, and somewhat
poorer at 5.8 $\mum$. The model predicts somewhat too few objects at
fainter fluxes in all of the IRAC bands.  Comparing the red and green
curves, we see that quiescent galaxies rather than bursts dominate the
counts at all observed fluxes in all of the IRAC bands, but especially
at the shorter wavelengths, consistent with the expectation that at
3.6 and 4.5 $\mum$, we are seeing mostly light from old stellar
populations. Comparing the solid and dashed blue lines, we see that
the effects of dust are small at 3.6 and 4.5 $\mum$, with a small
amount of extinction at faint fluxes (and thus higher average
redshifts), but negligible extinction for brighter fluxes (and thus
lower redshifts). On the other hand, dust has large effects at 8.0
$\mum$, with dust emission (due to strong PAH features at $\lambda
\sim 6-9 \mum$) becoming very important at bright fluxes (which
correspond to low average redshifts - see Fig.~\ref {fig:zmedian}(b)
in the Appendix). The 8.0 $\mum$ counts thus are predicted to be
dominated by dust emission from quiescently star-forming galaxies,
except at the faintest fluxes. The counts at 5.8 $\mum$ show behaviour
which is intermediate, with mild emission effects at bright fluxes and
mild extinction at faint fluxes. Comparing the solid blue and dashed
magenta lines, we see that the predicted number counts in the IRAC
bands are almost the same whether or not we assume a top-heavy IMF in
bursts, consistent with the counts being dominated by quiescent
galaxies.

Consider next the results for the MIPS bands, shown in
Fig.~\ref{fig:ncts-MIPS}. We again see remarkably good agreement of
the standard model with the observational data. The agreement is
especially good at faint fluxes (corresponding to higher redshifts).
In particular, the model matches well the observed 24 $\mum$ counts at
the ``bump'' around fluxes $S_{\nu} \sim 0.1-1 \mJy$. Accurate
modelling of the PAH emission features is obviously crucial for
modelling the 24 $\mum$ number counts, since the PAH features dominate
the flux in the 24 $\mum$ band as they are redshifted into the band at
$z \gsim 0.5$. On the other hand, the standard model overpredicts the
number counts at bright fluxes (corresponding to low redshifts) in all
three MIPS bands. The evolution at these wavelengths predicted by our
$\Lambda$CDM-based model thus seems to be not quite as strong as indicated by
observations.

In the MIPS bands, emission from galaxies is completely dominated by
dust, which is why no dashed blue lines are shown in
Fig.~\ref{fig:ncts-MIPS}. Comparing the red and green curves, we see
that quiescent (but star-forming) galaxies tend to dominate the number
counts in these bands at brighter fluxes, and bursts at fainter
fluxes. This reflects the increasing dominance of bursts in the mid-
and far-IR luminosity function at higher redshifts. Comparing the
solid blue and dashed magenta curves, we see that our standard model
with a top-heavy IMF in bursts provides a significantly better overall
fit to the observed 24 $\mum$ counts than the variant model with a
normal IMF in bursts (although at the brightest fluxes, the variant
model fits better). The faint number counts at 70 $\mum$ also favour
the top-heavy IMF model, while the number counts at 160 $\mum$ cover a
smaller flux range, and do not usefully distinguish between the two
variants of our model with different burst IMFs.

We can use our model to predict the flux levels at which sources
should become confused in the different \SPITZER\ bands. We estimate
the confusion limit using the {\em source density criterion}
\citep[e.g.][]{Vaisanen01,Dole03}: if the telescope has an FWHM
beamwidth of $\theta_{FWHM}$, we define the effective beam solid angle
as $\omega_{beam}= (\pi/(4\ln2))\,\theta_{FWHM}^2 =
1.13\theta_{FWHM}^2$, and then define the confusion limited flux
$S_{conf}$ to be such that $N(>S_{conf}) = 1/(N_{beam}\omega_{beam})$,
where $N(>S)$ is the number per solid angle of sources brighter than
flux $S$. We choose $N_{beam}=20$ for the number of beams per source,
which gives similar results to more detailed analyses
\citep[e.g.][]{Vaisanen01,Dole04b}. We use values of the beamsize
$\theta_{FWHM} = (1.66,1.72,1.88,1.98)$ arcsec for the four IRAC bands
\citep{Fazio04b} and $(5.6,16.7,35.2)$ arcsec for the three MIPS bands
\citep{Dole03}.  Our standard model then predicts confusion-limited
fluxes of $S_{conf}= (0.62,0.62,0.69,0.70) \muJy$ in the
$(3.6,4.5,5.8,8.0)\mum$ IRAC bands, and $S_{conf}= (0.072,2.6,43)
\mJy$ in the $(24,70,160)\mum$ MIPS bands. These confusion estimates
for the MIPS bands are similar to those of \citet{Dole04b}, which were
based on extrapolating from the observed counts. These values for the
confusion limits are indicated in Figs.~\ref{fig:ncts-IRAC} and
\ref{fig:ncts-MIPS} by vertical dashed lines.

\begin{figure*} 

\begin{minipage}{7cm}
\includegraphics[width=7cm]{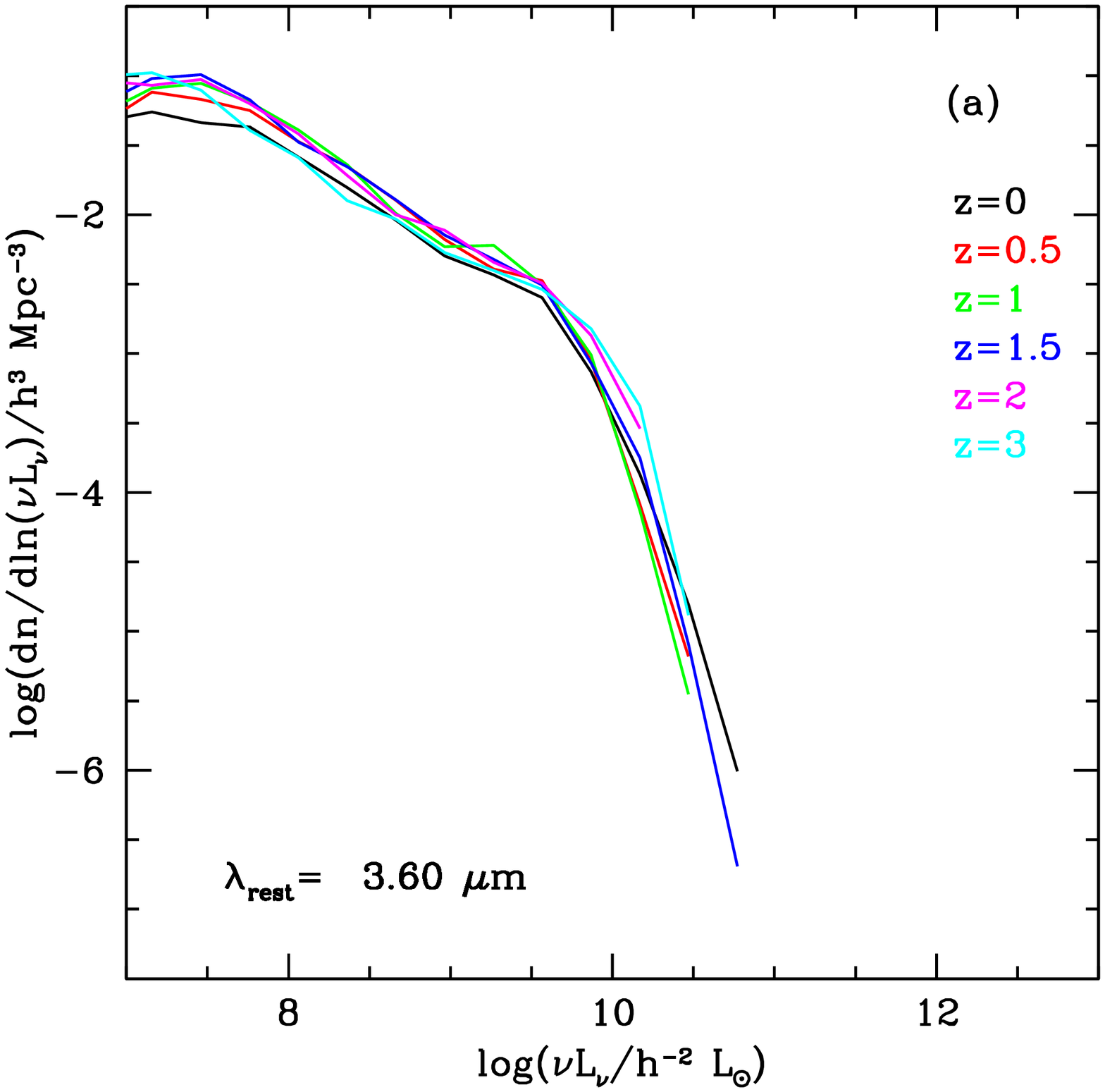}
\end{minipage}
\hspace{1cm}
\begin{minipage}{7cm}
\includegraphics[width=7cm]{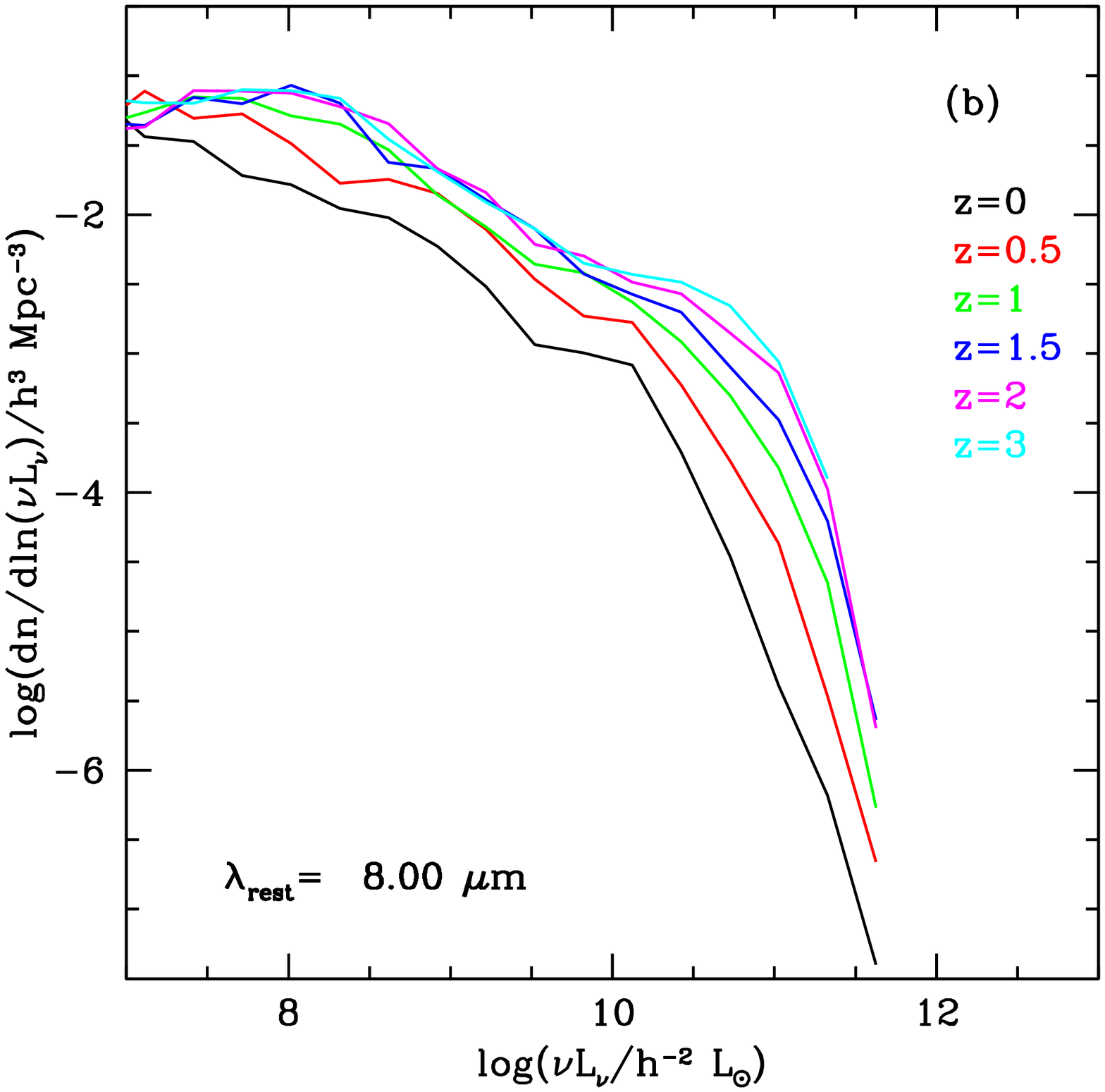}
\end{minipage}

\caption{Predicted evolution of the galaxy luminosity function in our
  standard model (including dust) at rest-frame wavelengths of (a) 3.6 
  and (b) 8.0 $\mum$ for redshifts $z=0,0.5,1,1.5,2$ and $3$, as
  shown in the key.}
\label{fig:lfNIR-evoln}
\end{figure*}

Our galaxy evolution model does not compute the contribution of AGN to
the IR luminosities of galaxies. On the other hand, the observed
number counts to which we compare include both normal galaxies, in
which the IR emission is powered by stellar populations, and AGN, in
which there is also IR emission from a dust torus, which is expected
to be most prominent in the mid-IR.  However, multi-wavelength studies
using optical, IR and X-ray data indicate that even at 24 $\mum$, the
fraction of sources dominated at that wavelength by AGN is only
10-20\% \citep[e.g.][]{Franceschini05}, and the contribution of
AGN-dominated sources in the other \SPITZER\ bands is likely to be
smaller. Therefore we should not make any serious error by comparing
our model predictions directly with the total number counts, as we
have done here.


\section{Evolution of the galaxy luminosity function}
\label{sec:lf-evoln}

While galaxy number counts provide interesting constraints on
theoretical models, it is more physically revealing to compare with
galaxy luminosity functions, since these isolate behaviour at
particular redshifts, luminosities and rest-frame wavelengths. In the
following subsections, we compare our model predictions with recent
estimates of luminosity function (LF) evolution based on \SPITZER\ data.

\subsection{Evolution of the galaxy luminosity function at 3-8 $\mum$}
\label{sec:lf_3-8}

\begin{figure}

\begin{center}

\includegraphics[width=7cm]{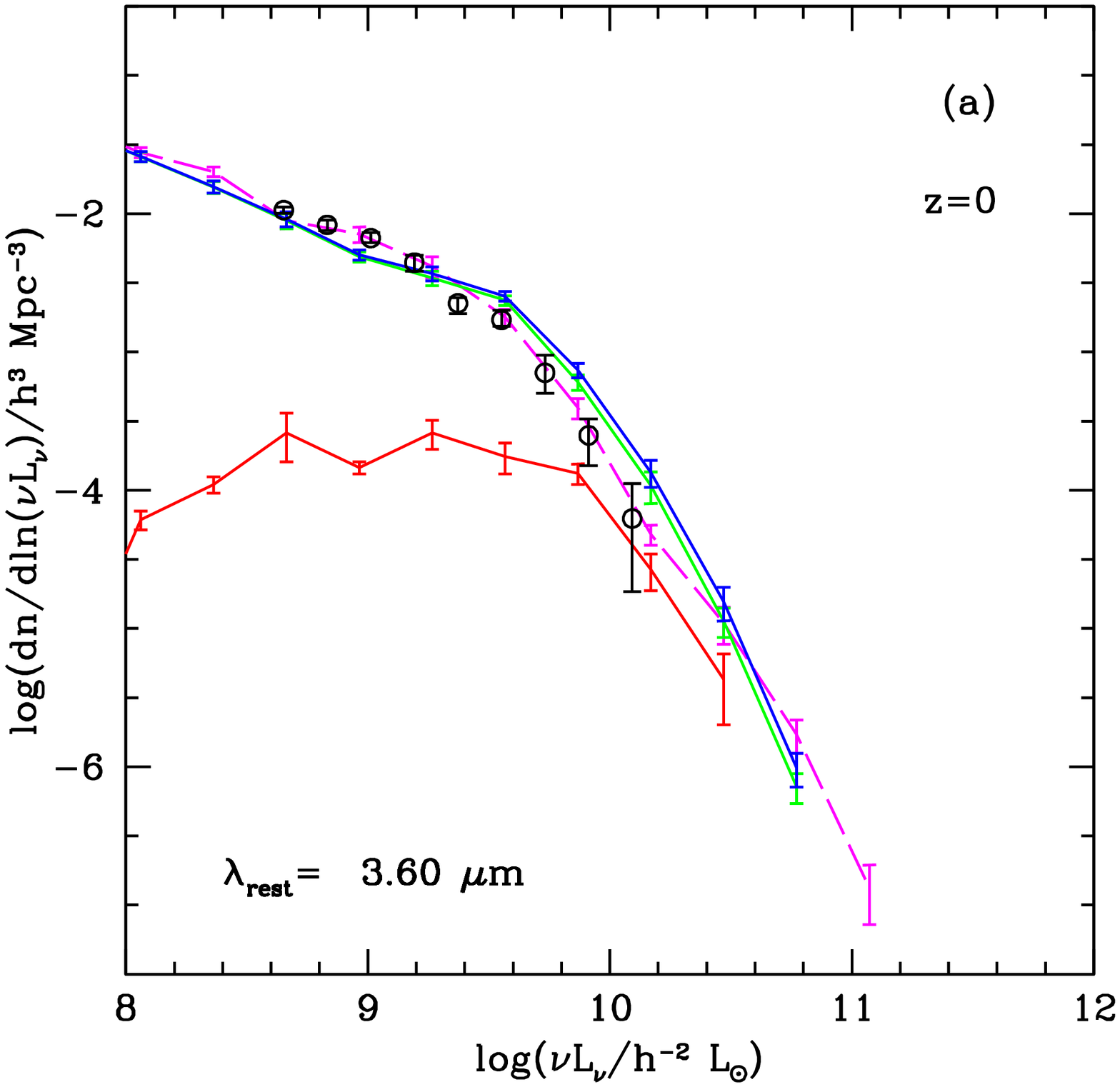}
\includegraphics[width=7cm]{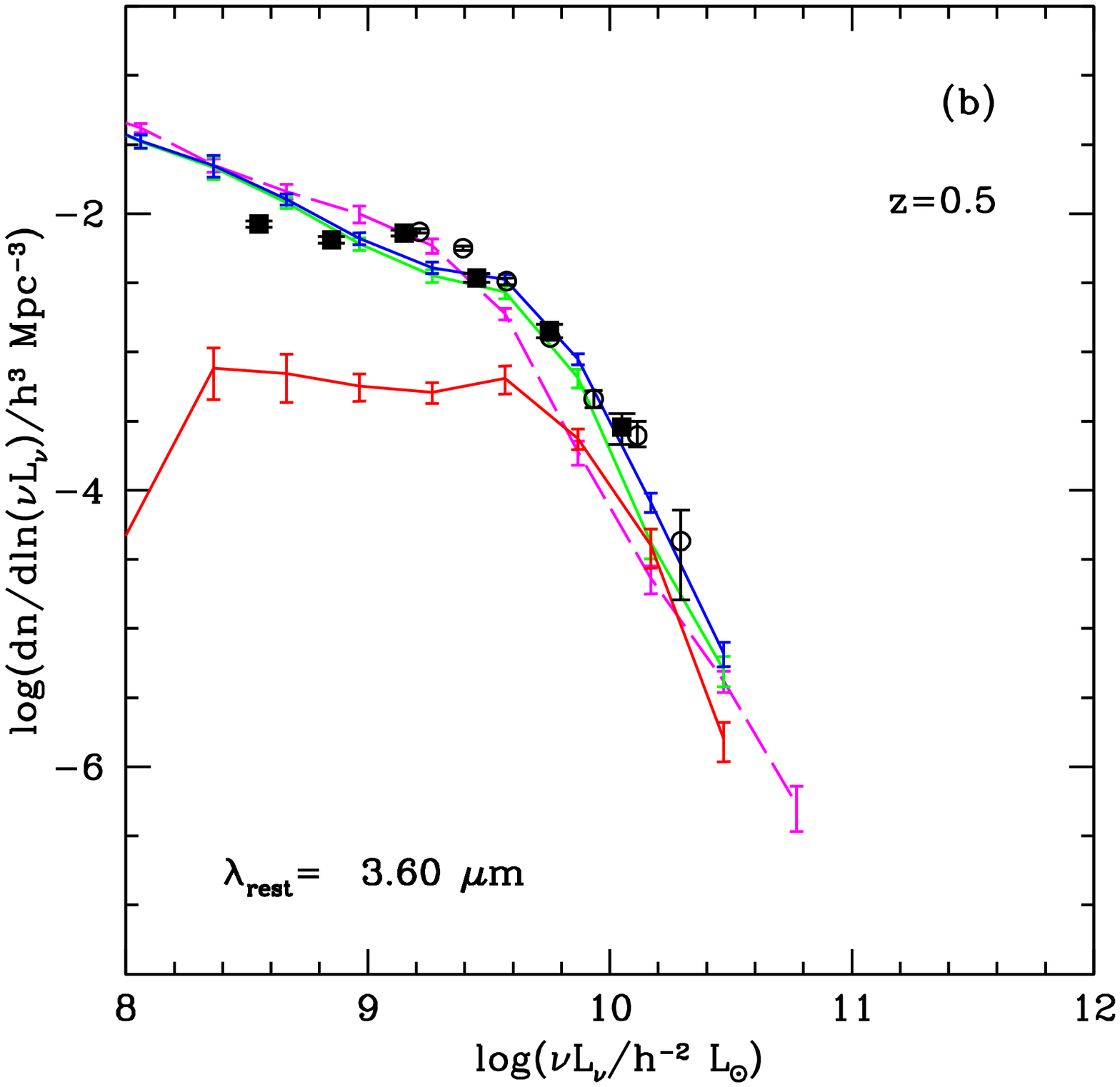}
\includegraphics[width=7cm]{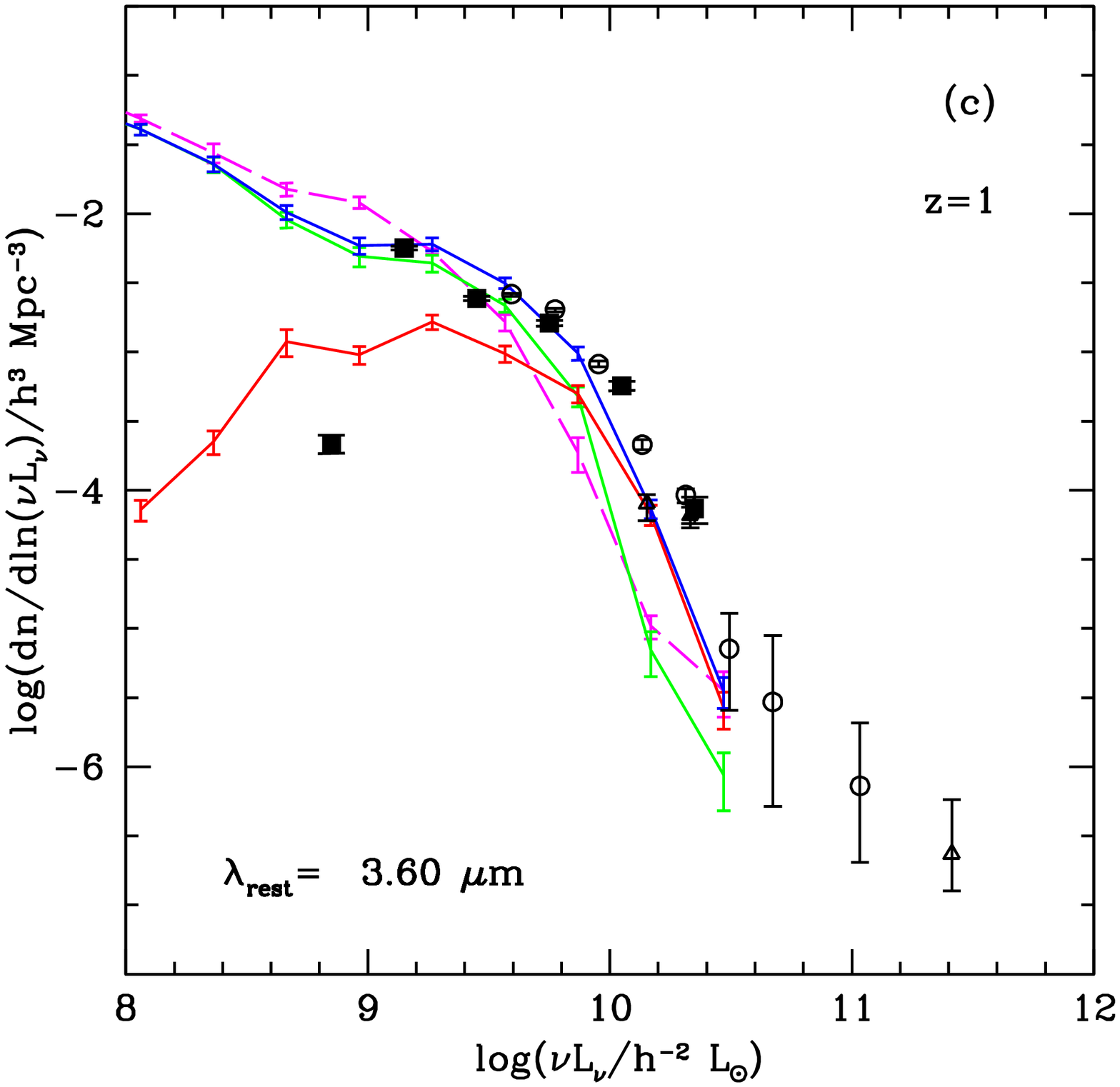}

\end{center}

\caption{Predicted evolution of the galaxy luminosity function at
  rest-frame 3.6 $\mum$ compared to observational data. The different
  panels show redshifts (a) $z=0$, (b) $z=0.5$ and (c) $z=1$. The
  predictions for our standard model are shown by the blue line, with
  the red and green lines showing the separate contributions from
  ongoing bursts and quiescent galaxies. The dashed magenta line shows
  the prediction for a variant model with a normal IMF for all stars.
  The error bars on the model lines indicate the Poisson uncertainties
  due to the finite number of galaxies simulated. The black symbols
  with error bars show observational data from \citet{Babbedge06}
  (open circles and triangles, for $z=0$, 0.5 and 1) and
  \citet{Franceschini06} (filled squares, for $z=0.5$ and 1).}
\label{fig:lf3.6-evoln-obs}
\end{figure}

\begin{figure}

\begin{center}

\includegraphics[width=7cm]{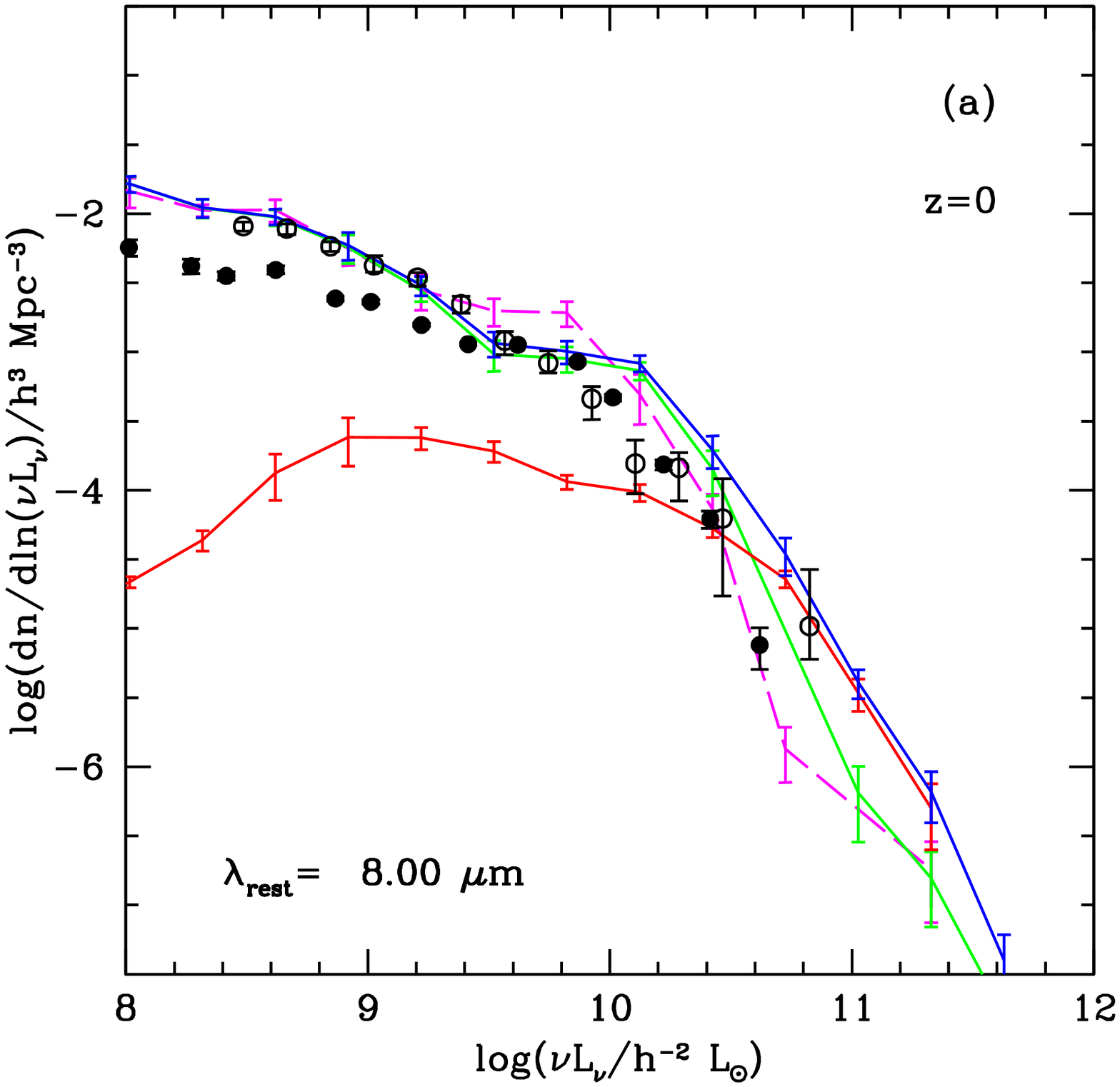}
\includegraphics[width=7cm]{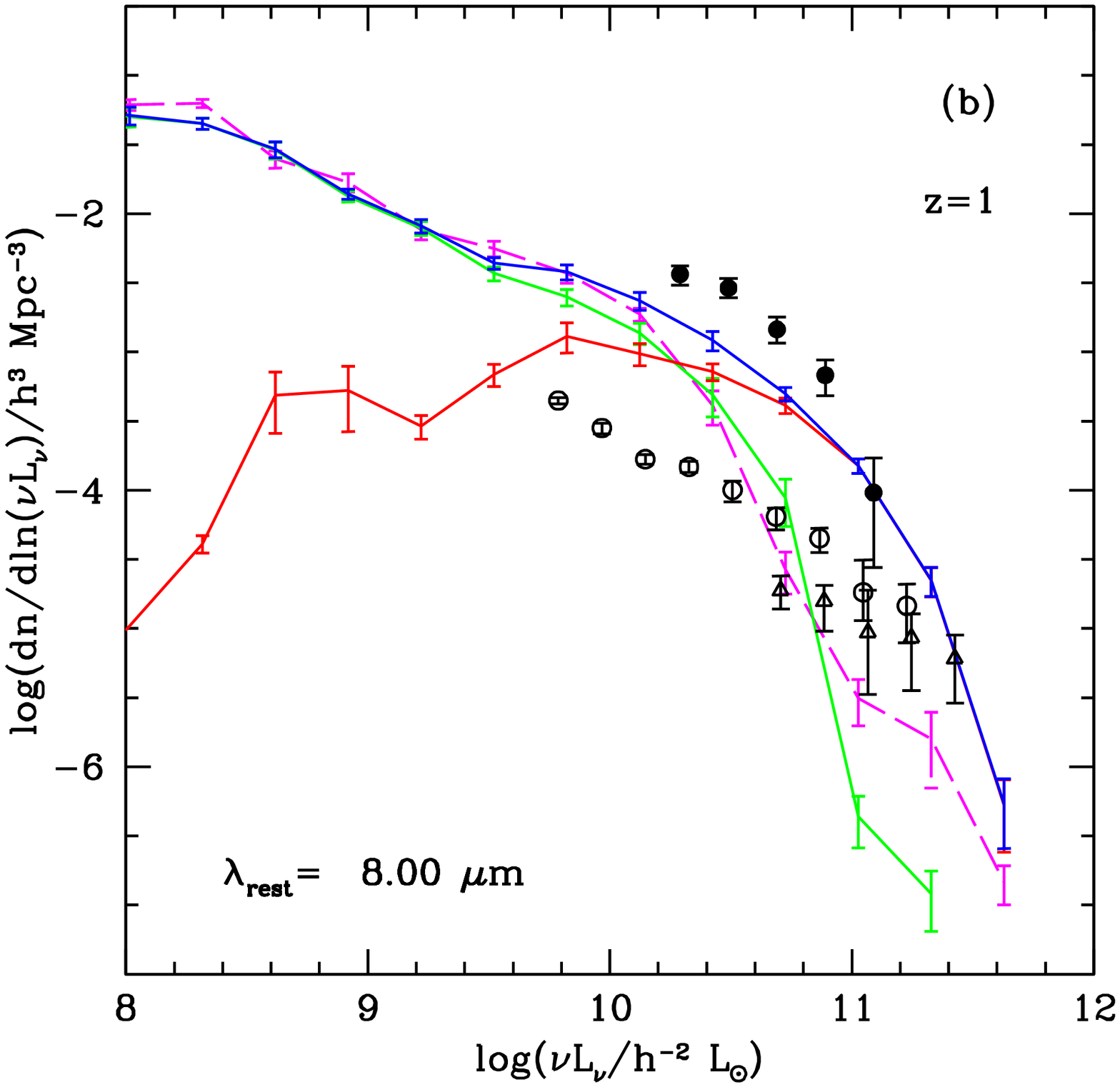}
\includegraphics[width=7cm]{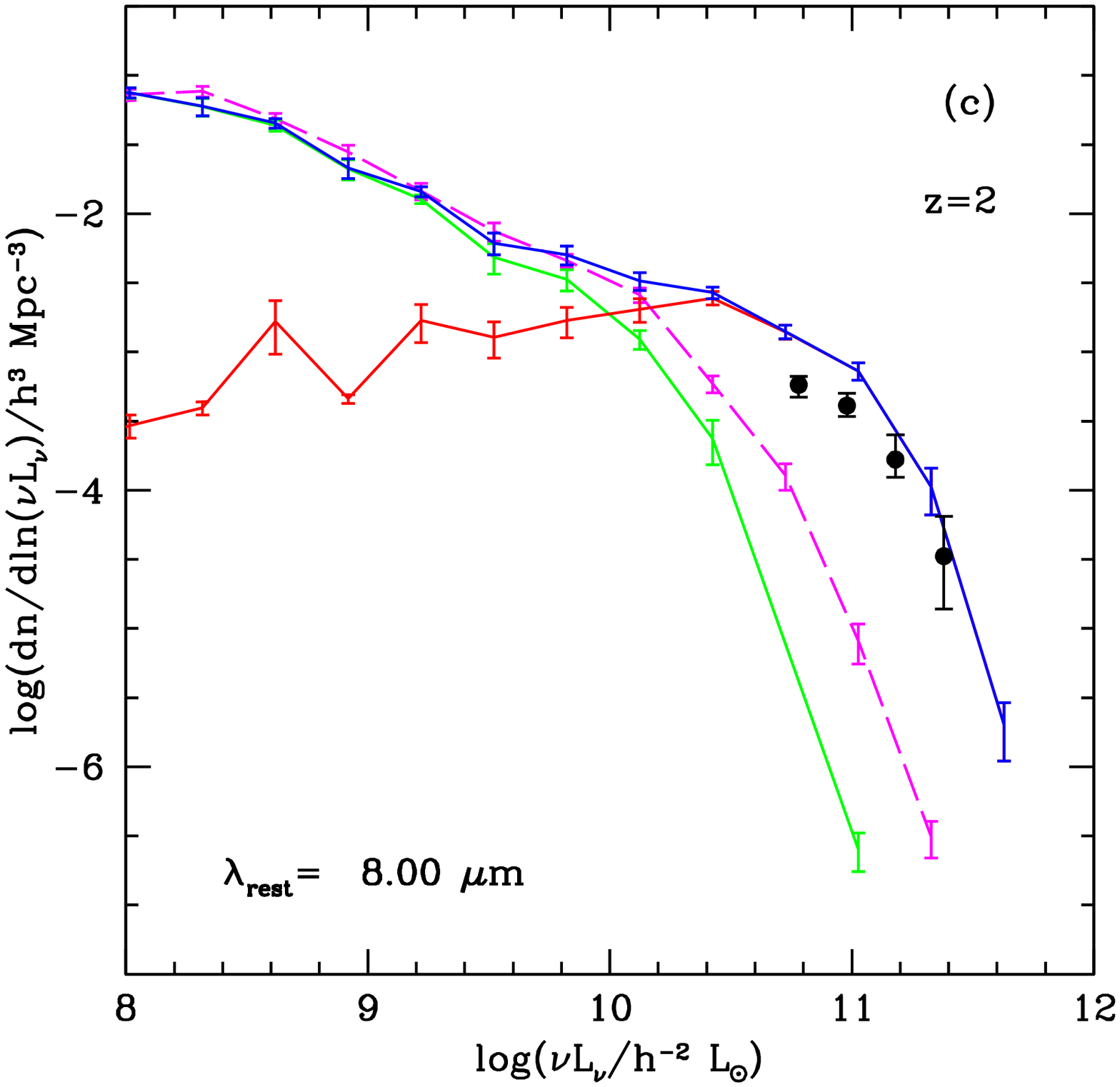}

\end{center}

\caption{Predicted evolution of the galaxy luminosity function at
  rest-frame 8.0 $\mum$ compared to observational data. The different
  panels show redshifts (a) $z=0$, (b) $z=1$ and (c) $z=2$. The
  coloured lines showing the model predictions have the same meaning
  as in Fig.~\ref{fig:lf3.6-evoln-obs}. The black symbols with error
  bars show observational data from \citet{Babbedge06} (open circles
  for $z=0$ and $0.7$, triangles for $z=1.2$), \citet{Huang07} (filled
  circles for $z=0$) and \citet{Caputi06b} (filled circles for $z=1$
  and 2). The observed LFs are for normal galaxies and exclude AGN.}
\label{fig:lf8.0-evoln-obs}
\end{figure}

\begin{figure}

\begin{center}

\includegraphics[width=7cm]{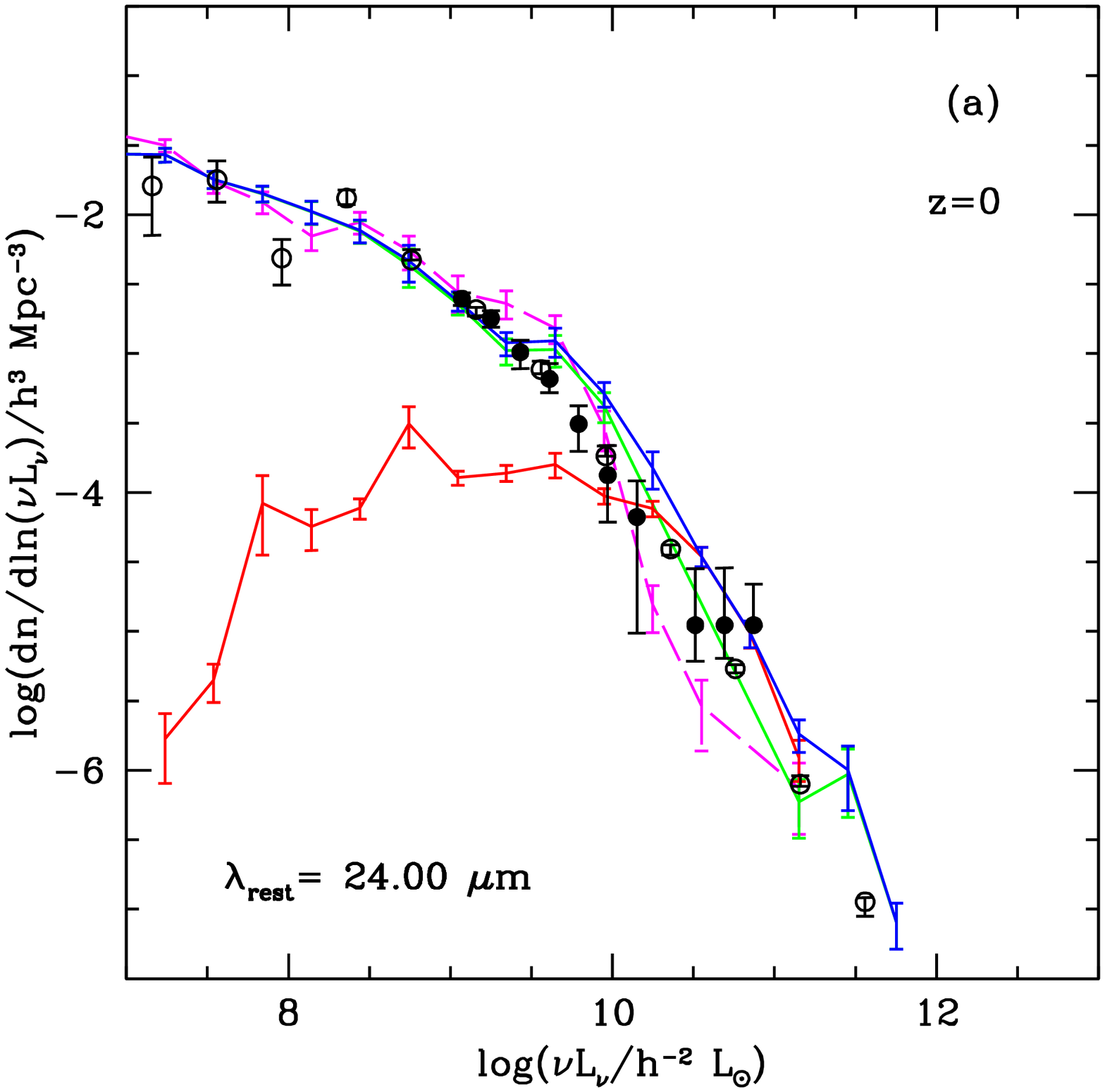}
\includegraphics[width=7cm]{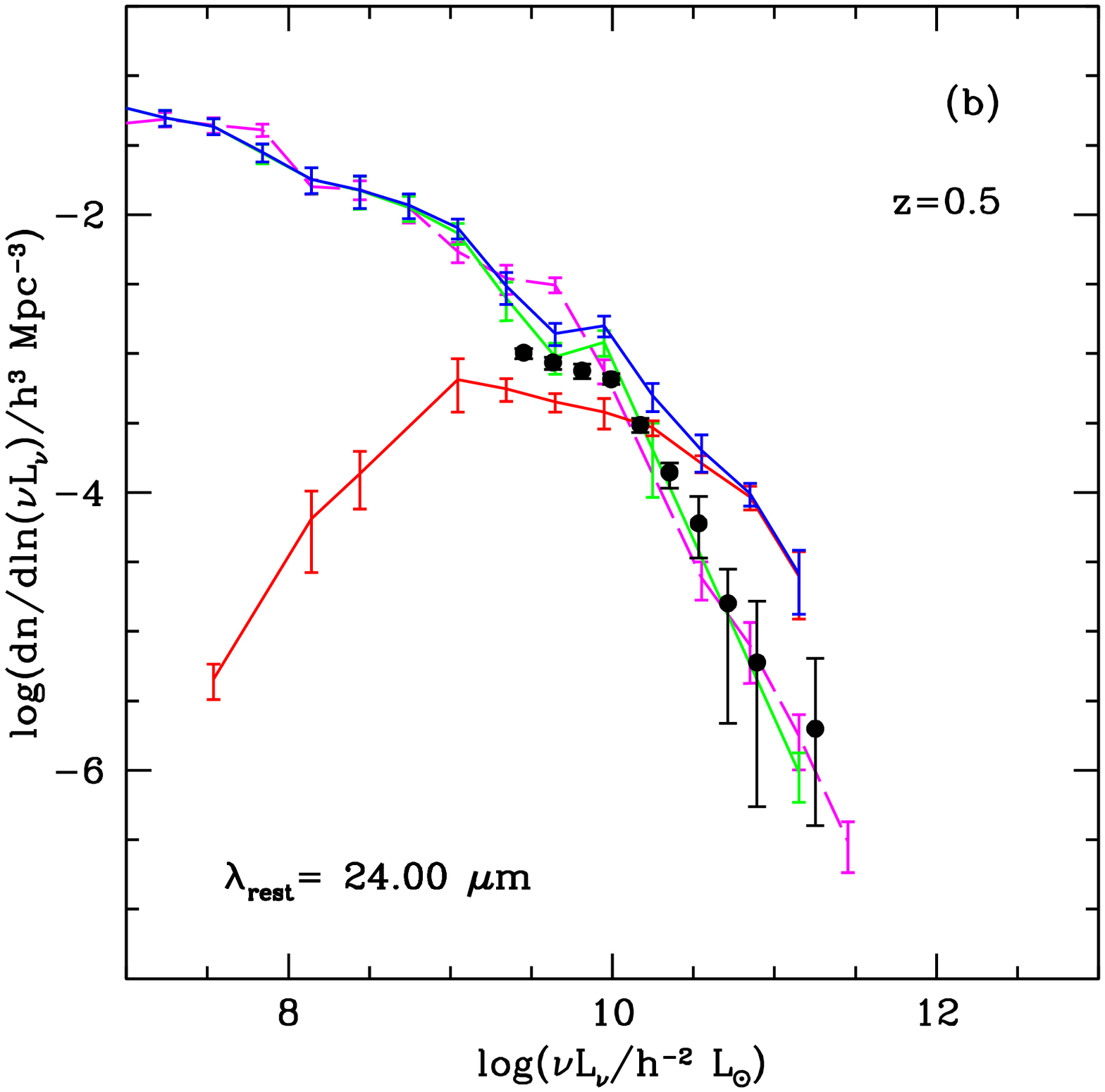}
\includegraphics[width=7cm]{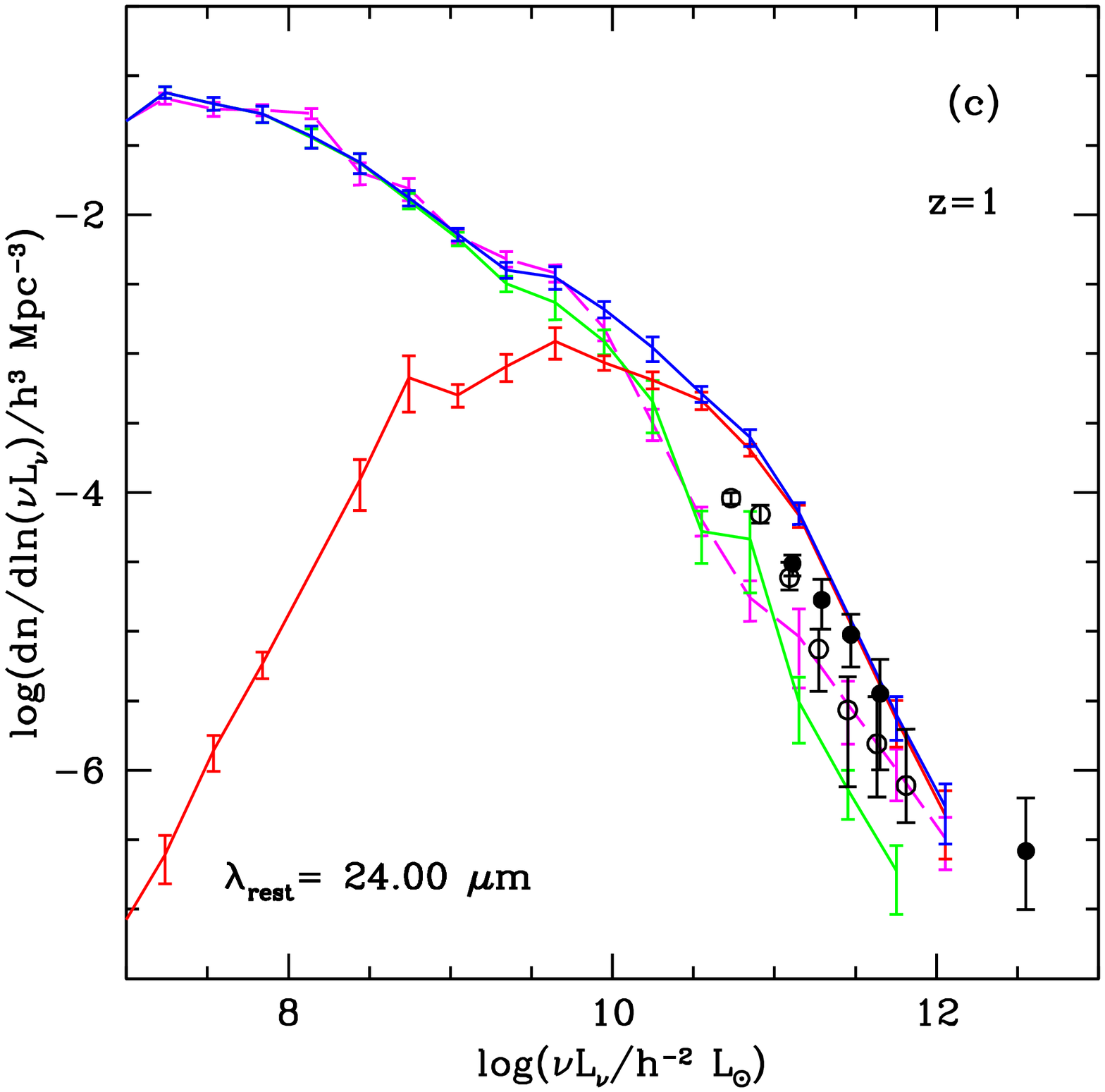}

\end{center}

\caption{Predicted evolution of the galaxy luminosity function at
  rest-frame wavelength 24 $\mum$ compared to observational data from
  \citet{Shupe98} (at $z=0$, open symbols) and from \citet{Babbedge06}
  (for the same redshifts as in Fig.~\ref{fig:lf3.6-evoln-obs}). The
  meaning of the curves showing the model predictions is the same as
  in Fig.~\ref{fig:lf3.6-evoln-obs}. (a) $z=0$, (b) $z=0.5$ and (c)
  $z=1$. }
\label{fig:lf24-evoln-obs}
\end{figure}

\begin{figure*} 

\begin{minipage}{7cm}
\includegraphics[width=7cm]{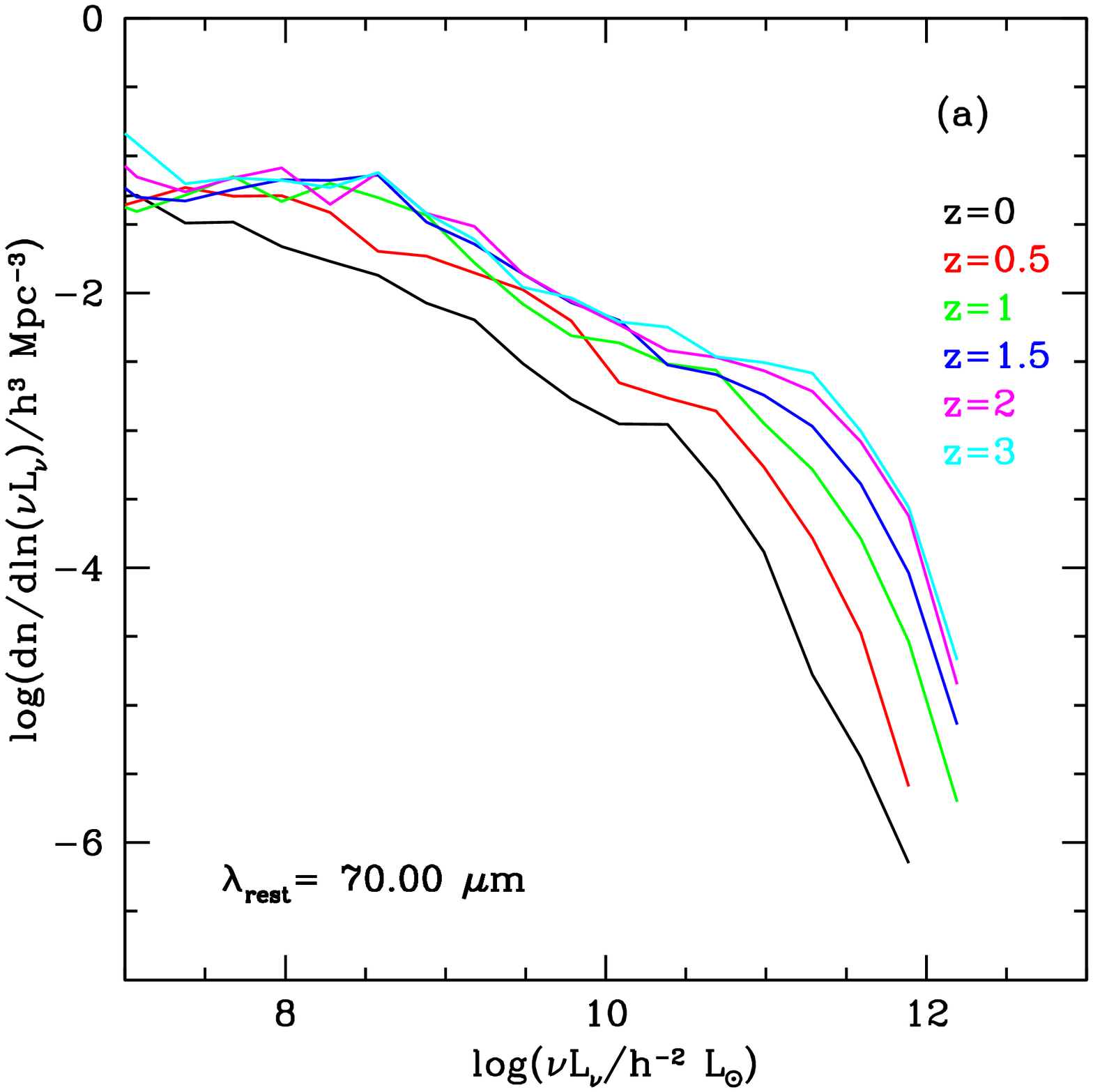}
\end{minipage}
\hspace{1cm}
\begin{minipage}{7cm}
\includegraphics[width=7cm]{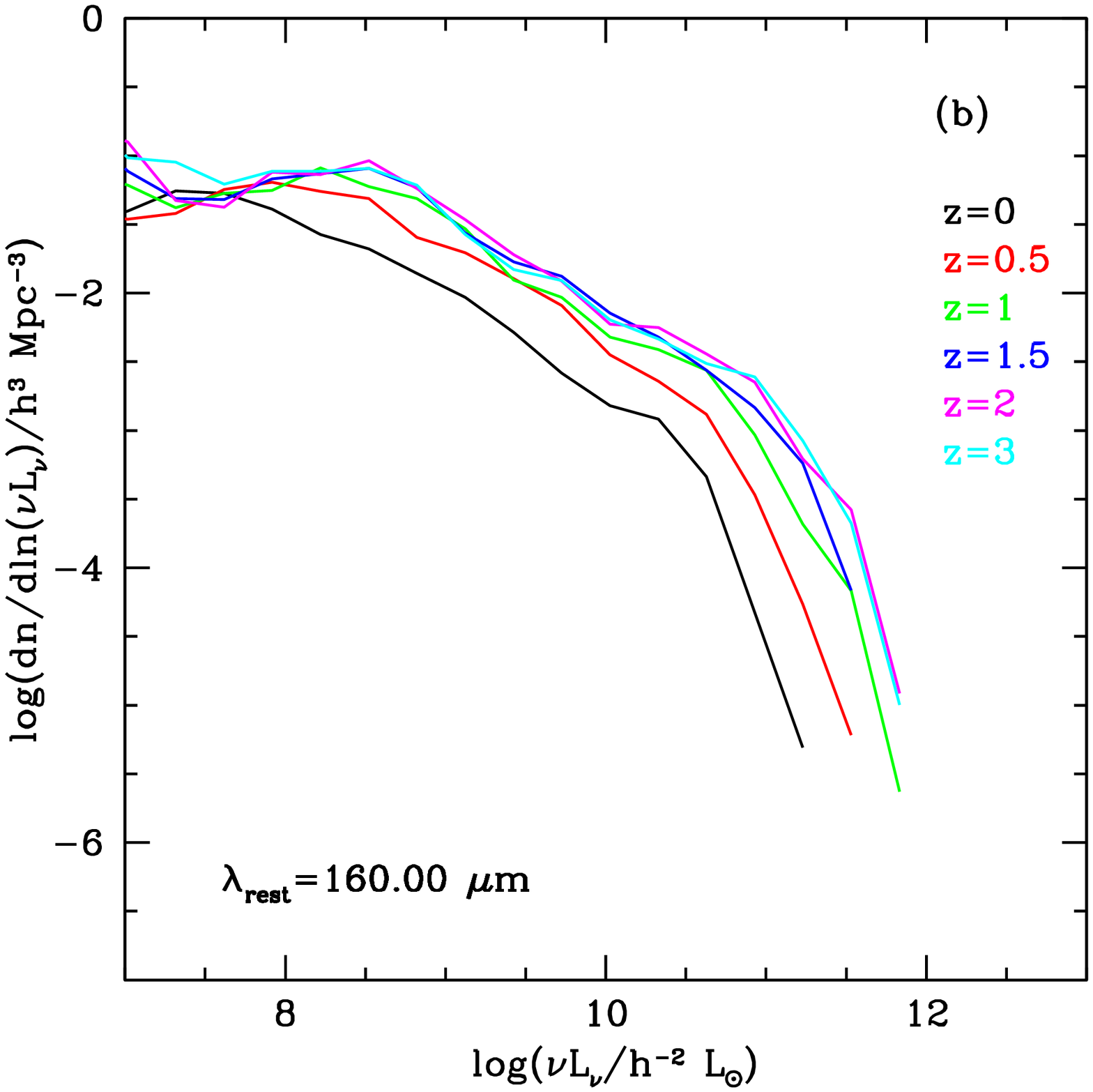}
\end{minipage}

\caption{Predicted evolution of the galaxy luminosity function in our
  standard model (including dust) at rest-frame wavelengths (a) 70
  $\mum$ and (b) 160 $\mum$, for redshifts $z=0,0.5,1,1.5,2$ and $3$,
  as shown in the key.}

\label{fig:lfFIR-evoln}
\end{figure*}

We consider first the evolution of the luminosity function in the
wavelength range covered by the IRAC bands, i.e. 3.6-8.0
$\mum$. Fig.~\ref{fig:lfNIR-evoln} shows what our standard model with
a top-heavy IMF in bursts predicts for LF evolution at rest-frame
wavelengths of 3.6 and 8.0 $\mum$ for redshifts $z=0-3$ \footnote{In
this figure, and in Figs.~\ref{fig:lf3.6-evoln-obs},
\ref{fig:lf8.0-evoln-obs}, \ref{fig:lf24-evoln-obs}, and
\ref{fig:lfFIR-evoln}, the luminosities $L_{\nu}$ are calculated
through the corresponding \SPITZER\ passbands.}. We see that at a
rest-frame wavelength of 3.6 $\mum$, the model LF hardly evolves at
all over the whole redshift range $z=0-3$. This lack of evolution
appears to be somewhat fortuitous. Galaxy luminosities at a rest-frame
wavelength of 3.6 $\mum$ are dominated by the emission from moderately
old stars, but the stellar mass function in the model evolves quite
strongly over the range $z=0-3$ (as we show in
\S\ref{sec:mstar-sfr}). The weak evolution in the 3.6 $\mum$ LF
results from a cancellation between a declining
luminosity-to-stellar-mass ratio with increasing time and increasing
stellar masses (see Figs.~\ref{fig:mstar-sfr}(a) and (e)). On the other
hand, at a rest-frame wavelength of 8.0~$\mum$, the model LF becomes
significantly brighter in going from $z=0$ to $z=3$. Galaxy
luminosities at a rest-frame wavelength of 8.0 $\mum$ are dominated by
emission from dust heated by young stars, so this evolution reflects
the increase in star formation activity with increasing redshift (see
Fig.~\ref{fig:mstar-sfr}(b) in \S\ref{sec:mstar-sfr}).

In Fig.~\ref{fig:lf3.6-evoln-obs}, we compare the model predictions
for evolution of the LF at 3.6 $\mum$ with observational estimates
from \citet{Babbedge06} and
\citet{Franceschini06}\footnote{\citet{Babbedge06} also compared their
measured LFs at 3.6, 8.0 and 24 $\mum$ with predictions from a
preliminary version of the model described in this paper}. The model
predictions are given for redshifts $z=0$, 0.5 and 1. For the
observational data, the mean redshifts for the different redshift bins
used do not exactly coincide with the model redshifts, so we plot them
with the model output closest in redshift \footnote{Specifically, for
$z=0$, we compare with the $z=0.1$ data from \citeauthor{Babbedge06},
for $z=0.5$ we compare with the $z=0.5$ data from
\citeauthor{Babbedge06} and $z=0.3$ data from
\citeauthor{Franceschini06}, and for $z=1$, we compare with the
$z=0.75$ (open symbols) and $z=1.25$ (filled symbols) data from
\citeauthor{Babbedge06} and $z=1.15$ data from
\citeauthor{Franceschini06}}. The observational estimates of the 3.6
$\mum$ LF rely on the measured redshifts. In the case of
\citet{Babbedge06}, these are mostly photometric, using optical and
NIR (including 3.6 and 4.5 $\mum$) fluxes, while for the
\citeauthor{Franceschini06} sample, about 50\% of the redshifts are
spectroscopic and the remainder photometric. In both samples, the
measured 3.6 $\mum$ fluxes were k-corrected to estimate the rest-frame
3.6 $\mum$ luminosities.

We see from comparing the blue curve with the observational data in
Fig.~\ref{fig:lf3.6-evoln-obs} that the 3.6 $\mum$ LF predicted by our
standard model is in very good agreement with the observations. In
particular, the observational data show very little evolution in the
3.6 $\mum$ LF over the redshift range $z=0-1$. The largest difference
seen is at $z=1$, where the \citeauthor{Babbedge06} data show a tail
of objects to very high luminosities, which is not seen in the model
predictions. However, this tail is not seen in the
\citeauthor{Franceschini06} data at the same redshift, and is also not
present in the observational data at the lower redshifts. More
spectroscopic redshifts are needed for the \citeauthor{Babbedge06}
sample to clarify whether this high-luminosity tail is real.
Comparing the red, green and blue lines for the standard model shows
that the model luminosity function is dominated by quiescent galaxies
at low luminosity, but the contribution of bursts becomes comparable
to that of quiescent galaxies at high luminosities. We have not shown
model LFs excluding dust extinction in this figure, since they are
almost identical to the predictions including dust. The dashed magenta
lines show the predicted LFs for the variant model with a normal IMF
in bursts. We see that these differ only slightly from our standard
model, but are a somewhat poorer fit to the observational data at
higher luminosities.

In Fig.~\ref{fig:lf8.0-evoln-obs} we show a similar comparison for the
LF evolution at a rest-frame wavelength of 8 $\mum$. The model
predictions are given for redshifts $z=0$, 1 and 2, and are compared
with observational estimates by \citet{Huang07} (for $z\sim0$),
\citet{Babbedge06} (for $z\sim 0$ and $z\sim 1$) and \citet{Caputi06b}
(for $z\sim 1$ and $z\sim 2$). These papers all classified objects in
their samples as either galaxies or AGN, and then computed separate
LFs for the two types of objects
\footnote{Note that a variety of criteria have been used for
classifying observed IR sources as AGN or normal galaxies, and these
do not all give equivalent results. Even if an object is classified as
an AGN, it is also not clear that in all cases the AGN luminosity
dominates over that of the host galaxy in all \SPITZER\ bands}. Our
model does not make any predictions for AGN, so we compare our model
predictions with the observed LFs for objects classified as galaxies
only. We see that for redshifts around $z=1$, the observed LFs from
\citeauthor{Babbedge06} and \citeauthor{Caputi06b} are in very poor
agreement with each other, with the \citeauthor{Caputi06b} LF being
around 10 times higher in number density at the same luminosity. This
differerence presumably results from some combination of: (a)
different methods of classifying objects as galaxies or AGN
(\citeauthor{Babbedge06} used only optical and IR fluxes to do this,
while \citeauthor{Caputi06b} also used X-ray data); (b) different
photometric redshift estimators; and (c) different methods for
k-correcting luminosities to a rest-frame wavelength of 8
$\mum$. There are smaller differences between the \citeauthor{Huang07}
and \citeauthor{Babbedge06} LFs at $z\sim 0$.  Futher observational
investigation appears to be necessary to resolve these issues. Our
standard model is in reasonable agreement with the
\citeauthor{Babbedge06} observed LF at $z\sim 0$, and with the
\citeauthor{Caputi06b} observed LFs at $z\sim1 $ and $z\sim 2$, but
not with the \citeauthor{Babbedge06} observed LF at $z\sim 1$. The
comparison with \citeauthor{Caputi06b} favours our standard model with
a top-heavy IMF in starbursts over the variant model with a normal
IMF.

\begin{figure*} 

\begin{minipage}{7cm}
\includegraphics[width=7cm]{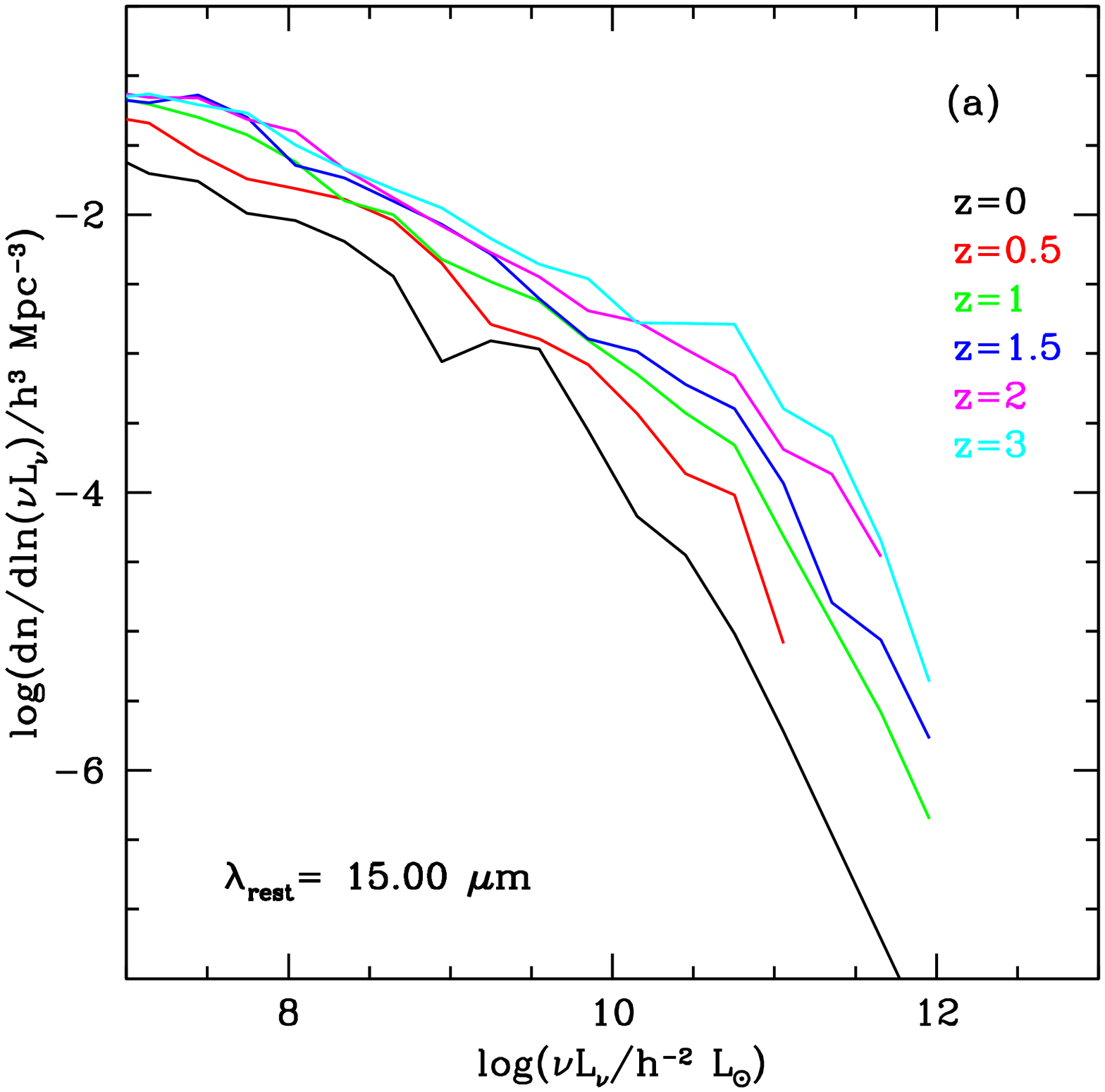}
\end{minipage}
\hspace{1cm}
\begin{minipage}{7cm}
\includegraphics[width=7cm]{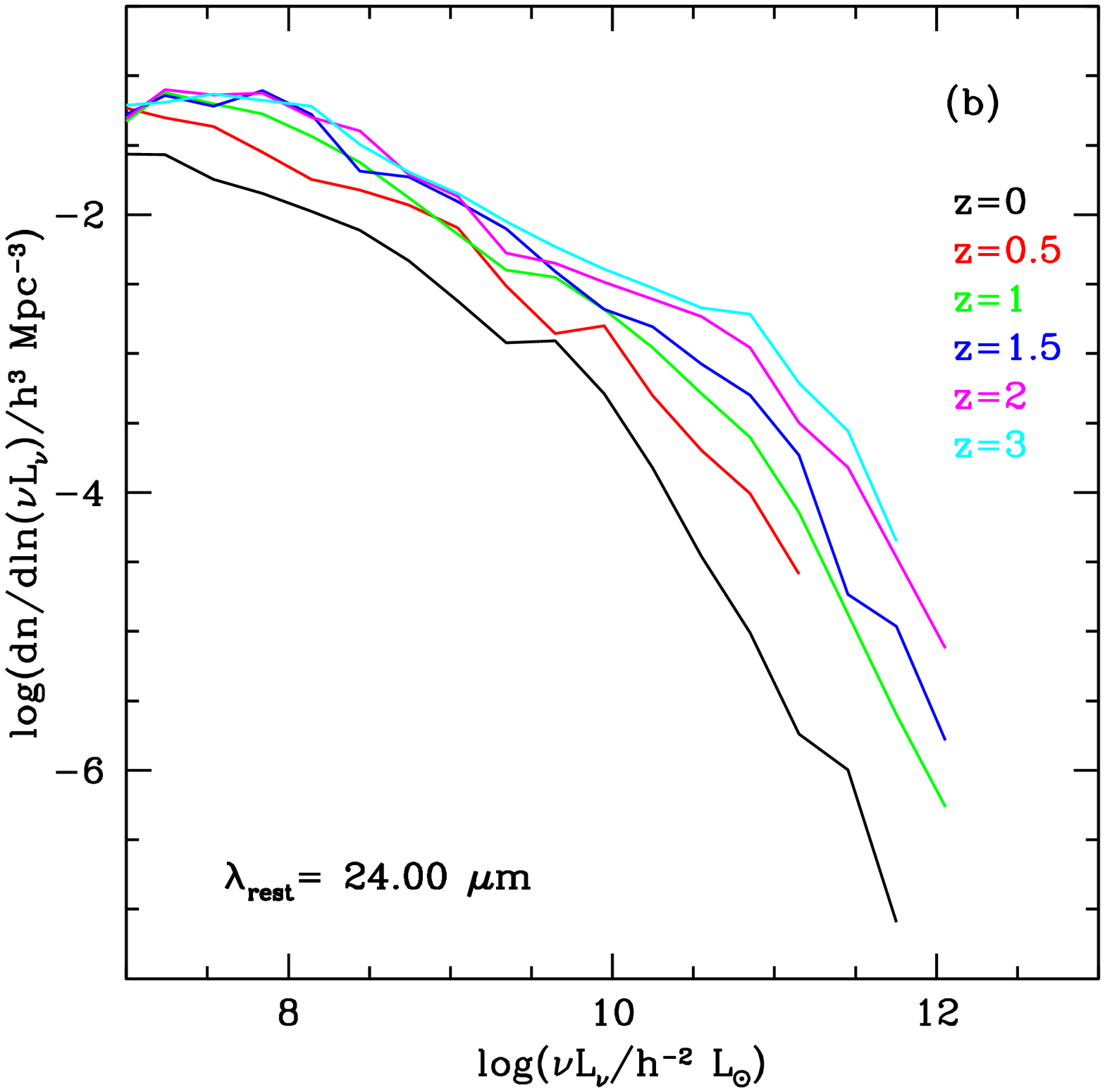}
\end{minipage}

\caption{Predicted evolution of the galaxy luminosity function in our
  standard model at rest-frame wavelengths(a) 15 $\mum$ (left) and
  (b) 24 $\mum$ (right) for redshifts $z=0,0.5,1,1.5,2$ and $3$, as
  shown in the key. }
\label{fig:lfMIR-evoln}
\end{figure*}

\begin{figure*}

\begin{center}

\begin{minipage}{7cm}
\includegraphics[width=7cm]{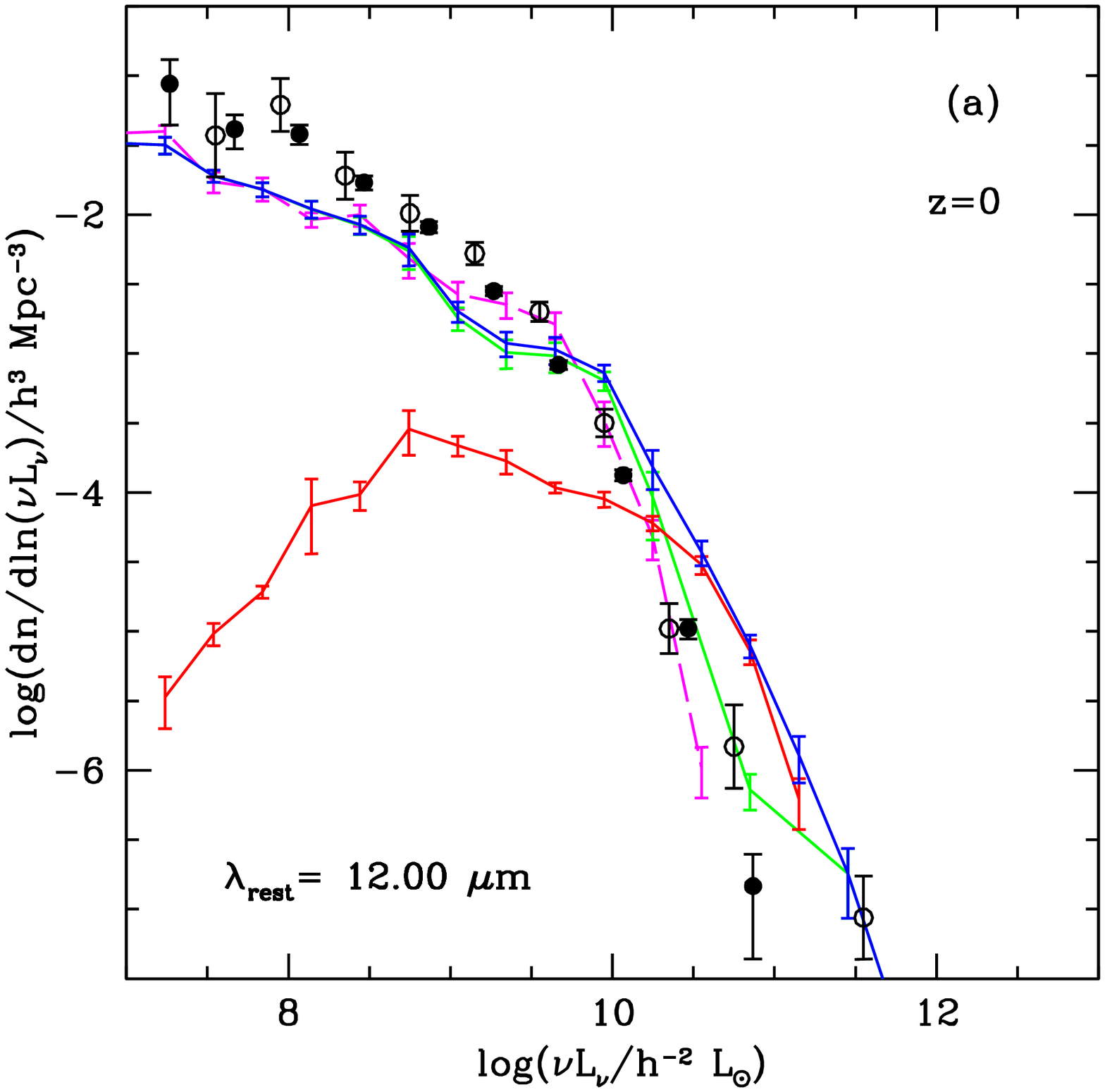}
\end{minipage}
\hspace{1cm}
\begin{minipage}{7cm}
\includegraphics[width=7cm]{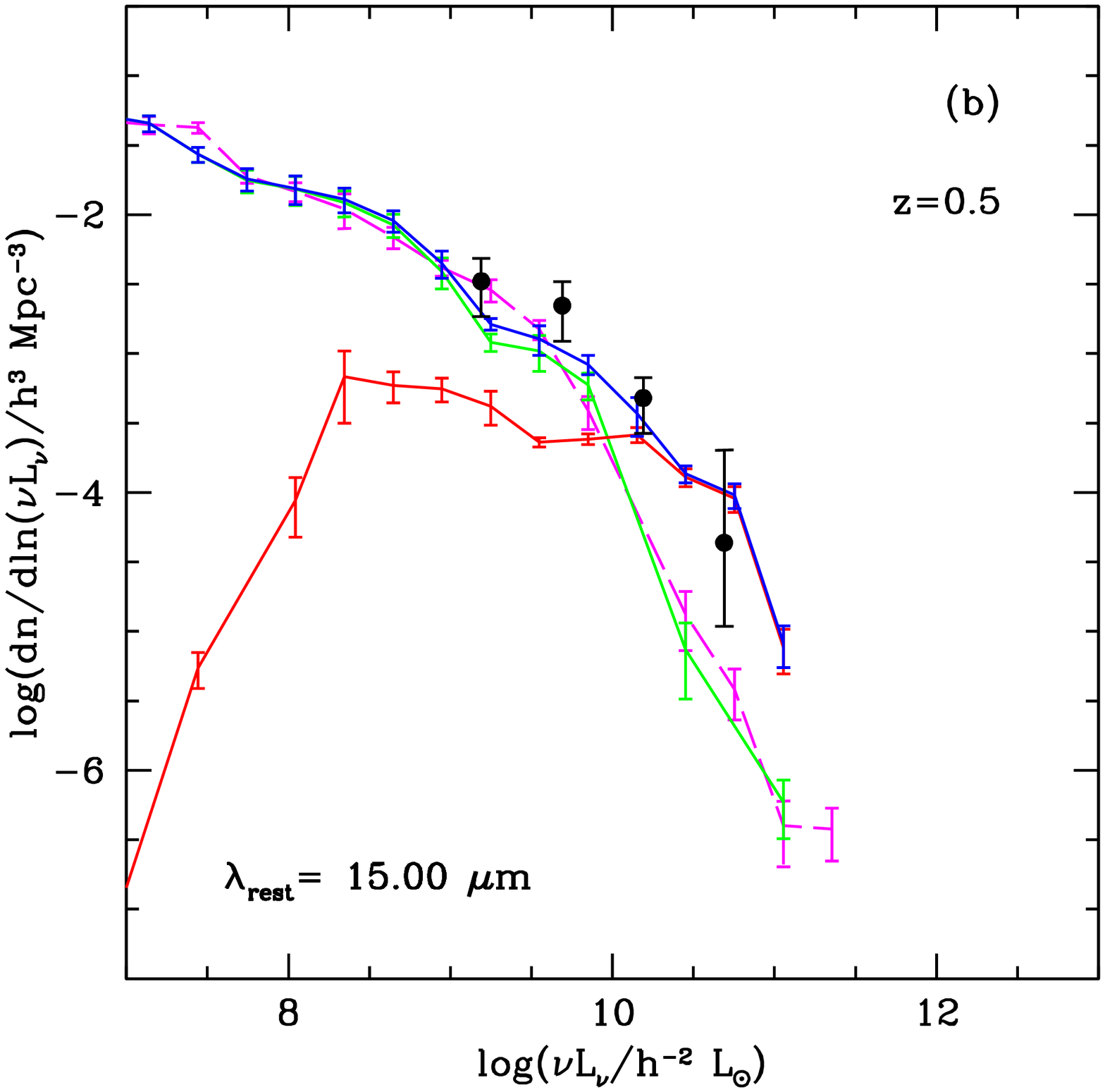}
\end{minipage}

\begin{minipage}{7cm}
\includegraphics[width=7cm]{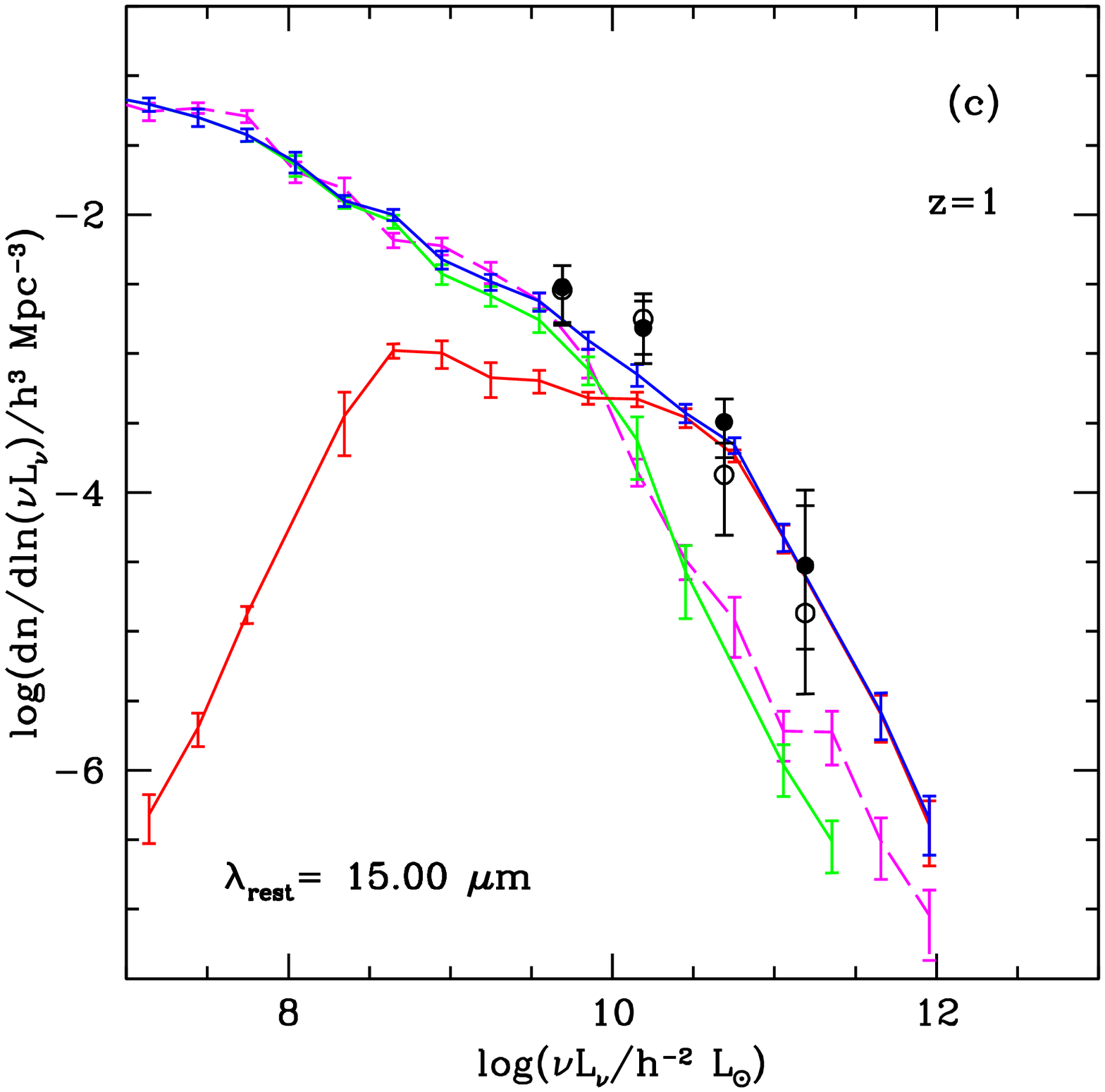}
\end{minipage}
\hspace{1cm}
\begin{minipage}{7cm}
\includegraphics[width=7cm]{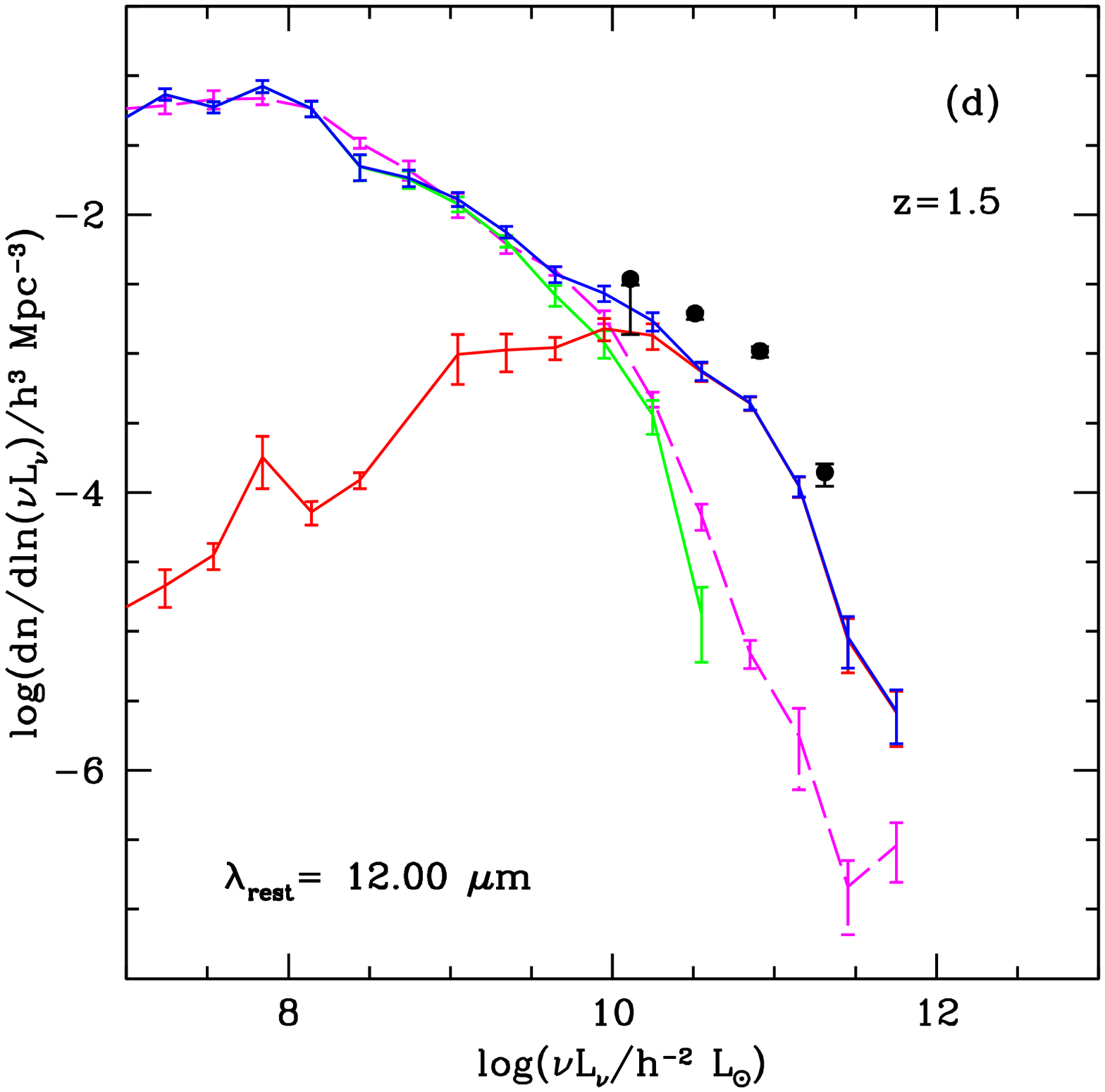}
\end{minipage}

\begin{minipage}{7cm}
\includegraphics[width=7cm]{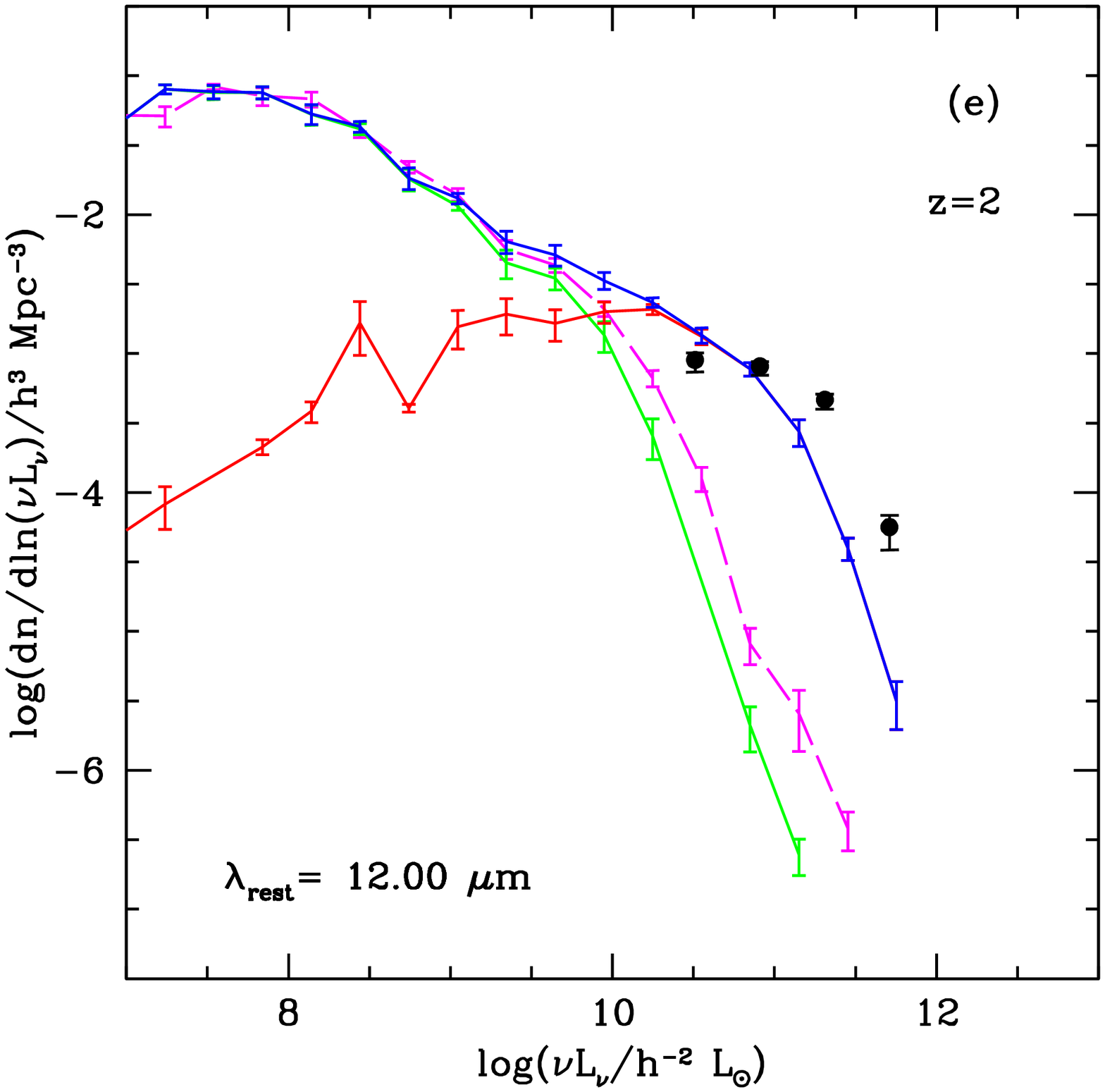}
\end{minipage}
\hspace{1cm}
\begin{minipage}{7cm}
\includegraphics[width=7cm]{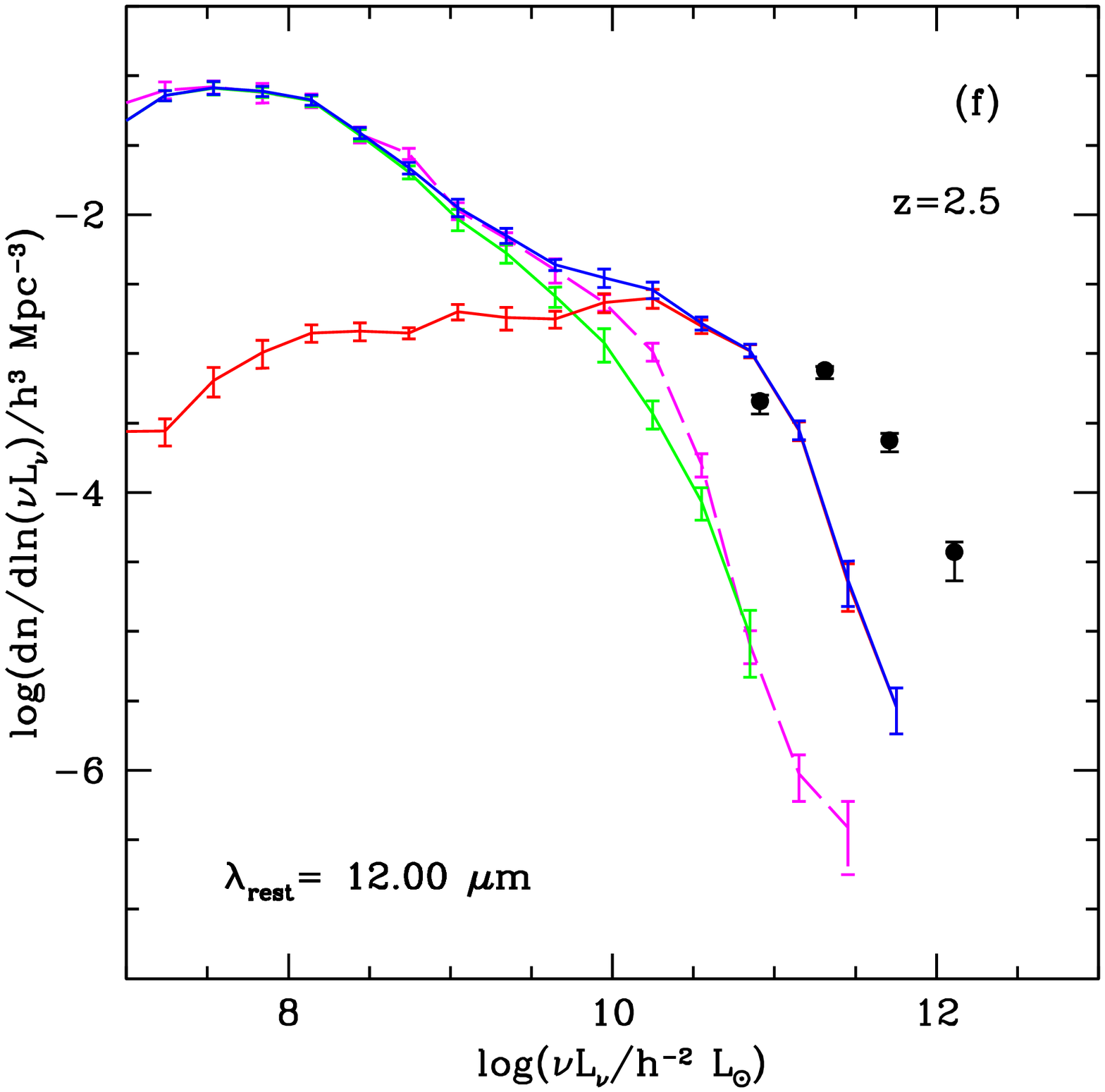}
\end{minipage}

\end{center}

\caption{Predicted evolution of the galaxy luminosity function at
  rest-frame wavelength 12 or 15 $\mum$ compared to observational
  data. The different panels show redshifts: (a) $z=0$, (b) $z=0.5$,
  (c) $z=1$, (d) $z=1.5$, (e) $z=2$ and (f) $z=2.5$. The meaning of
  the curves showing the model predictions is the same as in
  Fig.~\ref{fig:lf3.6-evoln-obs}. In panel (a), the predictions at
  12$\mum$ are compared to observational determinations from
  \citet{Soifer91} (open symbols) and \citet{Rush93} (filled symbols)
  based on \IRAS\ data. In panels (b) and (c), the predictions at
  15$\mum$ are compared to observational data from
  \citet{LeFloch05}. In panels (d), (e) and (f), the predictions at
  12$\mum$ are compared to observational data from \citet{Perez05}. }

\label{fig:lf15-evoln-obs}
\end{figure*}

\subsection{Evolution of the galaxy luminosity function at 12-24
  $\mum$}
\label{sec:lf_12-24}

In this subsection, we consider the evolution of the galaxy luminosity
function at mid-IR wavelengths, and compare with data obtained using
mainly the MIPS 24 $\mum$ band.

Fig.~\ref{fig:lfMIR-evoln} shows what our standard model with a
top-heavy IMF in bursts predicts for the evolution of the galaxy LF at
rest-frame wavelengths of 15 and 24 $\mum$ for redshifts $z=0-3$
\footnote{In this figure, and in Figs.~\ref{fig:lf15-evoln-obs} and
\ref{fig:lf24-evoln-obs}, the 24 $\mum$ luminosities are calculated
through the corresponding MIPS passband, while the 15 $\mum$
luminosities are calculated through a top-hat filter with a fractional
width of 10\% in wavelength.}. 
At rest-frame wavelengths of 15 and 24 $\mum$, galaxy luminosities are
typically dominated by the continuum emission from warm dust grains
heated by young stars (although PAH emission is also significant at
some nearby wavelengths). Fig.~\ref{fig:lfMIR-evoln} shows strong
evolution in the model LFs over the redshift range $z=0-3$ at both
wavelengths, reflecting both the increase in star formation activity
with increasing redshift (see Fig.~\ref{fig:mstar-sfr}(b)) and the
increasing dominance of the burst mode of star formation, for which
the top-heavy IMF further boosts the mid- and far-IR luminosities
compared to a normal IMF. Comparing Fig.~\ref{fig:lfMIR-evoln} with
Fig.~\ref{fig:lfNIR-evoln}(a), we also see a difference in the shape
of the bright end of the LF: at 3.6 $\mum$, where the LF is dominated
by emission from stars, the bright end cuts off roughly exponentially,
while at 15 and 24 $\mum$, where the LF is dominated by emission from
warm dust, the bright end declines more gradually, roughly as a
power-law. This difference reflects the difference in shape of the
galaxy stellar mass function (GSMF) and the galaxy star formation rate
distribution (GSFRD). The GSMF shows an exponential-like cutoff at
high masses, while the GSFRD shows a more gradual cutoff at high
SFRs because of starbursts triggered by galaxy mergers (see
Figs.~\ref{fig:mstar-sfr}(a) and (b) in \S\ref{sec:mstar-sfr}). This
difference was noticed earlier by observers comparing optical and
far-IR LFs of galaxies, but its origin was not understood
\citep{Lawrence86,Soifer87b}.

In Fig.~\ref{fig:lf15-evoln-obs}, we compare the model LFs at
rest-frame wavelengths 12 and 15 $\mum$ with observational
estimates. For $z=0$, we plot the observational estimates from
\citet{Soifer91} and \citet{Rush93}, based on \IRAS\ 12 $\mum$ data
(with AGN removed). For $z=0.5-1$ and $z=1.5-2.5$, we plot the data of
\citet{LeFloch05} and \citet{Perez05} respectively, which were
obtained from galaxy samples selected on \SPITZER\ 24 $\mum$ flux.
\citeauthor{LeFloch05} k-corrected their measured 24 $\mum$ fluxes to
15 $\mum$ rest-frame luminosities, while \citeauthor{Perez05}
k-corrected to 12 $\mum$ rest-frame\footnote{The exact passband used
for the model LF in each panel depends on which observational data we
are comparing with. For $z=0$, we use the IRAS 12 $\mum$ passband; at
$z=0.5$ and $z=1$ we use a top-hat passband centred at 15 $\mum$; and
at $z=1.5$, 2 and 2.5, we use a top-hat passband centred at 12 $\mum$
(both top-hat passbands having fractional width 10\% in
wavelength).}. \citeauthor{LeFloch05} obtained most of their redshifts
from photometric redshifts based on optical data, while
\citeauthor{Perez05} used a new photometric redshift technique based
on fitting empirical SEDs to all of the available broad-band data from
the far-UV to 24 $\mum$, and also removed ``extreme'' AGN from their
observed LF. Note that the redshifts for the observed LFs do not
exactly coincide with model redshifts in all cases, but are close.

We see from comparing the blue line to the observational datapoints in
Fig.~\ref{fig:lf15-evoln-obs} that our standard model with a top-heavy
IMF in bursts fits the observations remarkably well up to $z=2$.  In
particular, the model matches the strong evolution in the mid-IR LF
seen in the observational data. The model falls below the
observational data at $z=2.5$, but here both the photometric redshifts
and the k-corrections are probably the most uncertain. The standard
model also does not provide a perfect fit to the $z=0$ data,
predicting somewhat too many very bright galaxies and somewhat too few
very faint galaxies (though the latter discrepancy might be affected
by local galaxy clustering in the \IRAS\ data). Comparing
the red, green and blue lines for the standard model in the figure, we
see that the bright end of the 12 or 15 $\mum$ LF is dominated by
bursts at all redshifts. The figure also shows by a dashed magenta
line the predictions for the variant model with a normal IMF in
bursts. This latter model predicts much less evolution in the bright
end of the LF than is observed. This comparison thus strongly favours
the model with the top-heavy IMF in bursts.

Finally, in Fig.~\ref{fig:lf24-evoln-obs}, we carry out a similar
comparison of the evolution of predicted and observed LFs at a
rest-frame wavelength of 24 $\mum$ over the redshift range $z=0-1$, in
this case comparing with observational estimates from \citet{Shupe98}
(for $z=0$), based on \IRAS\ data, and from \citet{Babbedge06} (for
$z=0-1$), based on \SPITZER\ data\footnote{The model luminosities are
all computed through the \SPITZER\ 24 $\mum$ passband.}. The galaxy
redshifts for the \citeauthor{Babbedge06} data were obtained in the
same way as for the 3.6 $\mum$ LFs shown in
Fig.~\ref{fig:lf3.6-evoln-obs}, and the luminosities were k-corrected
from observer-frame 24 $\mum$ to rest-frame 24 $\mum$. The LF plotted
from \citeauthor{Babbedge06} is that for normal galaxies, with AGN
excluded.

The conclusions from comparing the model with the 24 $\mum$ LFs are
similar to those from the comparison with the 12 and 15 $\mum$
LFs. The data favour our standard model over the variant with a normal
IMF in bursts (except possibly for $z=0.5$), as the latter predicts
too little evolution at the bright end. At $z=0$, the model fits the
24 $\mum$ data rather better than for the corresponding comparison at
12 $\mum$. On the other hand, at $z=0.5$ and $z=1$, the model LF is a
somewhat worse fit to the observational data at 24 $\mum$ than at 15
$\mum$. These differences between the 12/15 and 24 $\mum$ comparisons
might result from the different photometric redshifts and
k-corrections used in the observational samples in the two
cases. Alternatively, they might result from problems in modelling the
dust SEDs in the complex mid-IR range.

\begin{figure}
\includegraphics[width=8cm]{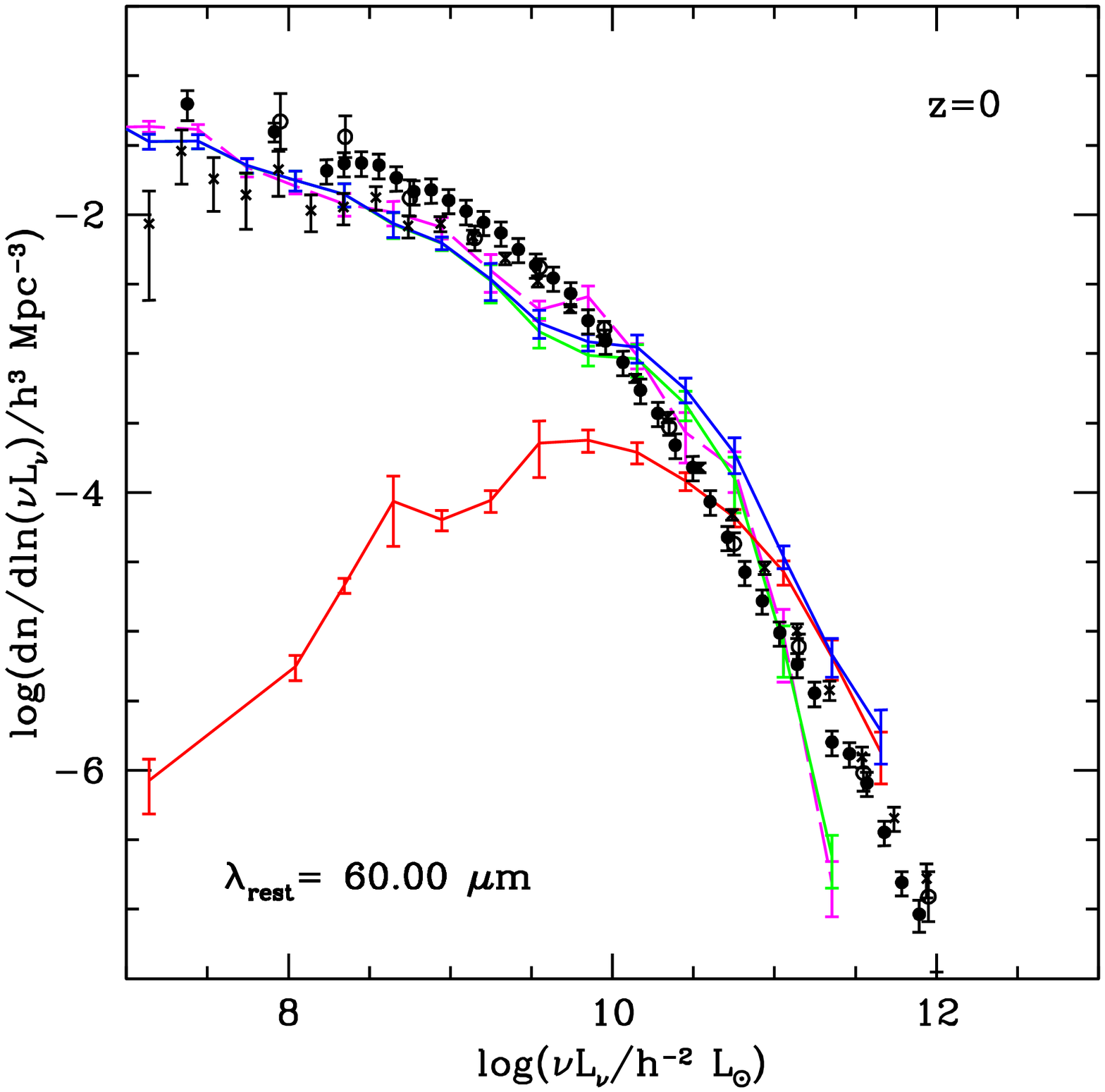}
\caption{The predicted galaxy luminosity function at 60 $\mum$
compared to observational data from IRAS. The meaning of the different
lines is the same as in Fig.~\ref{fig:lf3.6-evoln-obs}. The black
symbols show observational data from \citet{Saunders90} (crosses),
\citet{Soifer91} (open circles), and \citet{Takeuchi03} (filled
circles).}

\label{fig:lf60}
\end{figure}

\subsection{Evolution of the galaxy luminosity function at 70-160 $\mum$}

We now briefly consider the evolution of the luminosity function in
the far-IR. The far-IR is the wavelength range where most of the
luminosity from dust in normal galaxies is emitted. The local
60~$\mum$ luminosity function was very well measured by surveys with
IRAS, and so is commonly used as a starting point or benchmark for
modelling the evolution of the galaxy population in the far-IR. We
therefore present in Fig.~\ref{fig:lf60} the model prediction for the
60 $\mum$ luminosity function at $z=0$, compared with observational
data from \citet{Saunders90}, \citet{Soifer91} and
\citet{Takeuchi03}. As discussed in \citet{Baugh05}, the local 60
$\mum$ luminosity function was used as one of the primary constraints
in fixing the parameters of our galaxy formation model, and the figure
shows that our standard model provides a good match to the data. The
variant model with a normal IMF in bursts underpredicts the abundance
of the brightest 60~$\mum$ galaxies.

In Fig.~\ref{fig:lfFIR-evoln}, we show the model predictions for the
evolution of the luminosity function in the two longer wavelength MIPS
bands, at rest-frame wavelengths of 70 and 160 $\mum$, from $z=0$ to
$z=3$. At 70 $\mum$, the luminosity function at high luminosities is
predicted to brighten by about a factor 10 going from $z=0$ to
$z=2$. This is about a factor 2 less than the brightening predicted in
the mid-IR at 15 $\mum$ (compare to Fig.~\ref{fig:lfMIR-evoln}), but
nearly a factor 2 more evolution than is predicted at 160
$\mum$. These differences between the amount of evolution seen at
different IR wavelengths reflect evolution in the shapes of the SEDs
of the galaxies responsible for the bulk of the IR emission. No
observational estimates of the evolution of the luminosity function at
70 and 160 $\mum$ have yet been published, but they are expected to be
forthcoming from ongoing surveys with \SPITZER.

\begin{figure}
\includegraphics[width=8cm]{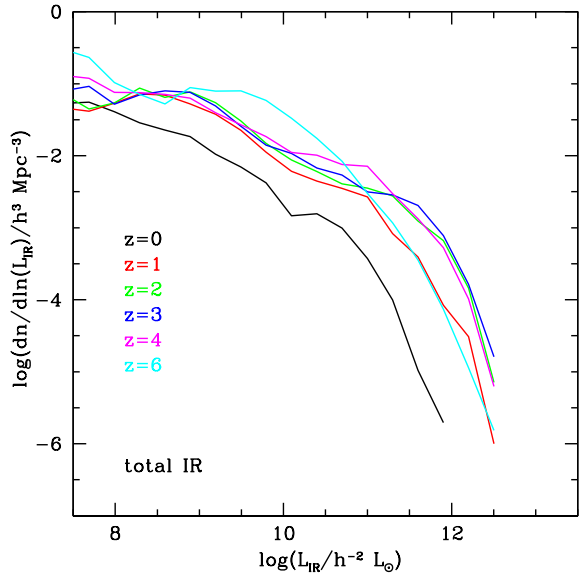}
\caption{Predicted evolution of the total mid+far-IR (8-1000 $\mum$)
  galaxy luminosity function for our standard model, for redshifts
  $z=0$, 1, 2, 3, 4 and 6, as shown in the key. }
\label{fig:lfIR-evoln}
\end{figure}

\begin{figure*}

\begin{center}

\begin{minipage}{7cm}
\includegraphics[width=7cm]{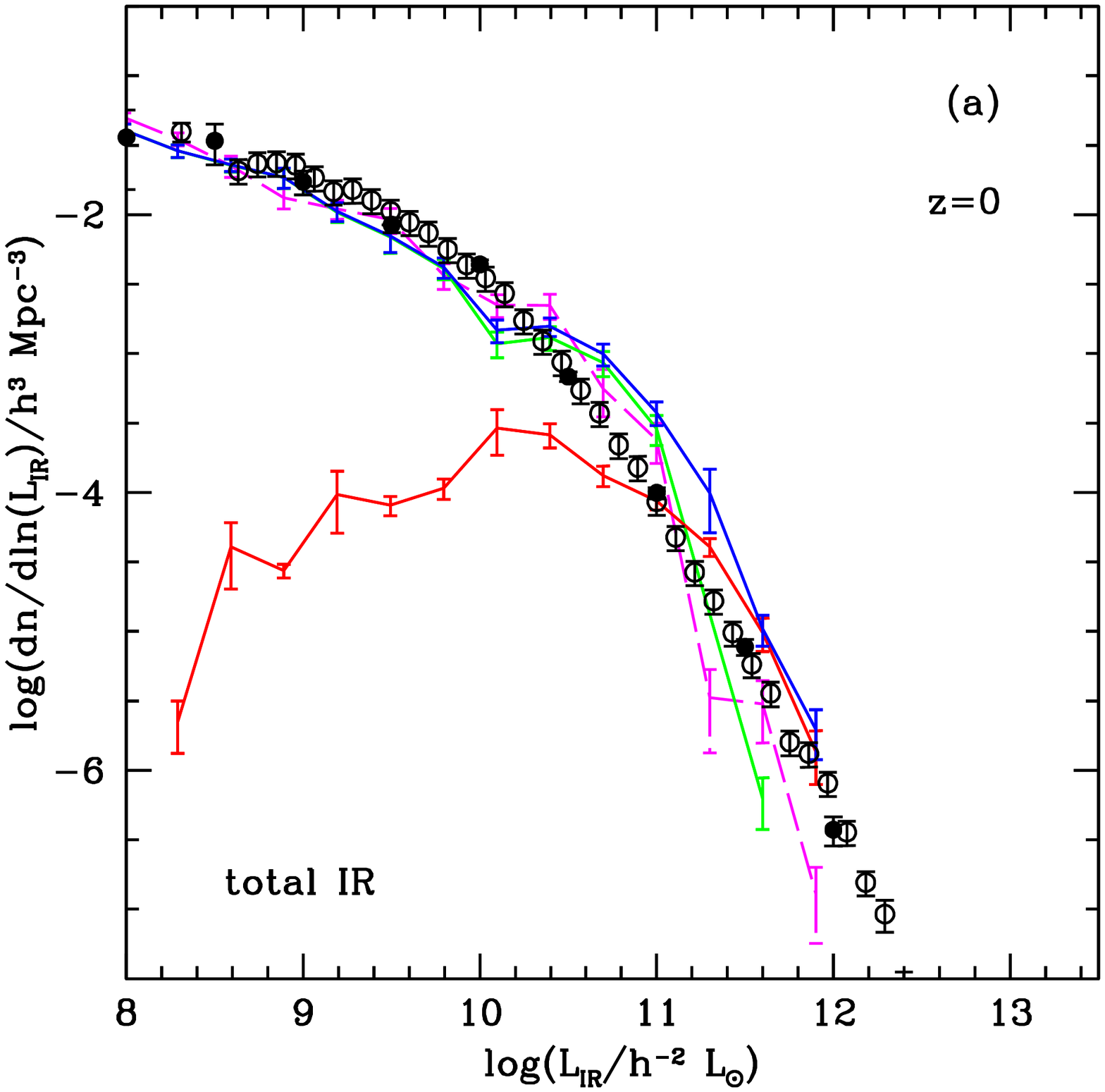}
\end{minipage}
\hspace{1cm}
\begin{minipage}{7cm}
\includegraphics[width=7cm]{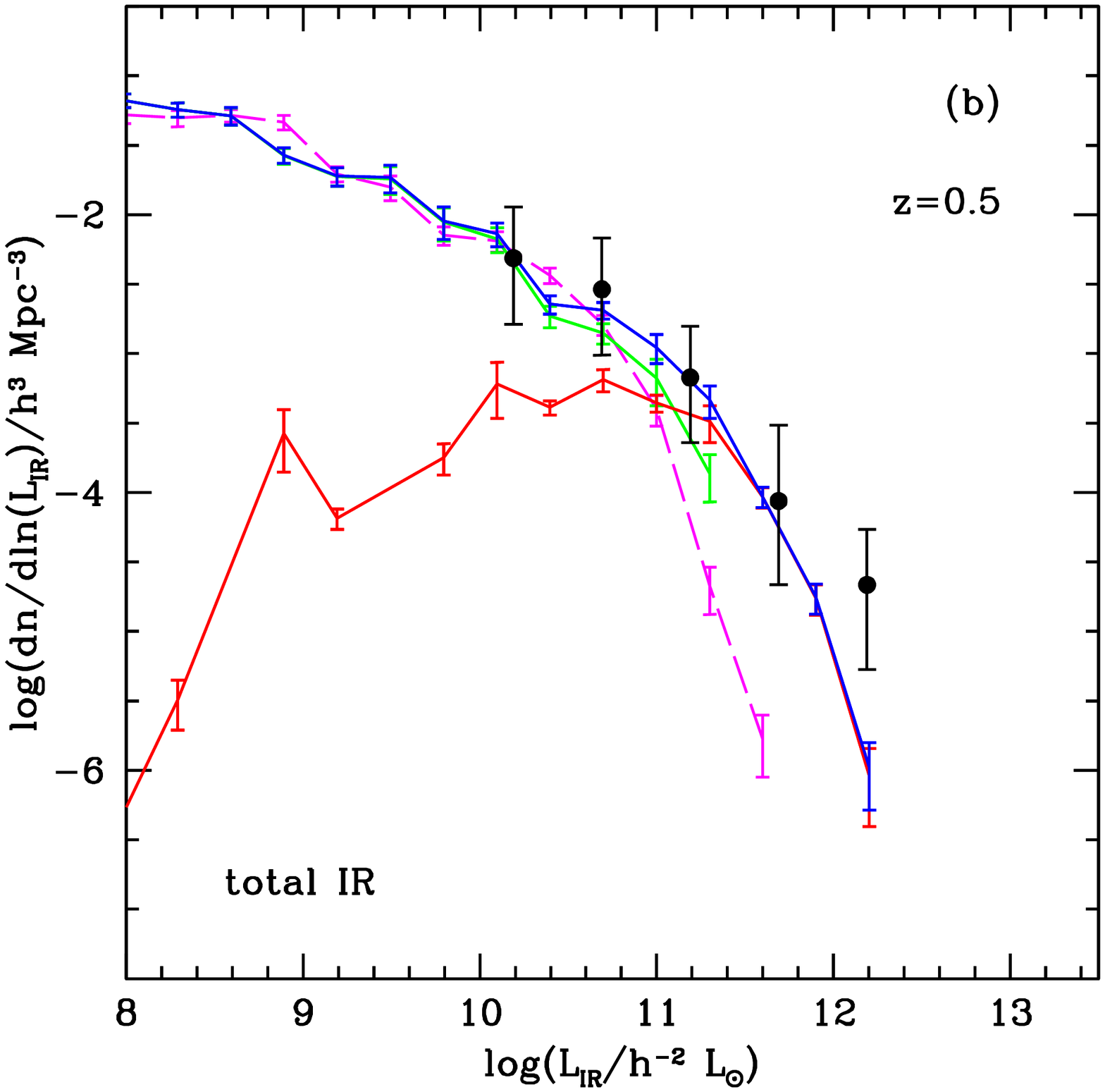}
\end{minipage}

\begin{minipage}{7cm}
\includegraphics[width=7cm]{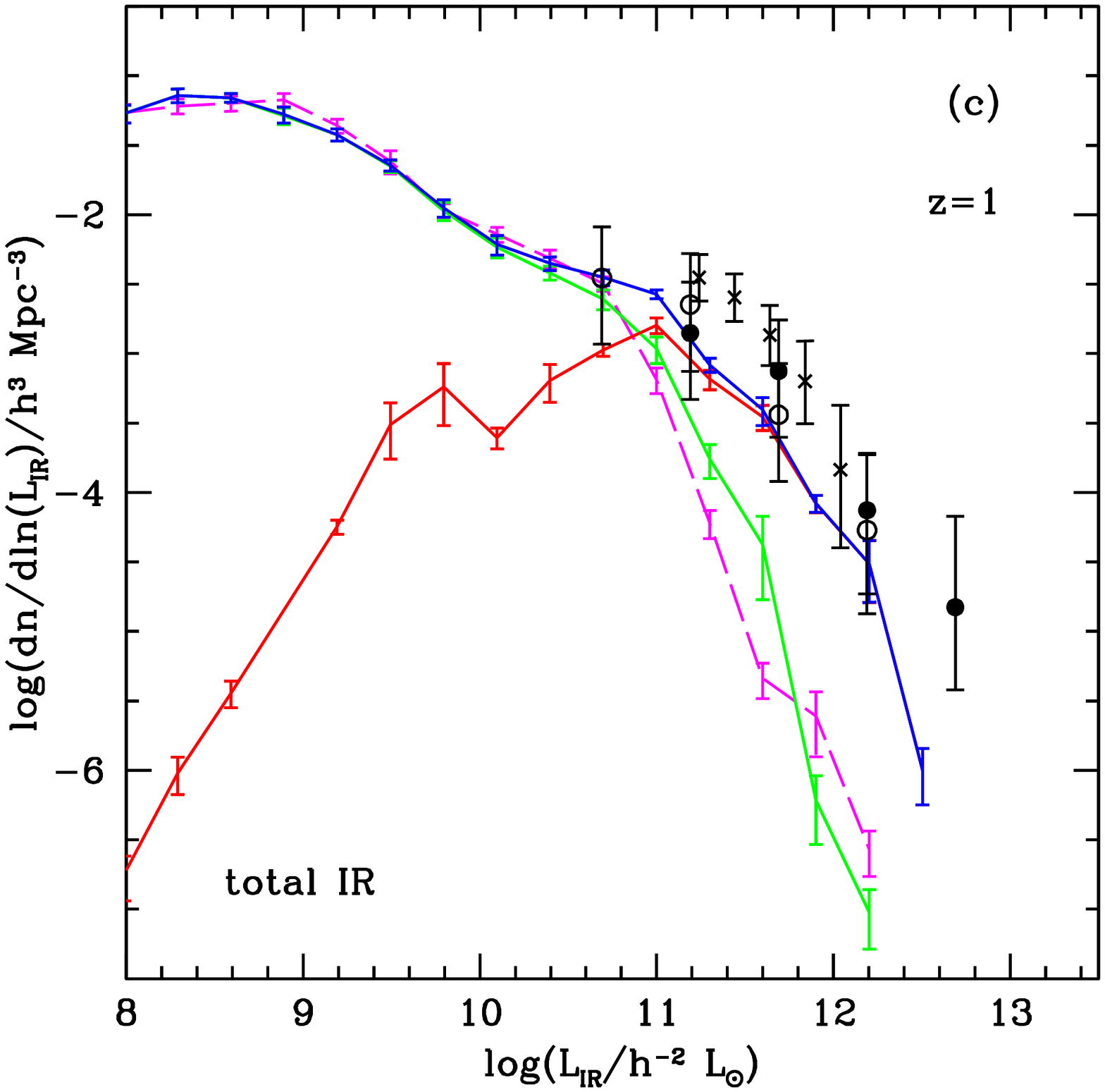}
\end{minipage}
\hspace{1cm}
\begin{minipage}{7cm}
\includegraphics[width=7cm]{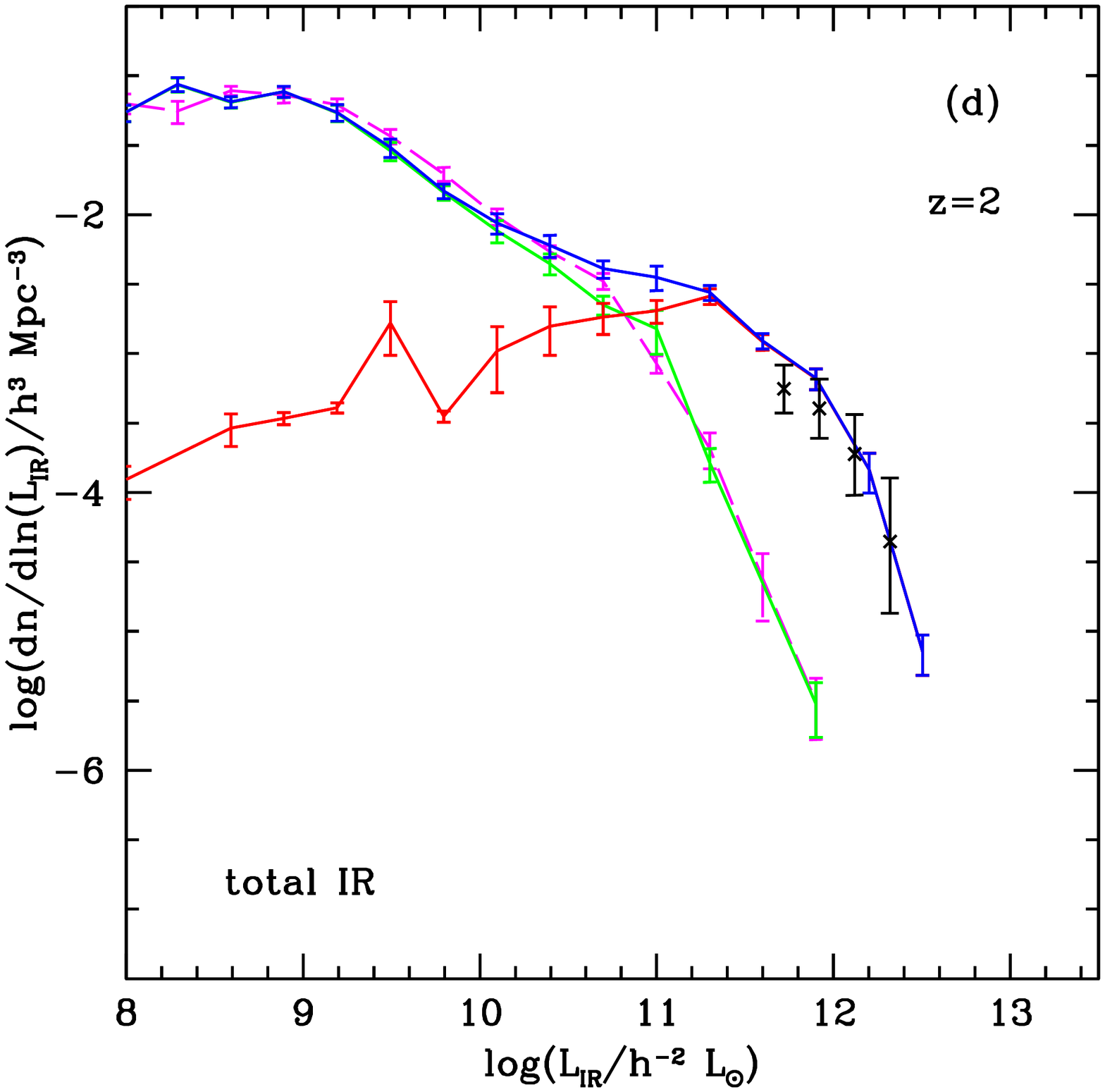}
\end{minipage}

\end{center}

\caption{Predicted evolution of the total mid+far IR (8-1000$\mum$)
galaxy LF compared to observational data. The different panels show
redshifts (a) $z=0$, (b) $z=0.5$, (c) $z=1$ and (d) $z=2$. For $z=0$,
we compare with observational data from \citet{Sanders03} (filled
symbols) and \citet{Takeuchi03} (open symbols, converting his 60
$\mum$ LF to a total IR LF assuming a constant conversion factor,
$L_{IR}/\nu L_{\nu}(60\mum) = 2.5$). We compare with data from
\citet{LeFloch05} for $z=0.5$ and $z=1$ (filled and open symbols), and
with \citet{Caputi06b} for $z=1$ and $z=2$ (crosses). }

\label{fig:lfIR-evoln-obs}
\end{figure*}

\subsection{Evolution of the total mid+far-IR luminosity function}

The total mid+far IR luminosity of a galaxy, $L_{IR}$, integrated over
the whole wavelength range 8-1000 $\mum$, is a very good approximation
to the total luminosity emitted by interstellar dust grains in all
galaxies except those with very small dust contents. In galaxies with
significant star formation, $L_{IR}$ is mostly powered by dust heated
by young stars, and so provides a quantitative indicator of the amount
of dust-obscured star formation which is independent of the shape of
the IR SED (though still subject to uncertainties about the IMF). The
evolution of the luminosity function in $L_{IR}$ is therefore a very
interesting quantity to compare between models and observations. We
show in Fig.~\ref{fig:lfIR-evoln} what our standard model predicts for
the evolution of the IR LF over the range $z=0-6$. We see that the
model predicts substantial evolution in this LF, with the high
luminosity end brightening by a factor $\sim 10$ from $z=0$ to $z=2$,
followed by a ``plateau'' from $z=2$ to $z=4$, and  a decline
from $z=4$ to $z=6$.

In Fig.~\ref{fig:lfIR-evoln-obs}, we compare our model predictions
with existing observational estimates of the total IR LF for
$z=0-2$. These observational estimates are only robust for $z=0$,
where they are based on \IRAS\ measurements covering the wavelength
range 12-100 $\mum$. At all of the higher redshifts plotted, the
observational estimates are based on measurements of the mid-IR
luminosity derived from \SPITZER\ 24 $\mum$ fluxes, converted to total
IR luminosities by assuming SED shapes for the mid- to far-IR
emission. The bolometric correction from the observed mid-IR
luminosity to the inferred total IR luminosity is typically a factor
$\sim 10$, and is significantly uncertain. Therefore, the most robust
way to compare the models with the observations is to compare them at
the mid-IR wavelengths where the measurements are actually made, as we
have done in \S\ref{sec:lf_3-8} and \S\ref{sec:lf_12-24}.
Nonetheless, if we take the observational determinations
at face value, then we see that observed evolution of the total IR LF
agrees remarkably well with the predictions of our standard model with
a top-heavy IMF. On the other hand, the variant model with a normal
IMF predicts far too few high $L_{IR}$ galaxies at higher z, and is
strongly disfavoured by the existing data.


\section{Inferring stellar masses and star formation rates from
  \SPITZER\ data}
\label{sec:mstar-sfr}

\begin{figure*}

\begin{center}

\begin{minipage}{7cm}
\includegraphics[width=7cm]{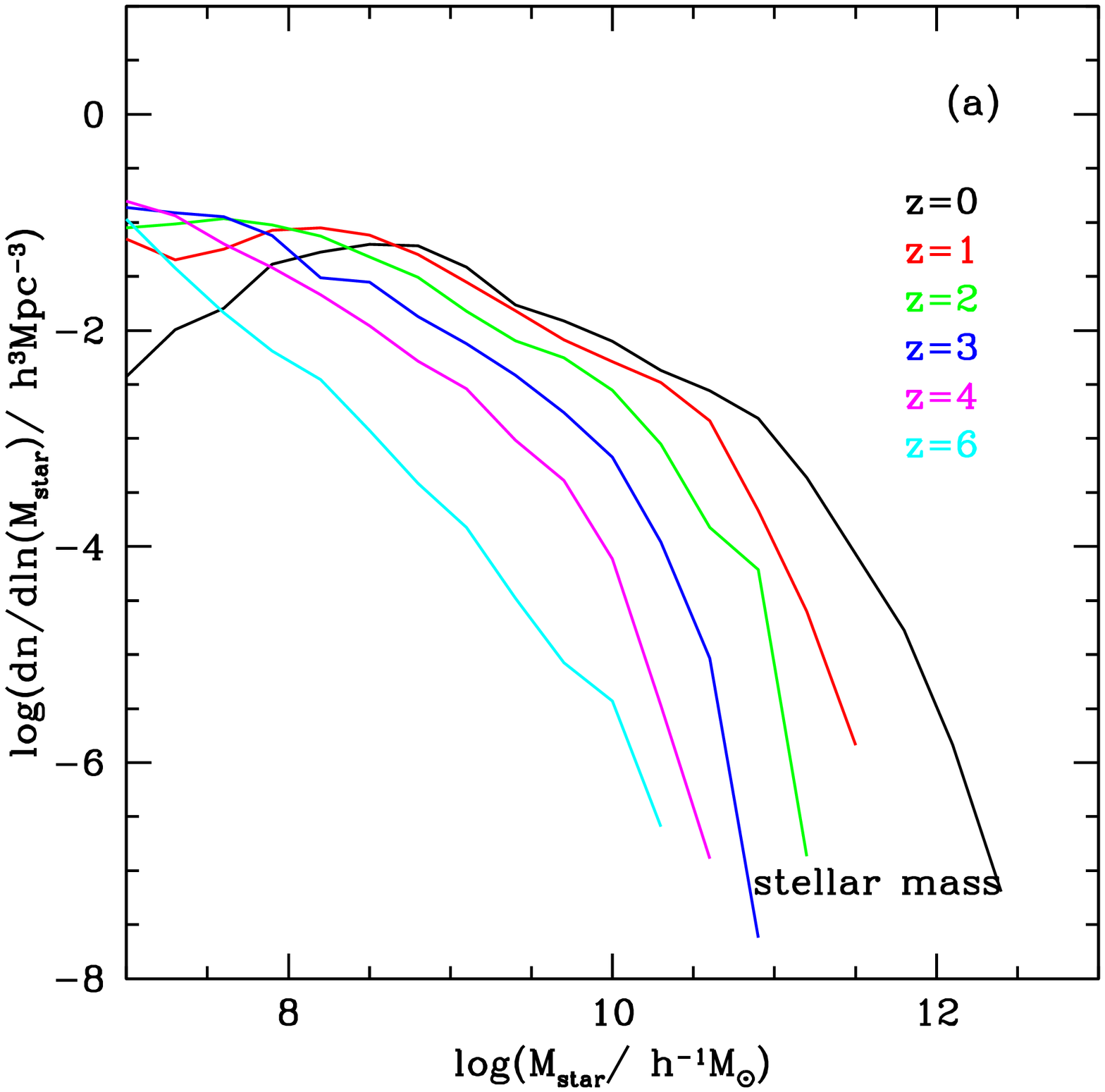}
\end{minipage}
\hspace{1cm}
\begin{minipage}{7cm}
\includegraphics[width=7cm]{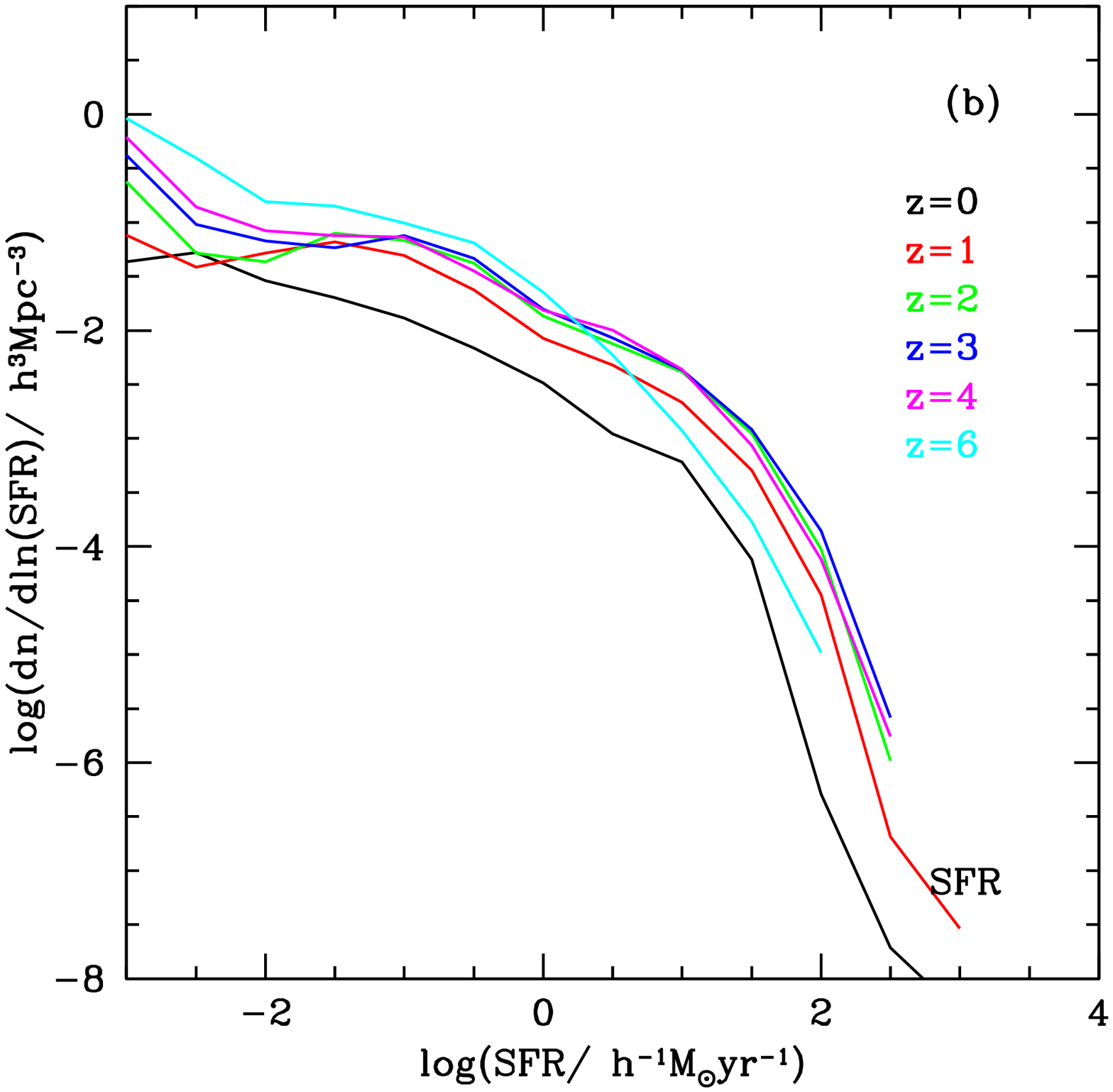}
\end{minipage}

\begin{minipage}{7cm}
\includegraphics[width=7cm]{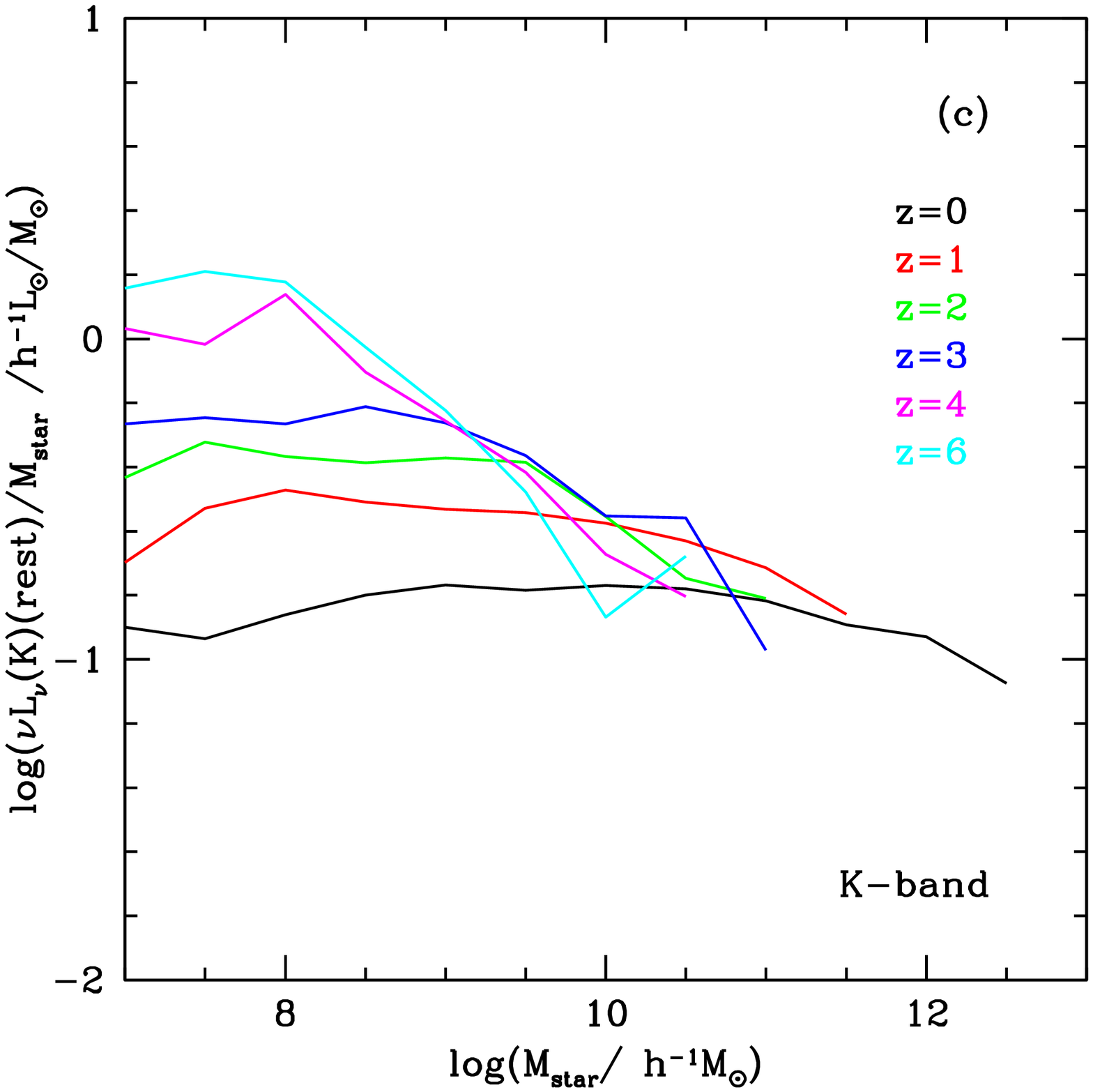}
\end{minipage}
\hspace{1cm}
\begin{minipage}{7cm}
\includegraphics[width=7cm]{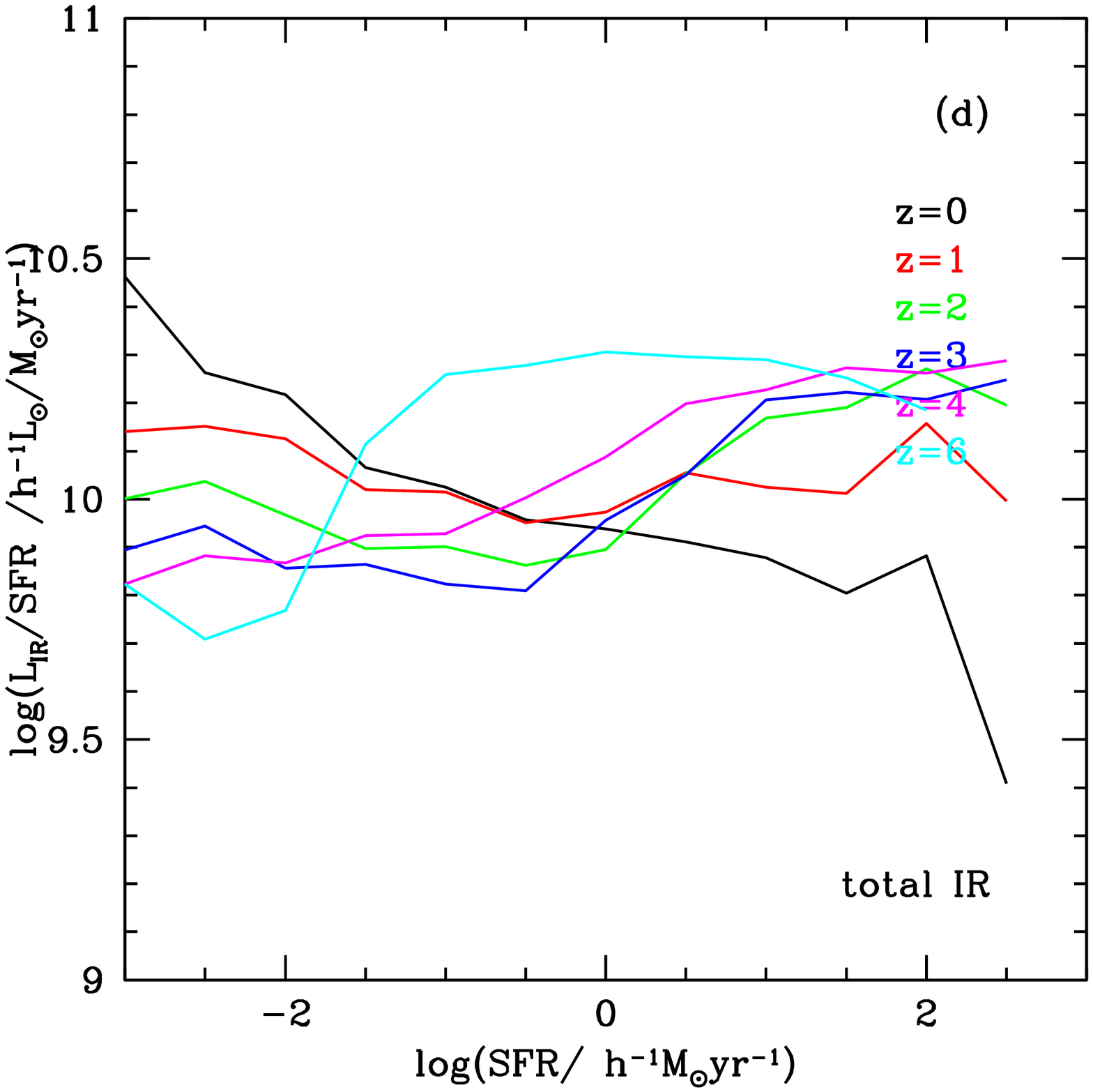}
\end{minipage}

\begin{minipage}{7cm}
\includegraphics[width=7cm]{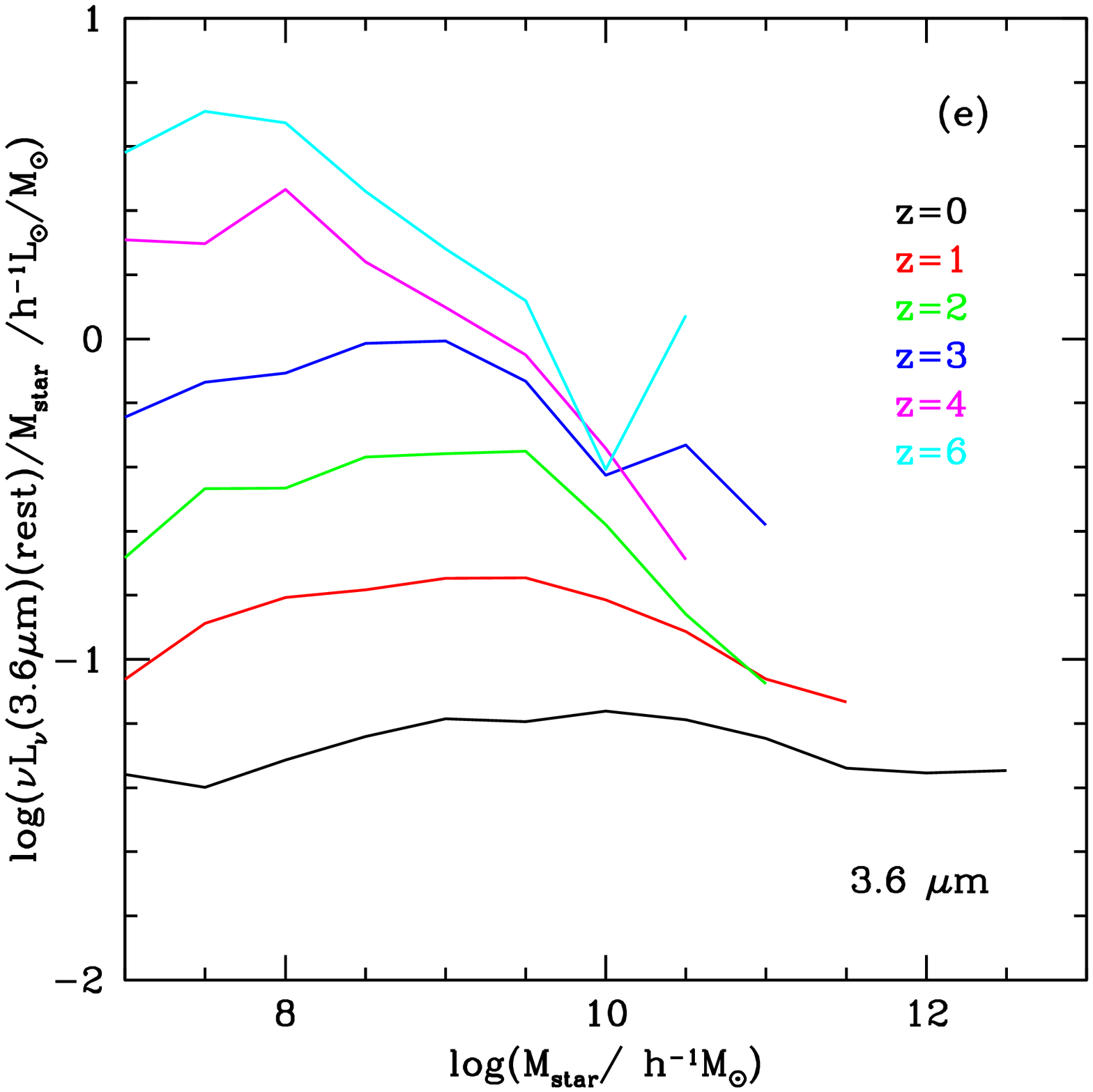}
\end{minipage}
\hspace{1cm}
\begin{minipage}{7cm}
\includegraphics[width=7cm]{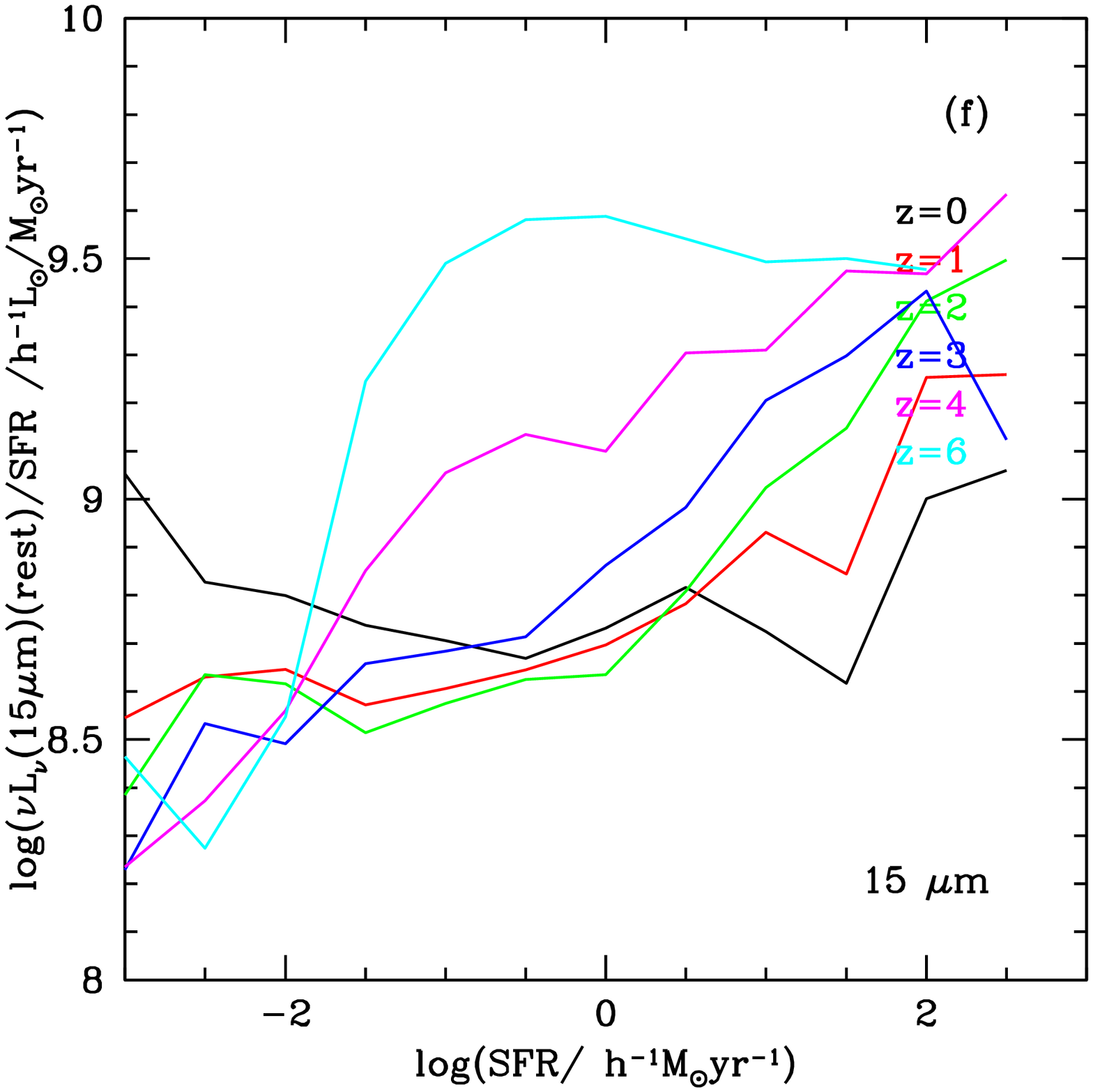}
\end{minipage}

\end{center}

\caption{Model predictions for properties related to stellar masses
  (left column) and star formation rates (right column), for redshifts
  $z=0$, 1, 2, 3, 4, and 6: (a) galaxy stellar mass function (GSMF);
  (b) galaxy star formation rate distribution (GSFRD); (c) mean ratio
  of rest-frame K-band luminosity to stellar mass, as a function of
  stellar mass; (d) mean ratio of total mid+far IR luminosity to SFR,
  as a function of SFR; (e) mean ratio of rest-frame 3.6 $\mum$
  luminosity to stellar mass, as a function of stellar mass; (f) mean
  ratio of rest-frame 15 $\mum$ luminosity to SFR, as a function of
  SFR. (The 15 $\mum$ luminosity is here calculated through top-hat
  filter with a fractional wavelength width of 10\%.)}

\label{fig:mstar-sfr}
\end{figure*}

In this section, we consider what the models imply about how well we
can infer the stellar masses and star formation rates (SFRs) in
galaxies from measurements of rest-frame IR luminosities. The top two
panels of Fig.~\ref{fig:mstar-sfr} show the predicted galaxy stellar
mass function (GSMF, left panel) and galaxy star formation rate
distribution (GSFRD, right panel), for redshifts $z=0-6$. We see that
the predicted stellar mass function shows dramatic evolution over this
redshift range, with a monotonic decline in the number of high-mass
galaxies with increasing redshift. On the other hand, the SFR
distribution shows much less dramatic evolution over this redshift
range, with a mild increase in the number of high-SFR objects up to
$z\sim 3$, followed by a decline above that. The lower four panels in
Fig.~\ref{fig:mstar-sfr} show the relation in the models between
stellar masses and SFRs and rest-frame luminosities at different IR
wavelengths. (Note that in all cases, luminosities are measured in
units of the {\em bolometric} solar luminosity.)  The middle and
bottom left panels respectively show the mean ratio of luminosity in
the rest-frame K ($2.2 \mum$) or 3.6 $\mum$ bands to stellar mass as a
function of stellar mass. The middle and bottom right panels
respectively show the mean ratio of total mid+far-IR ($8-1000 \mum$)
or rest-frame 15~$\mum$ luminosity to SFR as a function of SFR. (The
mean $L/M_*$ or $L/SFR$ ratios plotted are computed by dividing the
total luminosity by the total mass or SFR, in each bin of mass or
SFR.)

The near-IR luminosity is often used as a tracer of stellar mass. The
left panels of Fig.~\ref{fig:mstar-sfr} show that the $L/M_*$ ratio
varies strongly with redshift, reflecting the difference in the ages
of the stellar populations.  At higher redshifts it also shows a
significant dependence on stellar mass, presumably reflecting a trend
of age with mass. However, the variation of mean $L/M_*$ with redshift
is seen to be much smaller at 2.2 $\mum$ than at 3.6 $\mum$, implying
that the rest-frame K-band light should provide a more robust
estimator of stellar mass than the light at longer wavelengths. The
differences between $L/M_*$ values at 2.2 $\mum$ and 3.6 $\mum$ reflect
the larger contribution from AGB compared to RGB stars at the longer
wavelength. AGB stars have higher masses and younger ages than RGB
stars, and so are more sensitive to star formation at recent
epochs. The scatter in $L/M_*$ at a given mass is also found in the
models to increase with redshift. In the K-band, it increases from
$\sim 40\%$ at $z\sim 0$ to a factor $\sim 3$ at $z\sim 6$. The large
scatter at high redshifts results in part from having two different
IMFs.

The luminosity in the mid- and far-IR is widely used as a tracer of
dust-obscured star formation (although in galaxies with very low star
formation rates, the dust heating can be dominated by older
stars). The total mid+far-IR (rest-frame 8-1000 $\mum$) luminosity is
expected to provide a more robust tracer of star formation than the
luminosity at any single IR wavelength, since the shape of the SED of
dust emission depends on the dust temperature distribution (as well as
on the dust grain properties). This is borne out by our model
predictions. The middle right panel of Fig.~\ref{fig:mstar-sfr} shows
that the $L_{IR}/SFR$ ratio depends weakly on both SFR and
redshift. This behaviour results mostly from having different IMFs in
the model in quiescent and bursting galaxies, with the fractional
contribution of the bursts increasing both with SFR and with redshift.
If we look at quiescent and bursting galaxies separately, we find
roughly constant ratios $L_{IR}/SFR \approx 6\times 10^9
h^{-1}\Lsol/\Msol$ and $L_{IR}/SFR = 2\times 10^{10}
h^{-1}\Lsol/\Msol$ respectively, for galaxies where $L_{IR}$ is
powered mostly by young stars. However, there is also a trend at lower
redshift for $L_{IR}/SFR$ to be larger at lower SFR - this reflects
the larger fraction of dust heating from older stars in galaxies with
lower SFRs, which more than compensates for the lower average dust
obscuration in these galaxies. The lower right panel of
Fig.~\ref{fig:mstar-sfr} shows that the $L/SFR$ ratio in the mid-IR
(in this case at 15 $\mum$ in the rest-frame) shows more variation
with SFR and redshift than the ratio for the total IR luminosity. This
reflects the variation in the mid- to far-IR SED shapes in the
model. The scatter in the $L/SFR$ ratio is roughly a factor 2 around
the average relation for the total IR luminosity, but is larger for
the 15 $\mum$ luminosity.

The results of this section illustrate why it is not straightforward
to compare theoretical predictions for the evolution of the galaxy
stellar mass function and star formation rate distribution (or even
the stellar mass and star formation rate densities) with observational
estimates. In addition to assumptions about galaxy star formation
histories and metallicities (for stellar mass estimates), and about
the SED shapes for dust emission (for SFR estimates from IR and sub-mm
data), observational estimates all rest on some assumed form for the
IMF. If the IMF assumed in the observational analysis is different
from the true IMF, the observational estimates for stellar masses and
SFRs can be wrong by large factors. If the IMFs differ only below $ 1
\Msol$, then one can apply a simple rescaling to relate stellar mass
and SFR estimates for different IMFs. However, if our current galaxy
formation model is correct, stars form with different IMFs in
quiescent disks and in merger-driven bursts, and so no observational
estimate based on assuming a single IMF can give the correct GSMFs and
GSFRDs, nor the correct stellar mass and SFR densities. A direct
comparison of the GSMF and GSFRD evolution predicted by our model with
observational estimates is therefore not meaningful. Instead, the
comparison between models and observations must be made via directly
observable (rather than inferred) quantities, such as the K-band
luminosities to constrain stellar masses, and the total IR
luminosities to constrain SFRs.

\section{Conclusions}
\label{sec:conc}

We have computed predictions for the evolution of the galaxy
population at infrared wavelengths using a detailed model of
hierarchical galaxy formation and of the reprocessing of starlight by
dust, and compared these predictions with observational data from the
\SPITZER\ Space Telescope. We calculated galaxy formation in the
framework of the $\Lambda$CDM model using the \GALFORM\ semi-analytical
model, which includes physical treatments of the hierarchical assembly
of dark matter halos, shock-heating and cooling of gas, star
formation, feedback from supernova explosions and photo-ionization of
the IGM, galaxy mergers and chemical enrichment. We computed the IR
luminosities and SEDs of galaxies using the \GRASIL\ multi-wavelength
spectrophotometric model, which computes the luminosities of the
stellar populations in galaxies, and then the reprocessing of this
radiation by dust, including radiative transfer through a two-phase
dust medium, and a self-consistent calculation of the distribution of
grain temperatures in each galaxy based on a local balance between
heating and cooling. The \GRASIL\ model includes a treatment of the
emission from PAH molecules, which is essential for understanding the
mid-IR emission from galaxies.

Our galaxy formation model incorporates two different IMFs: quiescent
star formation in galaxy disks occurs with a normal solar
neighbourhood IMF, but star formation in bursts triggered by galaxy
mergers happens with a top-heavy $x=0$ IMF. In a previous paper
\citep{Baugh05}, we found that the top-heavy IMF in bursts was
required in order that the model reproduces the observed number counts
of the faint sub-mm galaxies detected at 850 $\mum$, which are
typically ultra-luminous starbursts at $z \sim 2$, with total IR
luminosities $L_{IR} \sim 10^{12} - 10^{13} \Lsol$. This conclusion
was arrived at following a search of a large grid of model parameters,
with the imposition of a variety of detailed observational
constraints. The parameters in the \citet{Baugh05} model were chosen
before the publication of any results from \SPITZER, without reference
to any IR data apart from the local 60 $\mum$ luminosity function and
the 850 $\mum$ galaxy counts. We have kept the same parameter values
in the present paper, in order to test what the same model predicts at
other wavelengths and other redshifts. By doing this, we hope to
address the criticism made of many semi-analytical models that they
have no predictive power, because their parameters are always adjusted
to match the observational data being analysed at that instant.

We first compared the predictions from our model with the galaxy
number counts measured in all 7 \SPITZER\ bands, from 3.6 to 160
$\mum$. We found broad agreement between the model and the
observations. In the 4 IRAC bands (3.6-8.0 $\mum$), where the counts
are mostly dominated by emission from older stellar populations, we
found that the predicted counts were insensitive to whether we had a
top-heavy or normal IMF in bursts. On the other hand, in the MIPS
bands (24-160 $\mum$), where the counts are dominated by emission from
dust in star-forming galaxies, the predicted counts are more sensitive
to the choice of IMF, and the counts are fit better by the model with
a top-heavy IMF. We next investigated the evolution of the galaxy
luminosity function at IR wavelengths, where several groups have now
used \SPITZER\ data to try to measure the evolution of the galaxy
luminosity function over the redshift range $z \sim 0-2$, at
rest-frame wavelengths from 3.6 to 24 $\mum$.

Our model predicts that at mid- and far-IR rest-frame wavelengths, the
luminosity function evolution is very sensitive to the choice of IMF
in bursts. We found that our standard model with a top-heavy IMF in
bursts fits the measured evolution of the mid-IR luminosity function
remarkably well (when allowance is made for complexity of predicting
dust emission in the mid-IR), without any adjustment of the
parameters. On the other hand, a model with a normal IMF in bursts
predicts far too little evolution in the mid-IR luminosity function
compared to what is observed. We made a similar comparison with the
evolution of the total IR luminosity function, where in the case of
the observations, the total IR luminosities at high redshifts have
been inferred from the 24 $\mum$ fluxes by fitting SEDs, and reached
the same conclusion. The evolution of the galaxy luminosity function
in the mid-IR found by \SPITZER\ thus supports our original conclusion
about the need for a top-heavy IMF in bursts, which was based only on
the sub-mm counts. This conclusion will be further tested by ongoing
\SPITZER\ surveys at longer wavelengths. To assist this, we have also
presented  predictions for the evolution of the luminosity
function in the \SPITZER\ 70$\mum$ and 160$\mum$ bands.

We have also presented predictions for the evolution of the stellar
mass function and star formation rate distribution of galaxies. We
investigated how the $L/M_*$ and $L/SFR$ ratios varied with galaxy
mass, SFR and redshift in different IR wavelength ranges, and
considered the implications for observational estimates of stellar
masses and SFRs from IR observations. Even in the near-IR, the
predicted variations in $L/M_*$ with mass and redshift can be
surprisingly large. The variations in $L/M_*$ are much larger at a
rest-frame wavelength of 3.6 $\mum$ than at 2.2 $\mum$, implying that
the 2.2 $\mum$ luminosity is a more robust tracer of stellar mass.

Finally, we have presented in an Appendix the predictions of our model
for the redshift distributions of galaxies selected at different IR
fluxes in the \SPITZER\ bands. 

One significant limitation of our model is that it does not include
the effects of AGN. Two effects are relevant here. The first is
feedback from AGN on galaxy formation. In several recent galaxy
formation models, AGN feedback is invoked to prevent the formation of
too many massive galaxies at the present day. In the model presented
here, we instead posit feedback from supernova-driven galactic
superwinds, which perform a similar role to AGN feedback in
suppressing the formation of very massive galaxies. Both the superwind
and AGN feedback models include free parameters which are tuned to
give a match to the present-day optical galaxy luminosity
function. However, the redshift dependence of the feedback will be
different between our superwind model and the various AGN feedback
models, so in general they will all predict different evolution of the
galaxy population with redshift. We will investigate galaxy evolution
in the IR in a model with AGN feedback in a future paper. The second
effect of AGN which we have not included is the emission from AGN and
their associated dust tori. In order to compensate for this, we have
wherever possible compared our model predictions with observations
from which the AGN contribution has been subtracted out. This was
possible for most of our comparisons of luminosity function
evolution. This was not possible for the number counts comparisons,
but in this case the contribution from AGN is thought (based on
observations) to be a small fraction of the total over the flux range
explored by \SPITZER, even in the mid-IR where the dust tori are the
most prominent. We therefore believe that emission from AGN does not
seriously affect our conclusions about the IR evolution of
star-forming galaxies. We hope to include AGN emission directly into
our models in the future.

We have thus shown that \SPITZER\ data provide a stringent test of
galaxy formation theory, by probing galaxy evolution, constraining
star formation rates and the role of dust to $z\sim 2$. We find that
an {\em ab initio} $\Lambda$CDM model gives an acceptable fit to the
\SPITZER\ data provided that $\sim 10\%$ of the stars in galaxies
today formed in bursts of star formation with a top-heavy IMF. Future
facilities like {\em Herschel}, {\em SPICA}, {\em JWST} and {\em ALMA}
will continue to exploit the valuable information on galaxy formation
contained in the IR part of the electromagnetic spectrum.

\section*{Acknowledgements} 
We thank T. Babbedge, K. Caputi, A. Franceschini, E. Le Floch, and
P. Perez-Gonzalez, for providing us with their observational data in a
convenient form.  CMB acknowledges the receipt of a Royal Society
University Research Fellowship. CSF is the recipient of a Royal
Society Wolfson Research Merit Award. This work was also supported by
the PPARC rolling grant for extragalactic astronomy and cosmology at
Durham.



\appendix

\section{Redshift distributions}
\label{sec:dndz}

\begin{figure*}

\begin{center}

\begin{minipage}{7cm}
\includegraphics[width=7cm]{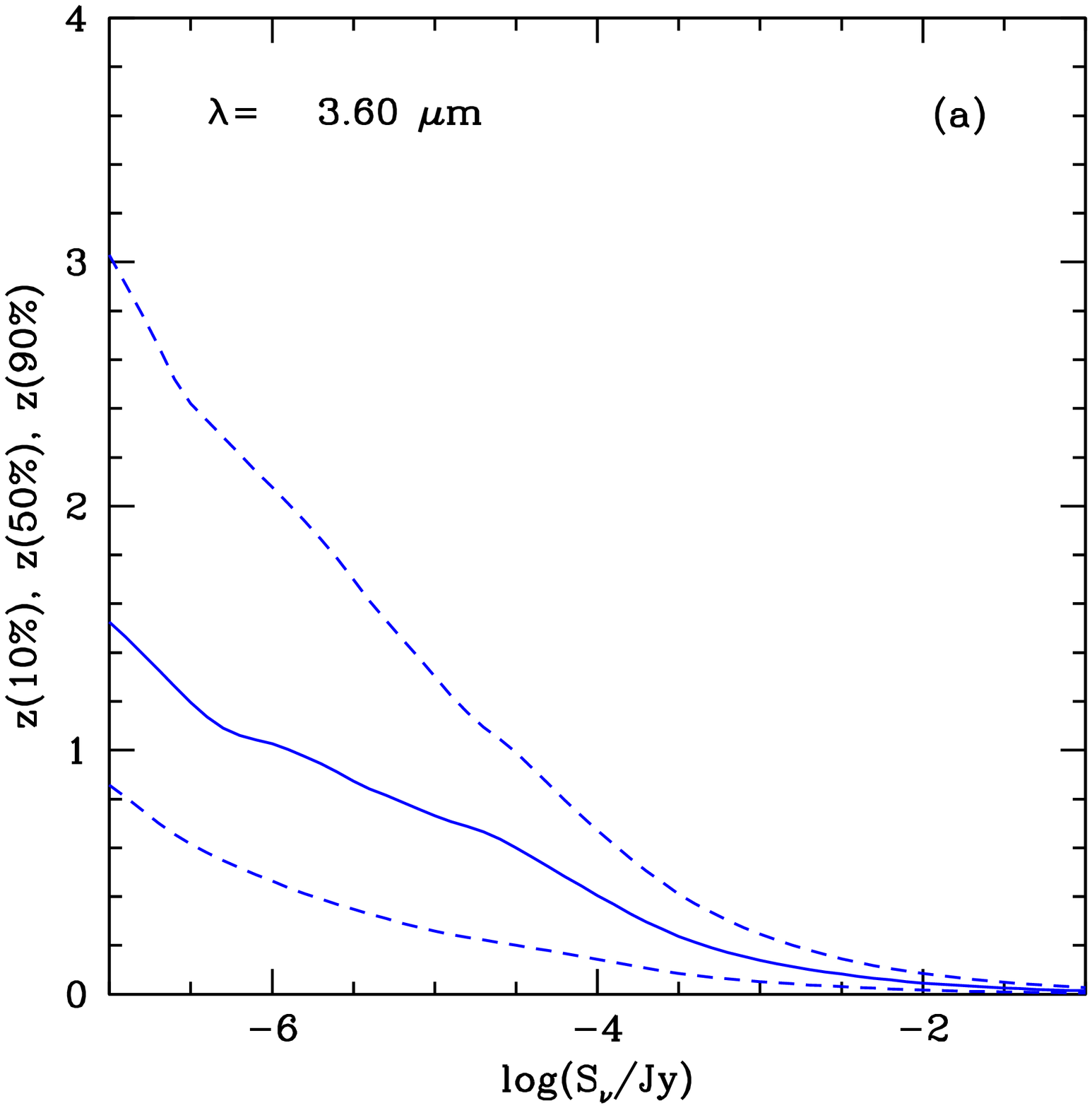}
\end{minipage}
\hspace{1cm}
\begin{minipage}{7cm}
\includegraphics[width=7cm]{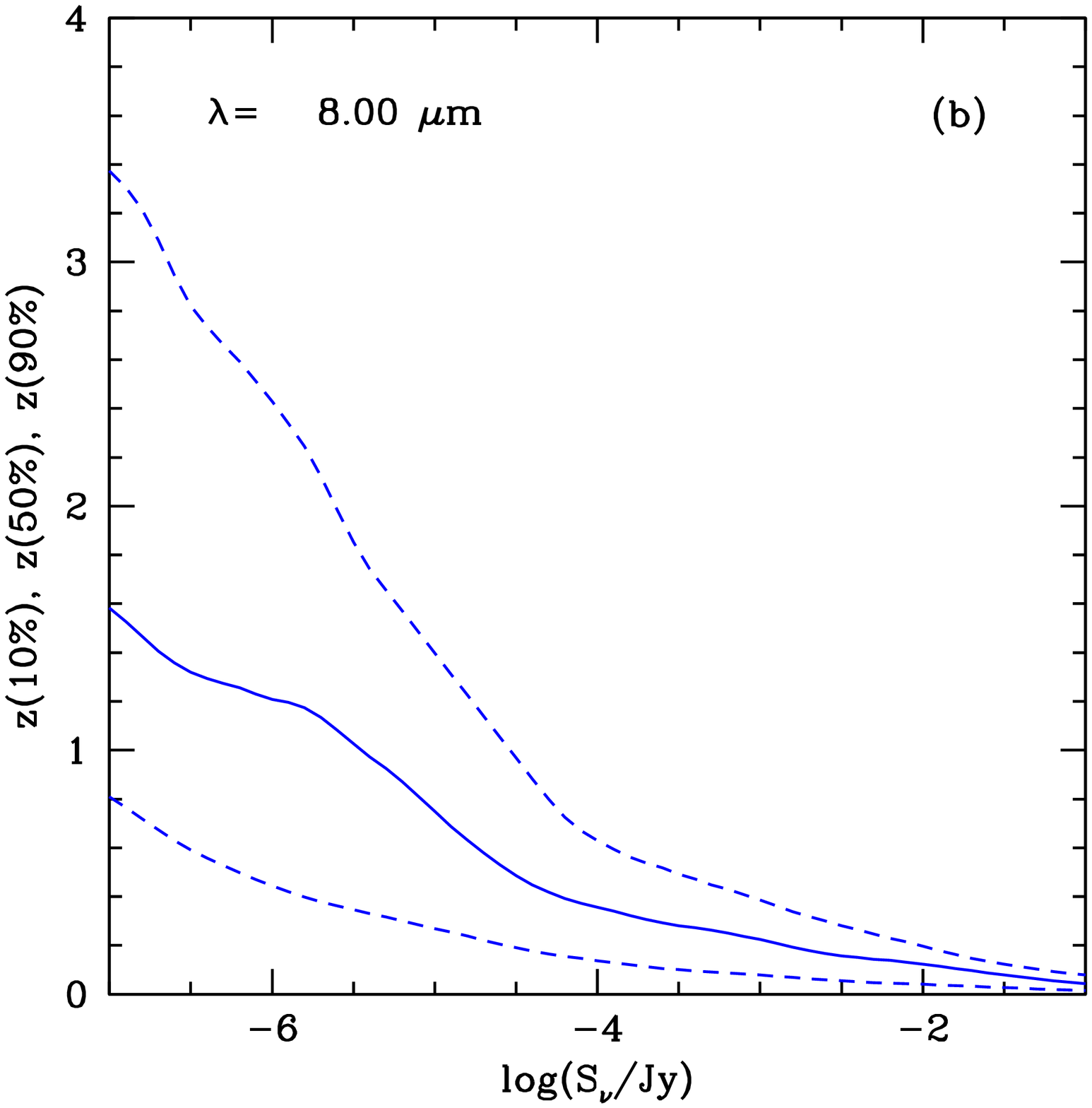}
\end{minipage}

\begin{minipage}{7cm}
\includegraphics[width=7cm]{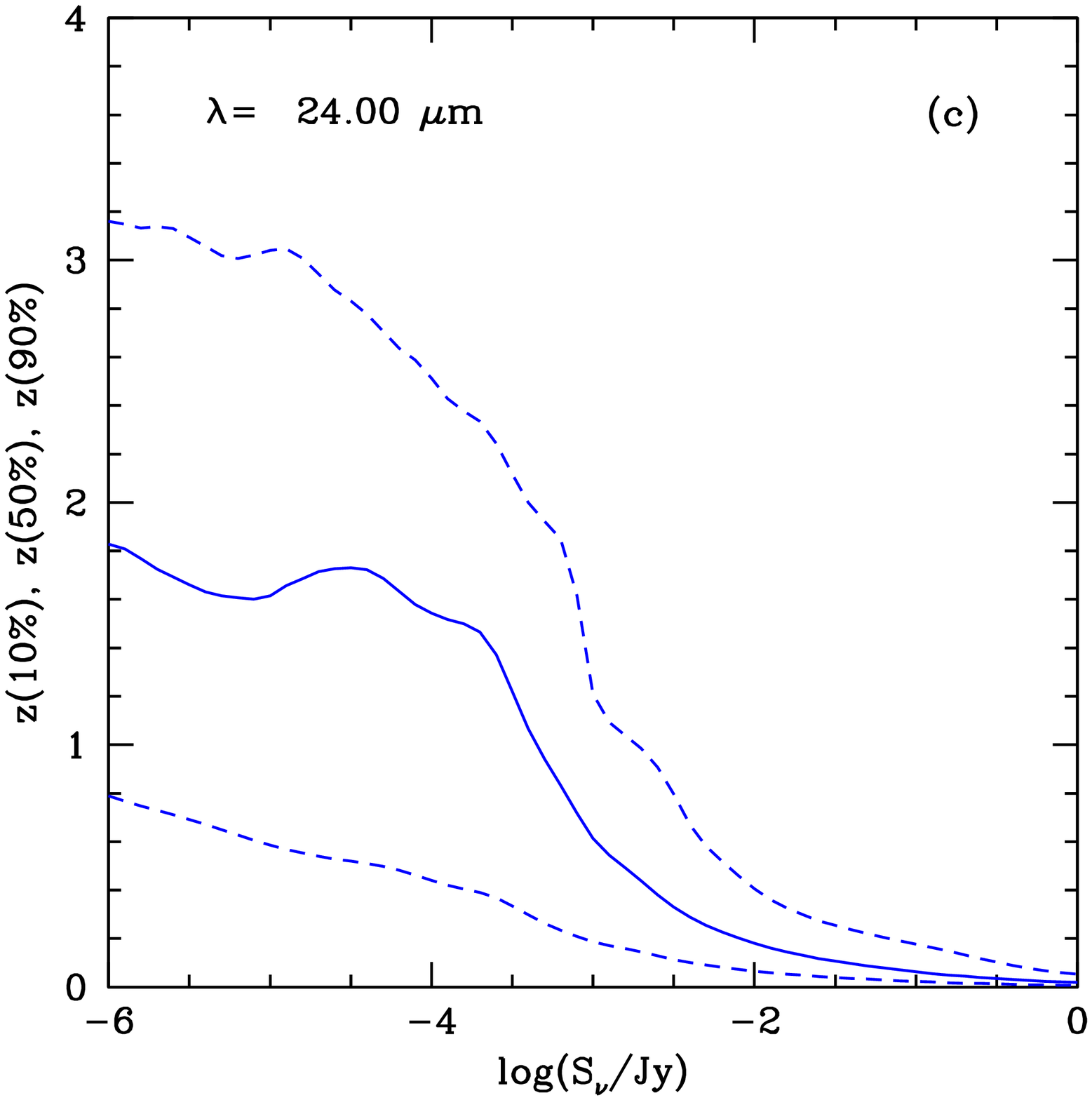}
\end{minipage}
\hspace{1cm}
\begin{minipage}{7cm}
\includegraphics[width=7cm]{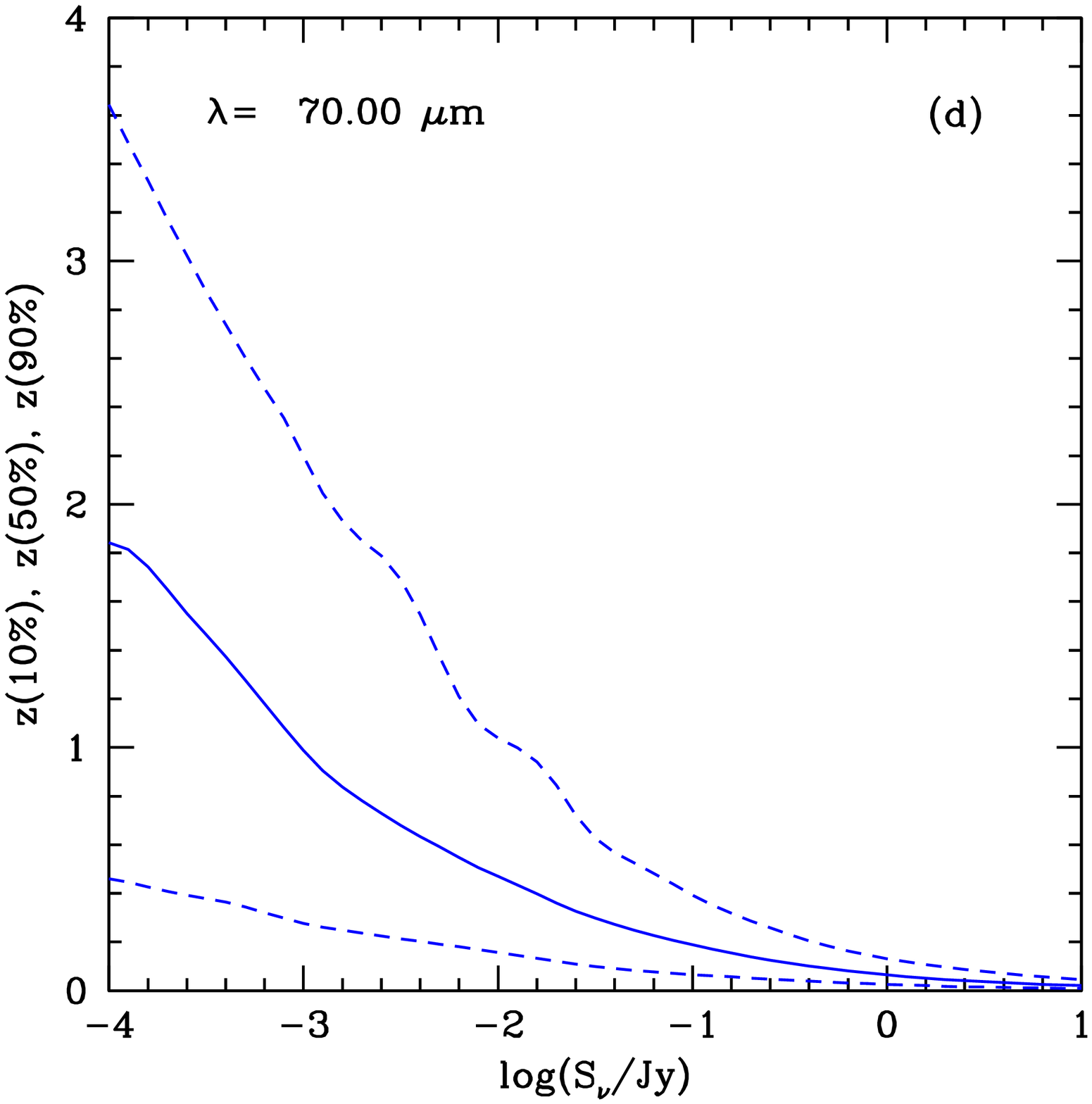}
\end{minipage}

\end{center}

\caption{Model predictions for the median redshift as a function of
  flux in four \SPITZER\ bands. (a) 3.6 $\mum$, 8.0 $\mum$, (c) 24
  $\mum$, (d) 70 $\mum$. In each panel, the median redshift for
  galaxies at each flux is shown by a solid line, and the 10- and
  90-percentiles are shown by dashed lines. }

\label{fig:zmedian}
\end{figure*}

In this Appendix, we present some predictions from our standard model
for the redshift distributions of galaxies selected at different
fluxes in the \SPITZER\ bands. This is principally for completeness,
to assist in interpreting data from current surveys, and to assist in
planning future surveys based on \SPITZER\ data. The set of redshift
distributions at all observed fluxes in principle contains equivalent
information to that in the luminosity functions at different
wavelengths and redshifts. However, comparing models with observations
via luminosity functions is more physically transparent than making
the comparison via redshift distributions, which is why we have
presented our results on luminosity functions in the main part of the
paper, and why we make only a limited direct comparison with observed
redshift distributions in this Appendix. In addition, if one only
compares the predicted and observed redshift distributions for
galaxies above a single flux limit (e.g. the flux limit of a survey),
this has less information than comparing the luminosity functions at
different redshifts.

We first show in Fig.~\ref{fig:zmedian} how the median redshift, and
the 10-90 percentile range, are predicted to change with flux for
galaxies selected in one of the four \SPITZER\ bands 3.6, 8.0, 24 or
70 $\mum$. While at most wavelengths the median redshift is predicted
to increase smoothly and monotonically with decreasing flux, this is
not true at 24 $\mum$, where there is a bump around $S_{\nu} \sim 100
\muJy$. The structure seen for the 24 $\mum$ band as compared to the
other wavelengths results from different PAH emission features moving
through the band with increasing redshift.

\begin{figure*}

\begin{center}

\begin{minipage}{7cm}
\includegraphics[width=7cm]{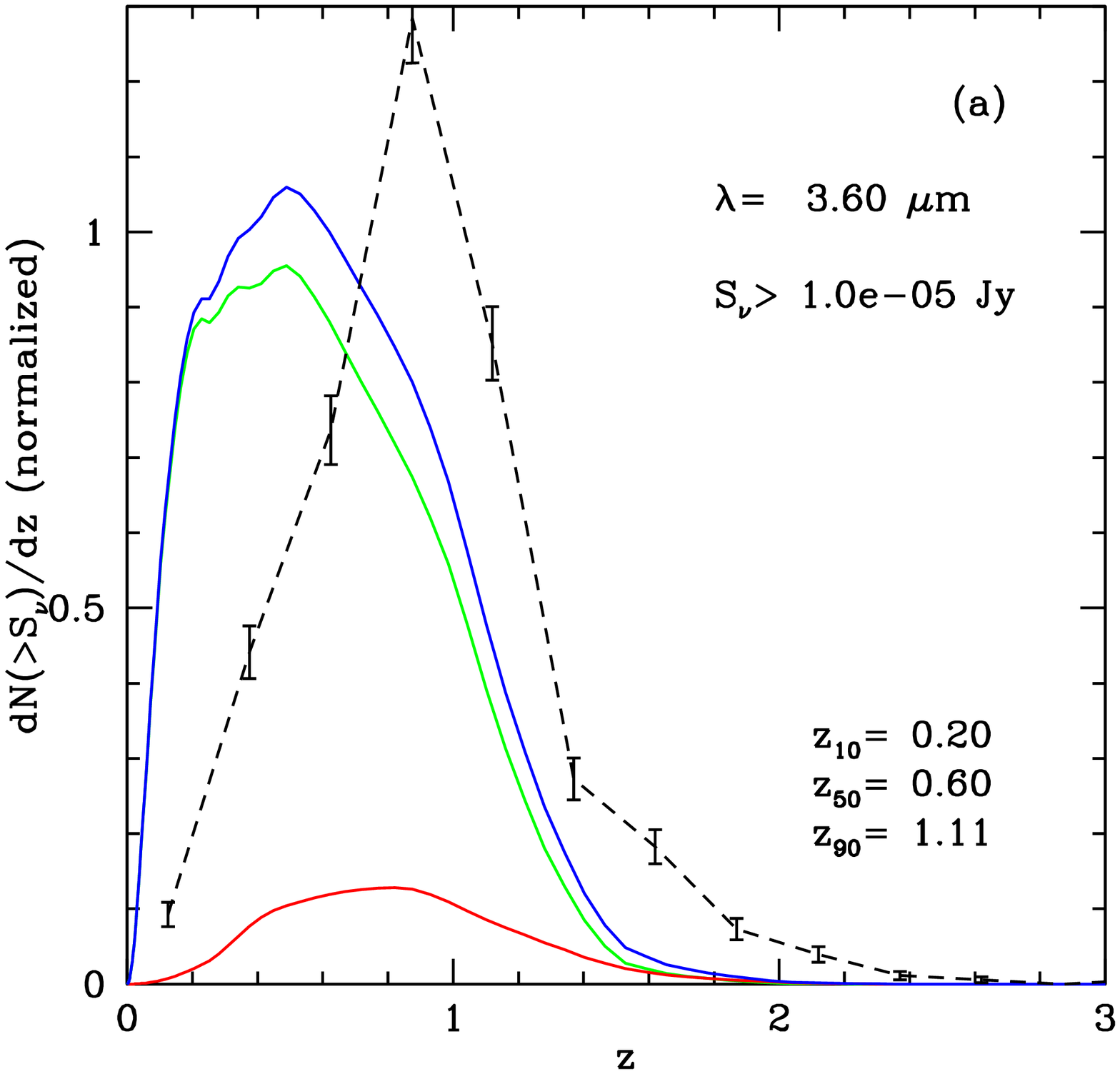}
\end{minipage}
\hspace{1cm}
\begin{minipage}{7cm}
\includegraphics[width=7cm]{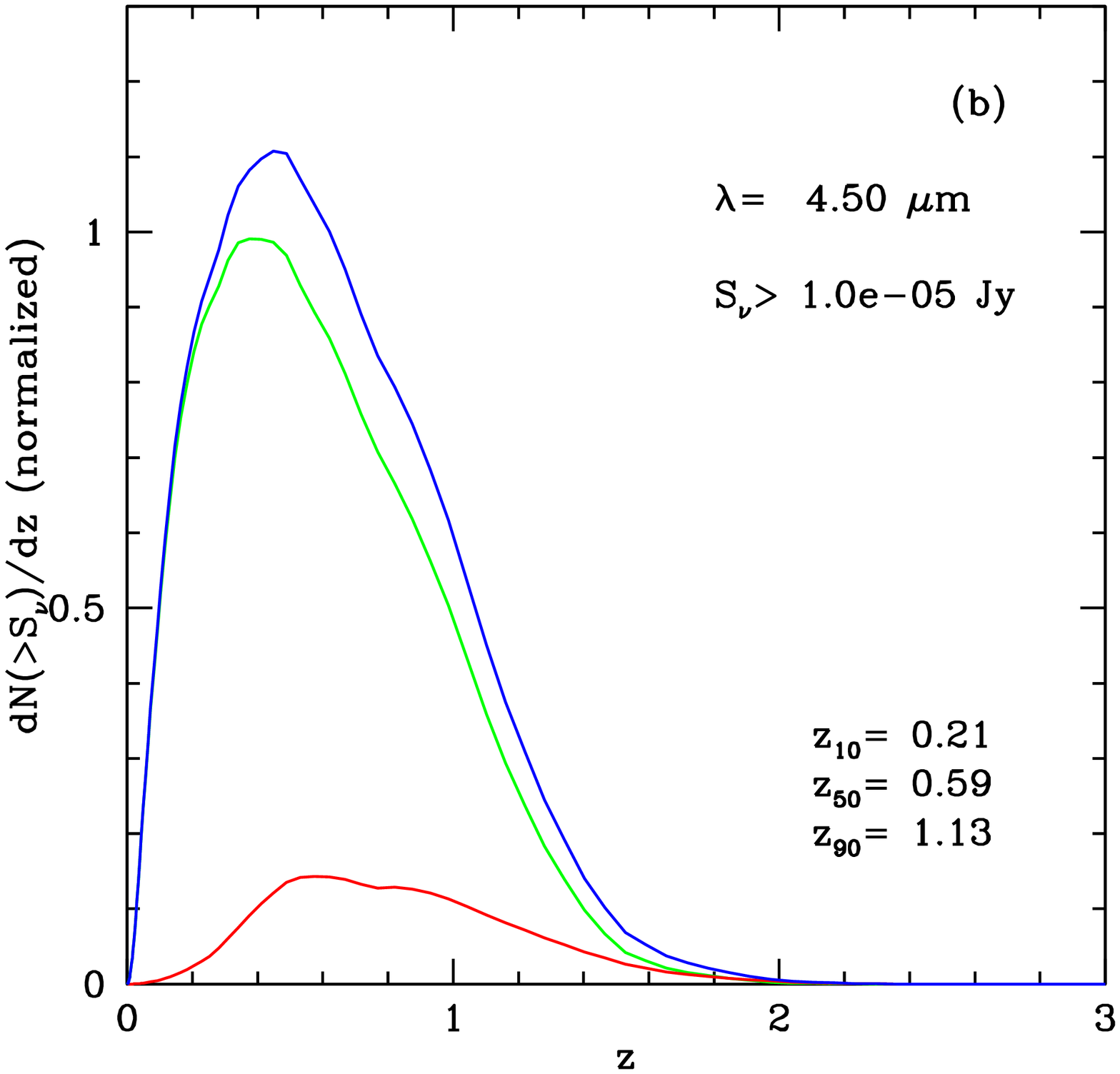}
\end{minipage}

\begin{minipage}{7cm}
\includegraphics[width=7cm]{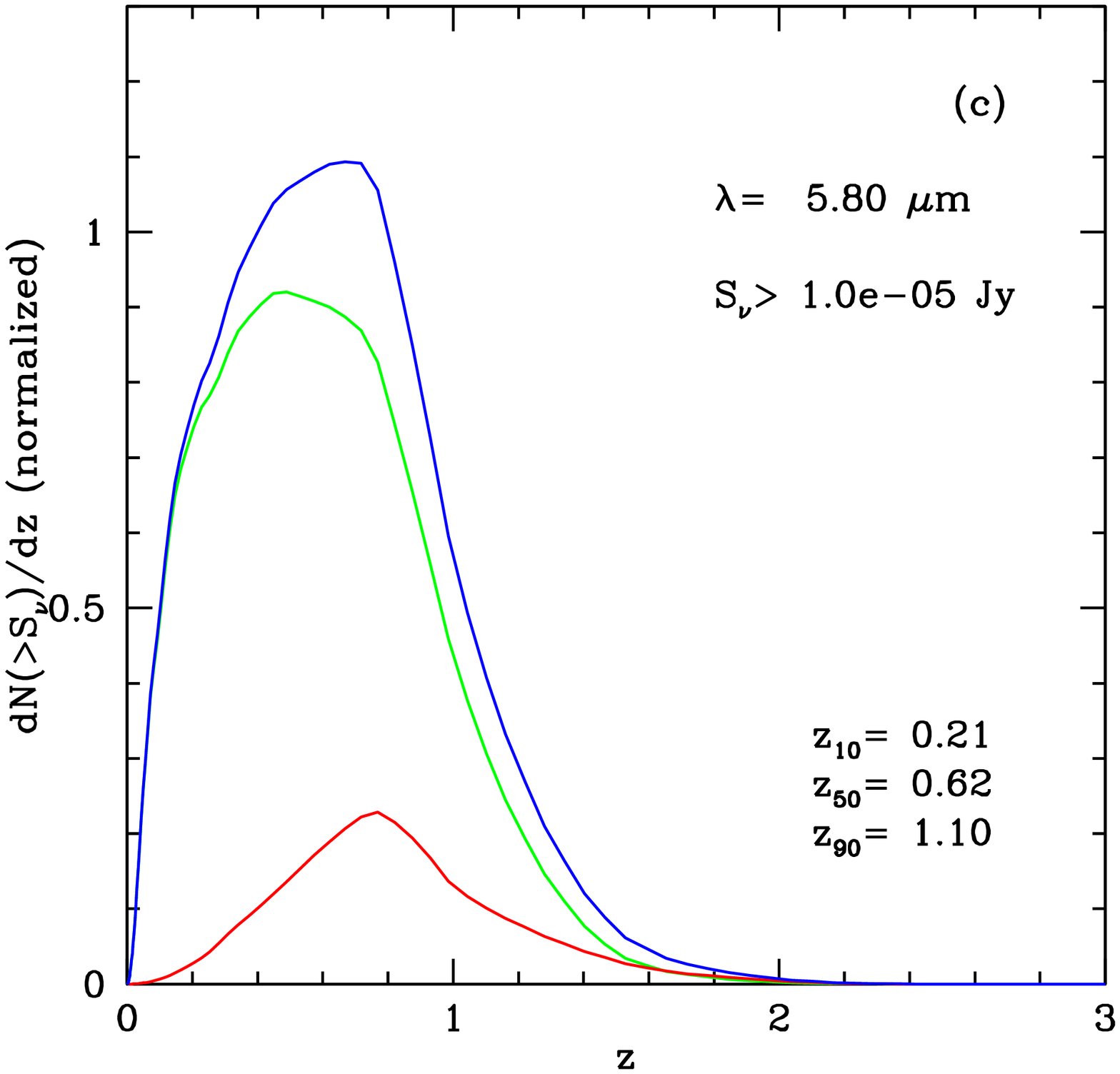}
\end{minipage}
\hspace{1cm}
\begin{minipage}{7cm}
\includegraphics[width=7cm]{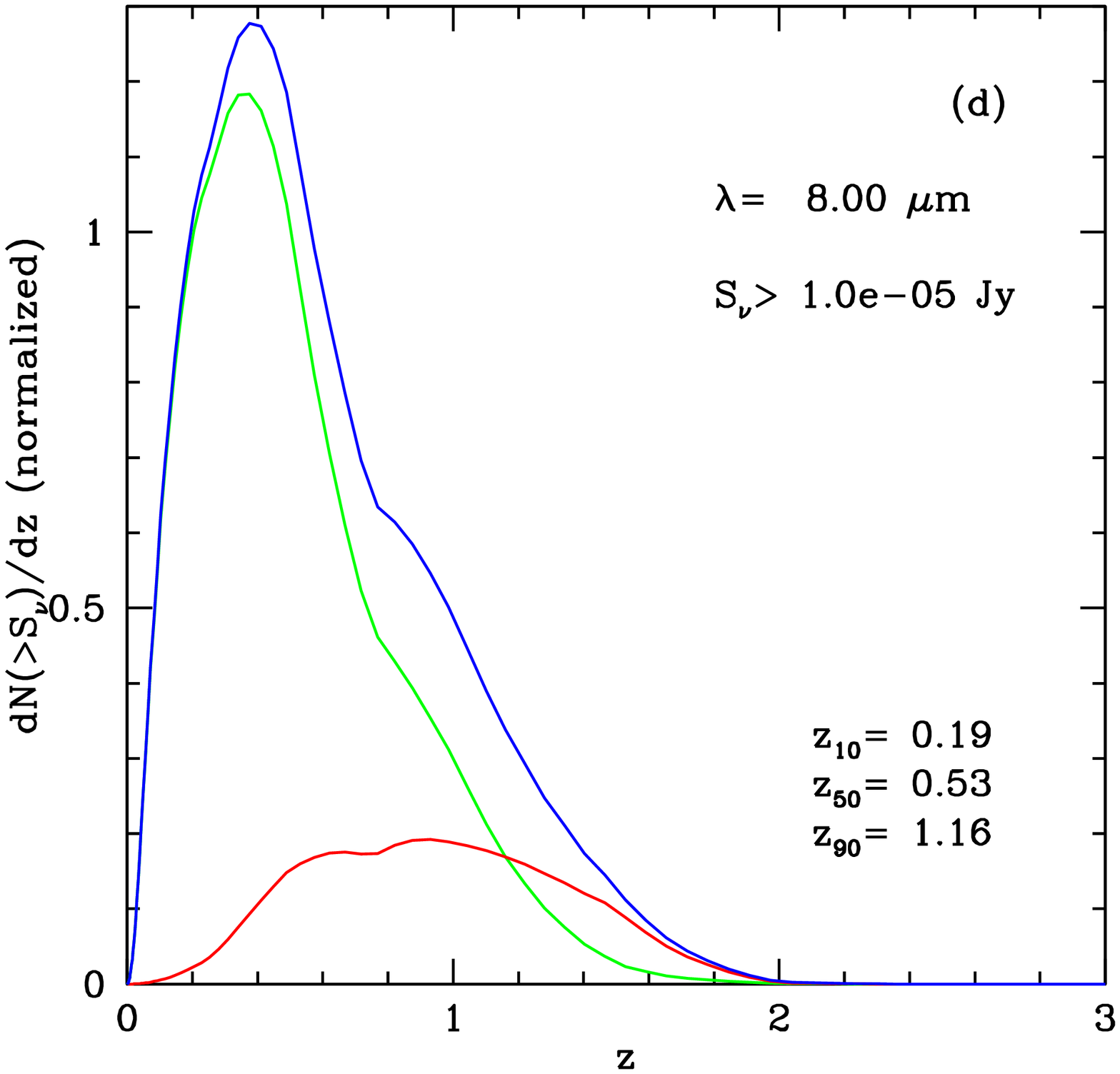}
\end{minipage}

\end{center}

\caption{Predicted galaxy redshift distributions in the four IRAC
  bands, for galaxies brighter than $S_{\nu}= 10 \muJy$. (a) 3.6
  $\mum$, (b) 4.5 $\mum$, (c) 5.8 $\mum$, and (d) 8.0 $\mum$.  The
  model curves (which all include the effects of dust) are as follows
  - blue: total; red: ongoing bursts; green: quiescent galaxies. The
  curves are normalized to unit area under the curve for the total
  counts. The median ($z_{50}$) and 10- and 90-percentile ($z_{10}$,
  $z_{90}$) redshifts for the total counts in each band are also given
  in each panel. For 3.6 $\mum$, the model predictions are compared
  with observational data from \citet{Franceschini06} (black dashed
  line), normalized to unit area as for the models. The error bars
  plotted on the observational data include Poisson errors only.}

\label{fig:dndz-IRAC}
\end{figure*}

In Fig.~\ref{fig:dndz-IRAC}, we show the predictions from our standard
model for the redshift distributions of galaxies in the four IRAC
bands. For each band, we show the redshift distribution for galaxies
selected to be brighter than $S_{\nu}> 10 \muJy$ in that band. The
flux limit $S_{\nu}> 10 \muJy$ has been chosen to match that in the
observed deep sample selected at 3.6$\mum$ by
\citet{Franceschini06}. In each panel, the blue curve shows the
predicted $dN/dz$ for all galaxies, normalized to unit area under the
curve, and the red and green curves show the separate contributions of
bursting and quiescent galaxies to the total. For 3.6$\mum$, the black
line shows the observed redshift distribution from
\citet{Franceschini06}, which has also been normalized to unit area
under the curve. We see that the observed redshift distribution peaks
at a slightly higher redshift than in the model. However, the
luminosity function evolution derived from this same sample is in
reasonable agreement with the model, as was already shown in
Fig.~\ref{fig:lf3.6-evoln-obs}. \citet{Franceschini06} note that the
peak seen in their data at $z\sim 0.8$ is partly contributed by
large-scale structures in the CDFS field.

\begin{figure*}

\begin{center}

\begin{minipage}{7cm}
\includegraphics[width=7cm]{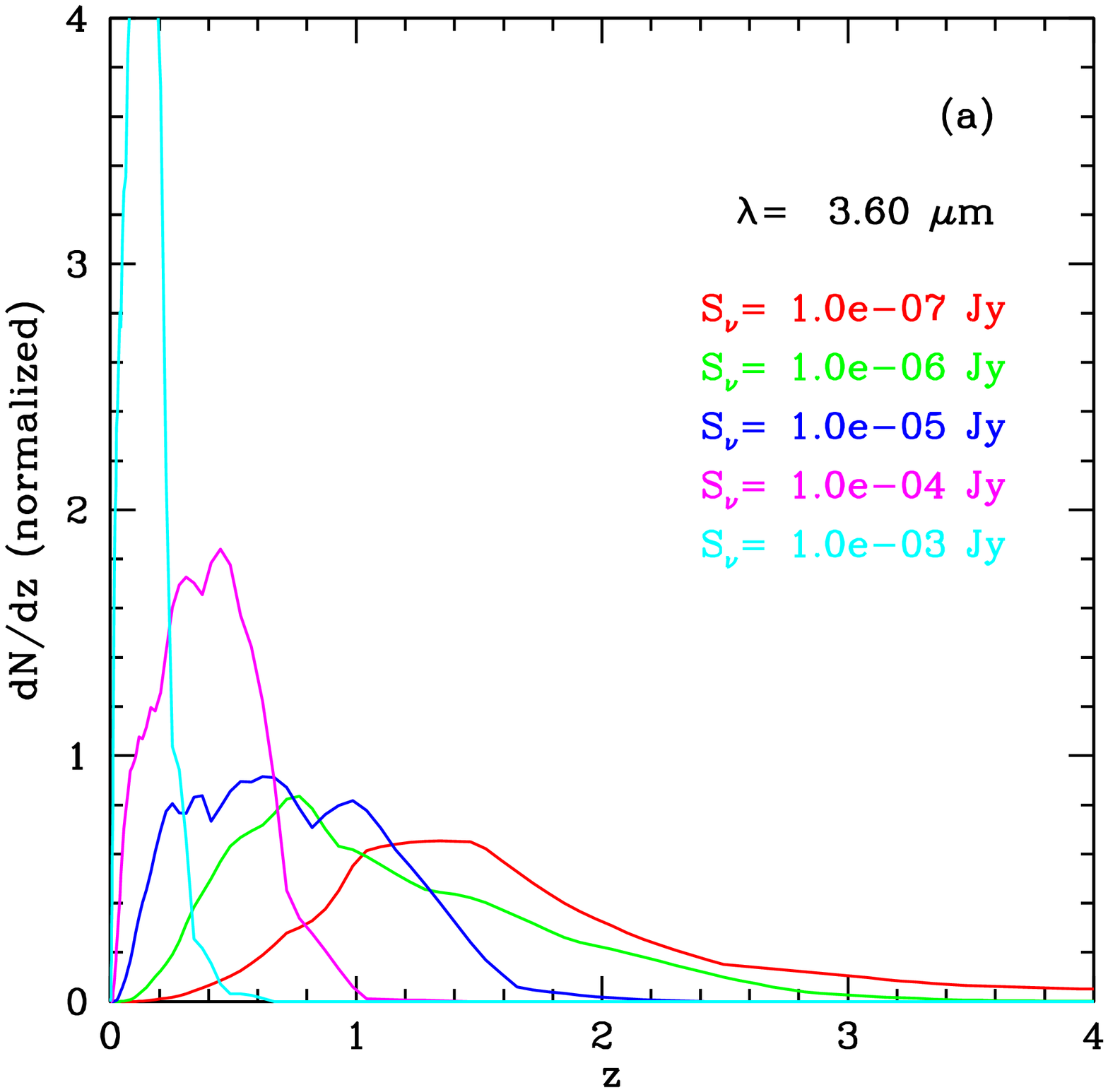}
\end{minipage}
\hspace{1cm}
\begin{minipage}{7cm}
\includegraphics[width=7cm]{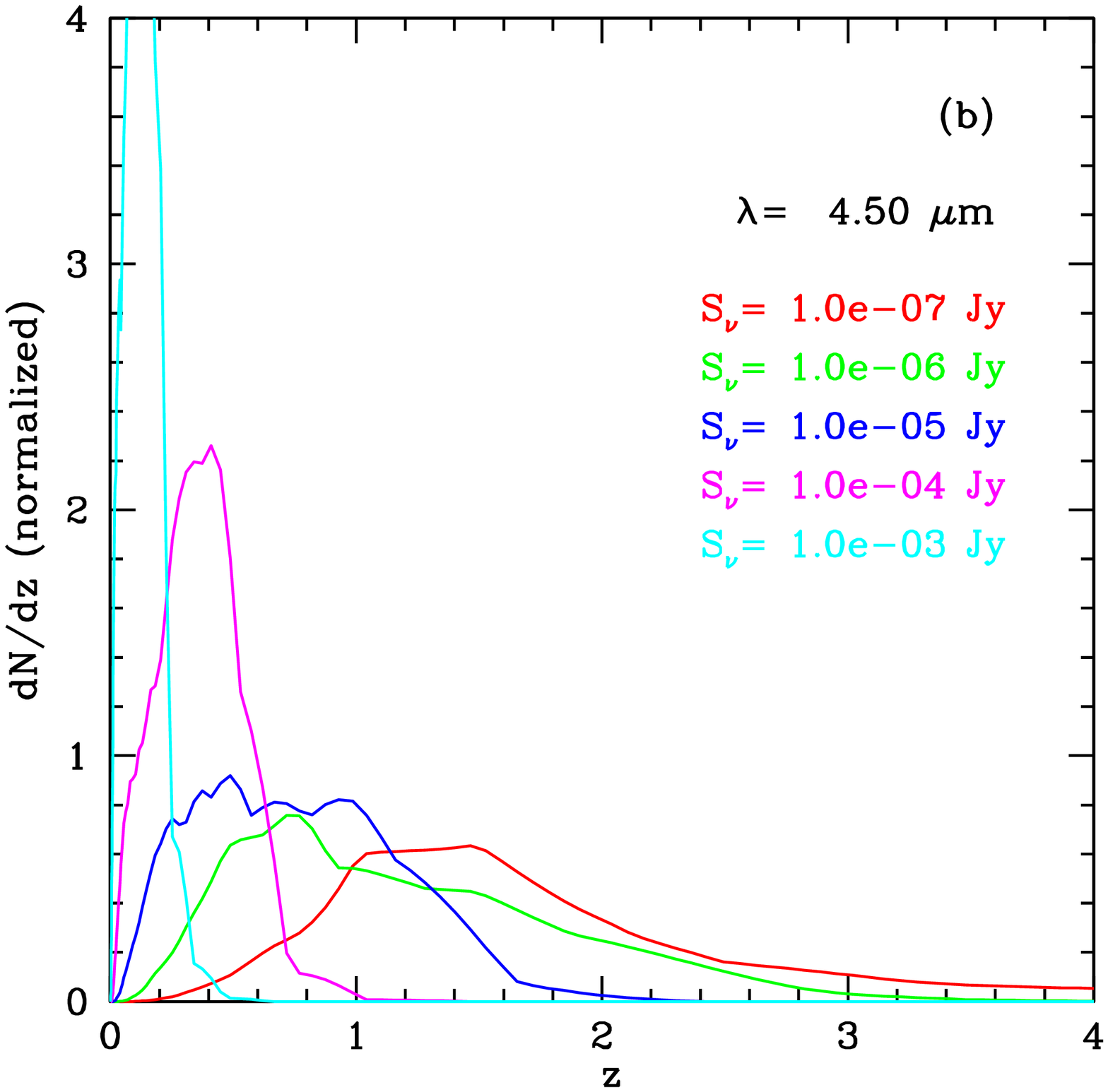}
\end{minipage}

\begin{minipage}{7cm}
\includegraphics[width=7cm]{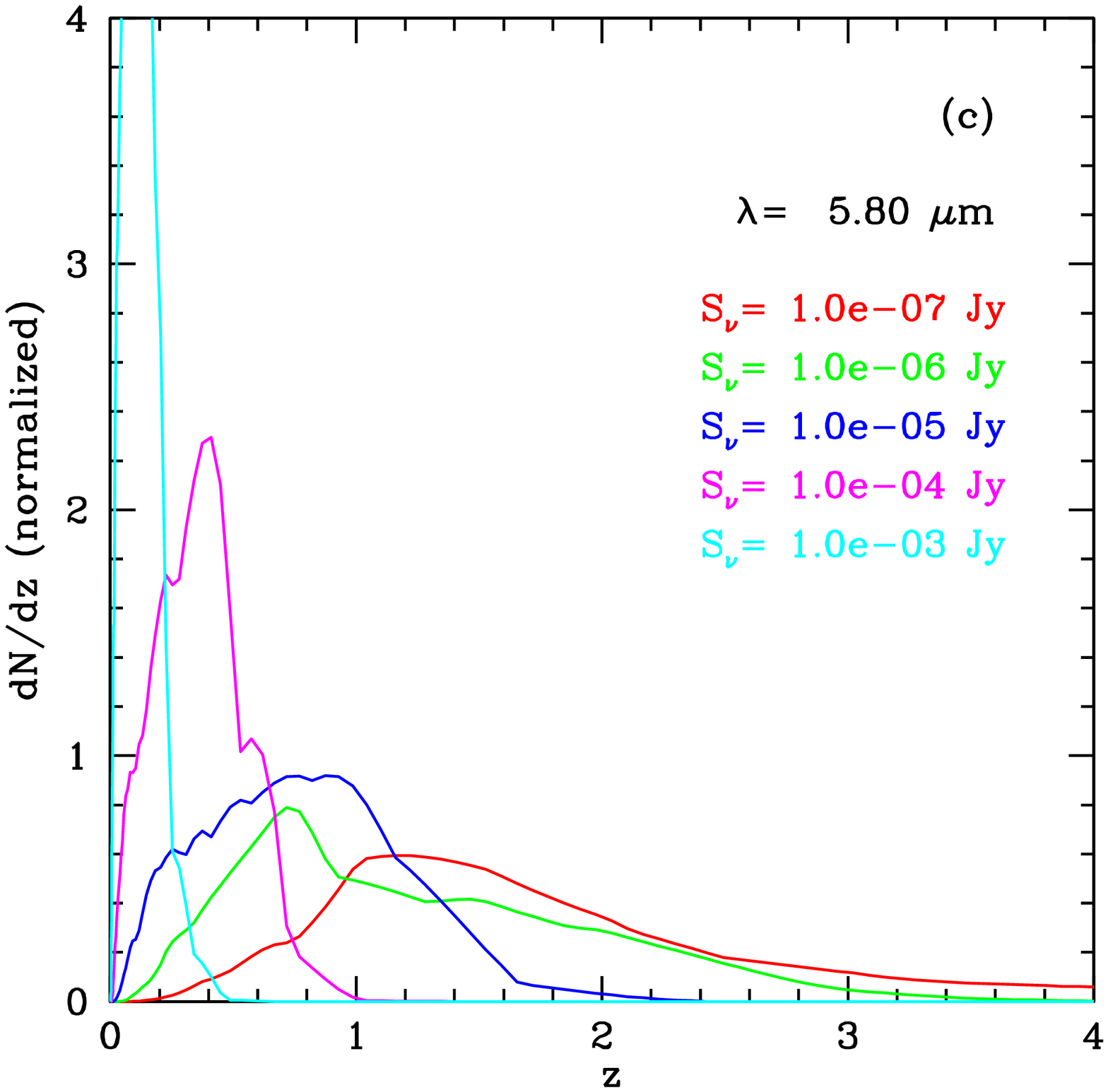}
\end{minipage}
\hspace{1cm}
\begin{minipage}{7cm}
\includegraphics[width=7cm]{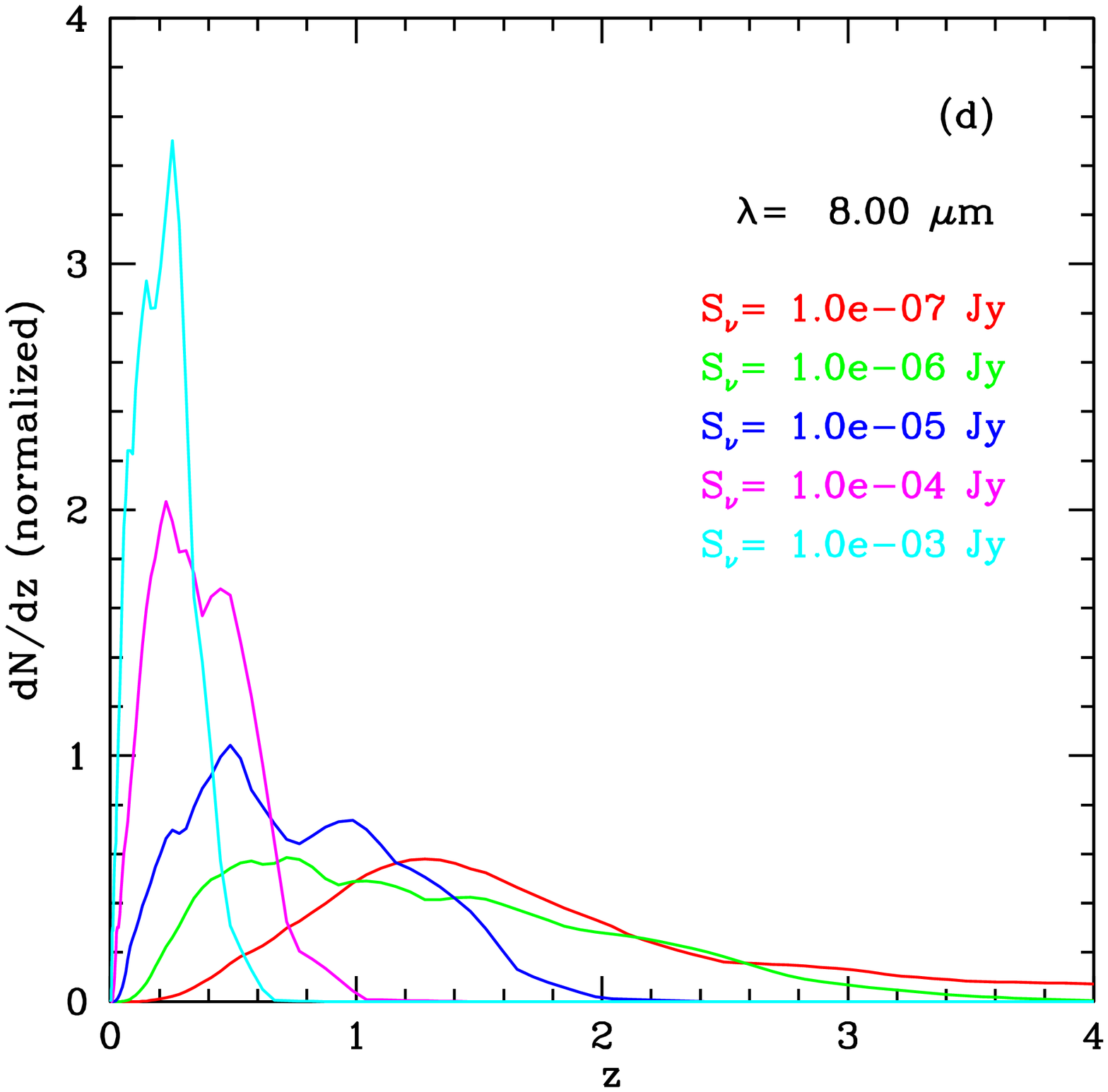}
\end{minipage}

\end{center}

\caption{Predicted galaxy redshift distributions in the four IRAC
  bands, for different fluxes. (a) 3.6 $\mum$, (b) 4.5 $\mum$, (c) 5.8
  $\mum$, and (d) 8.0 $\mum$. In this figure, the redshift
  distributions are for galaxies at a particular flux. Predictions are
  shown for fluxes $S_{\nu} = 0.1$, 1, 10, 100 and 1000 $\muJy$, as
  shown in the key. In all cases, the model curves are normalized to
  unit area, and include the effects of dust.}

\label{fig:dndzs-IRAC}
\end{figure*}

In Fig.~\ref{fig:dndzs-IRAC}, we show predicted redshift distributions
for galaxies selected to be at a set of different fluxes in the four
IRAC bands. The curves for the different fluxes are all normalized to
have unit area as before, but in this figure the galaxies are selected
to be at a particular flux, rather than being brighter than a certain
flux. As one would expect, the typical redshift increases as the flux
decreases.

\begin{figure}

\begin{center}

\includegraphics[width=6.5cm]{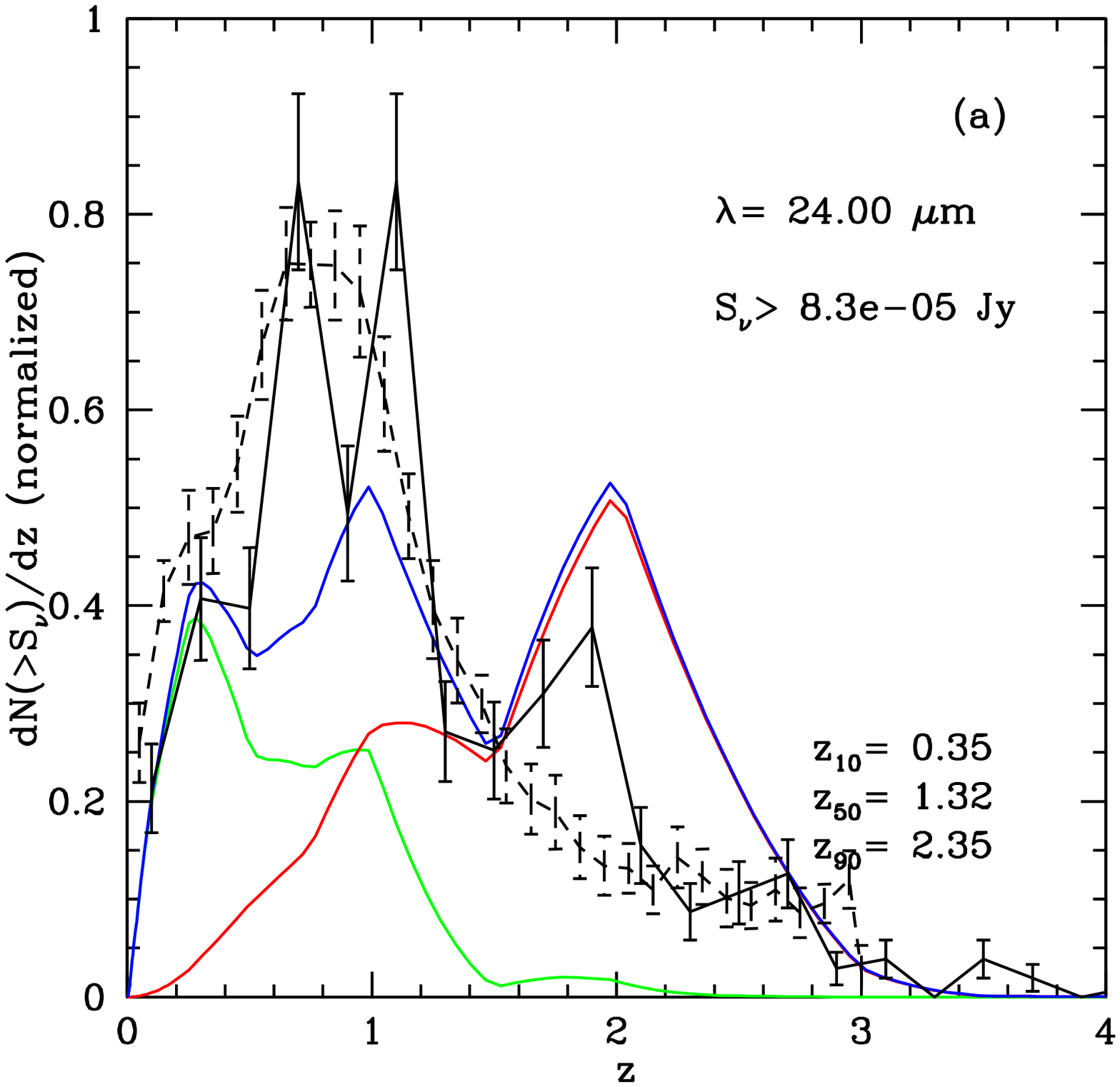}
\includegraphics[width=6.5cm]{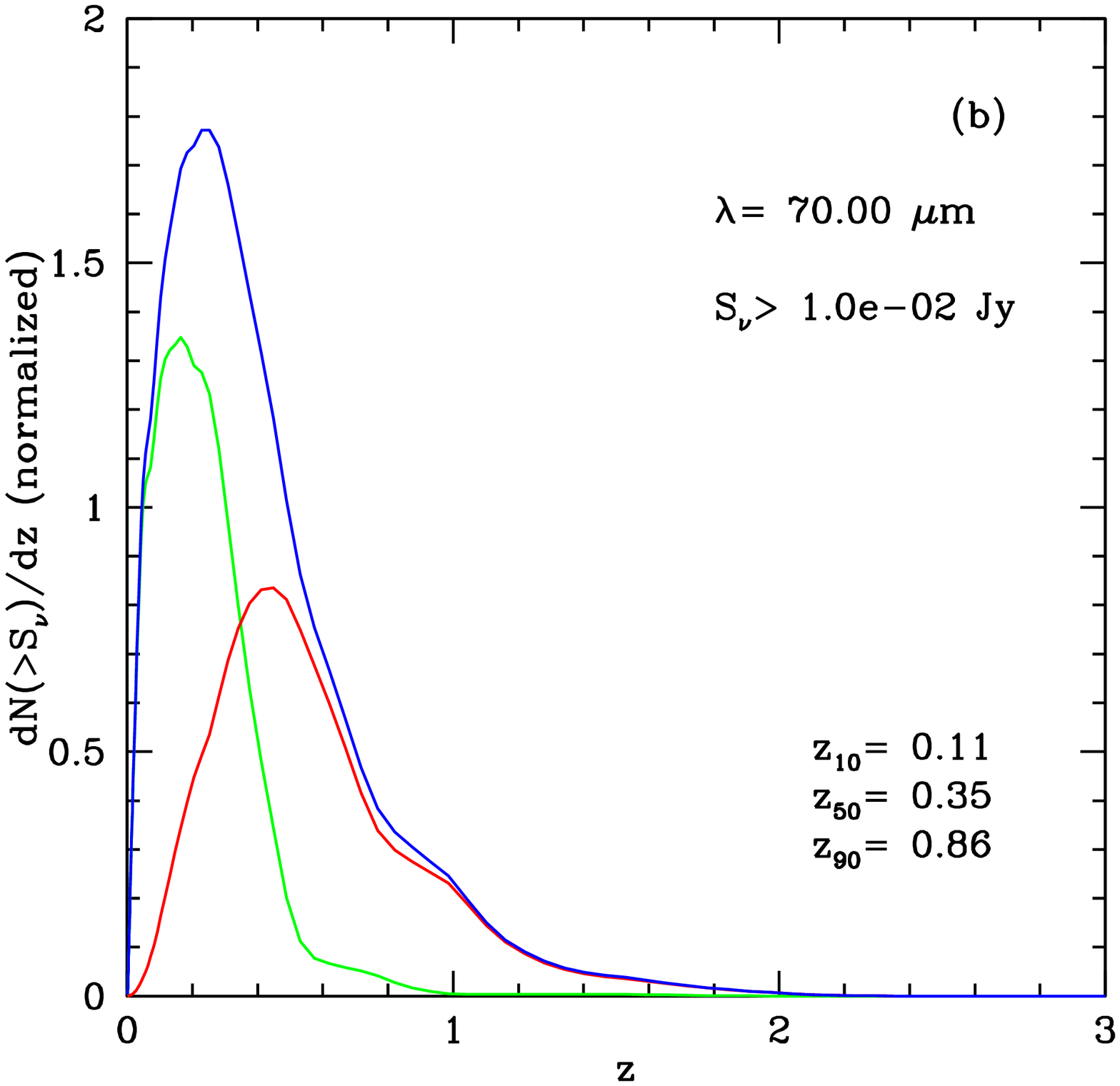}
\includegraphics[width=6.5cm]{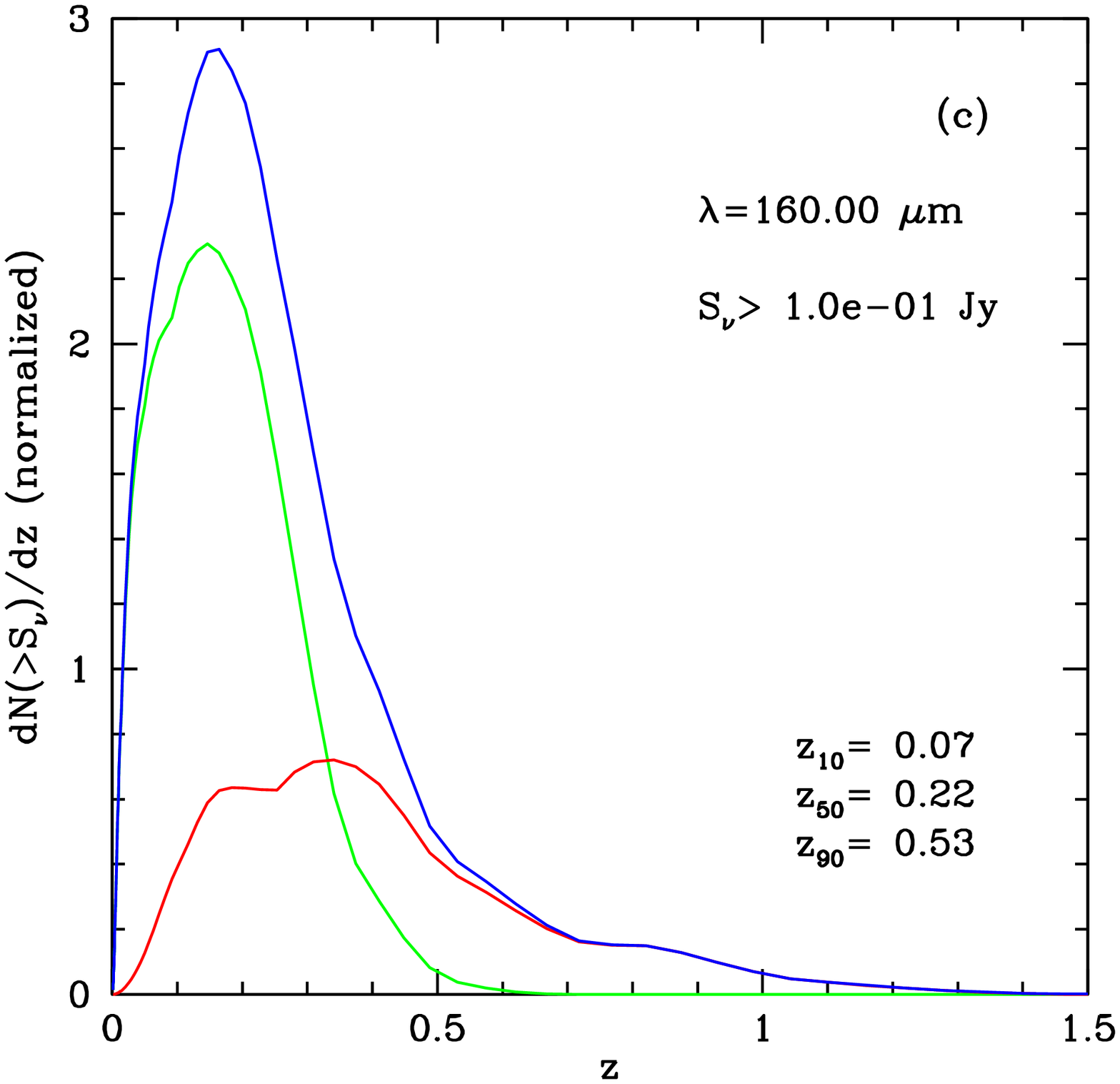}

\end{center}

\caption{Predicted galaxy redshift distributions in the three MIPS
  bands, for galaxies brighter than a specified flux. (a) 24 $\mum$,
  $S_{\nu}> 83 \muJy$, (b) 70 $\mum$, $S_{\nu}> 10 \mJy$, and (c) 160
  $\mum$, $S_{\nu}> 100 \mJy$. The model curves are as follows - blue:
  total; red: ongoing bursts; green: quiescent galaxies. The curves
  are normalized to unit area under the curve for the total
  counts. The median ($z_{50}$) and 10- and 90-percentile ($z_{10}$,
  $z_{90}$) redshifts for the total counts in each band are also given
  in each panel. For 24 $\mum$, the model predictions are compared
  with observational data from \citet{Caputi06} (solid black line) and
  \citet{Perez05} (dashed black line), normalized to unit area as for
  the models. The error bars plotted on the observational data include
  Poisson errors only for \citeauthor{Caputi06}, but also include
  errors in photometric redshifts for \citeauthor{Perez05} }

\label{fig:dndz-MIPS}
\end{figure}

\begin{figure}

\begin{center}

\includegraphics[width=6.5cm]{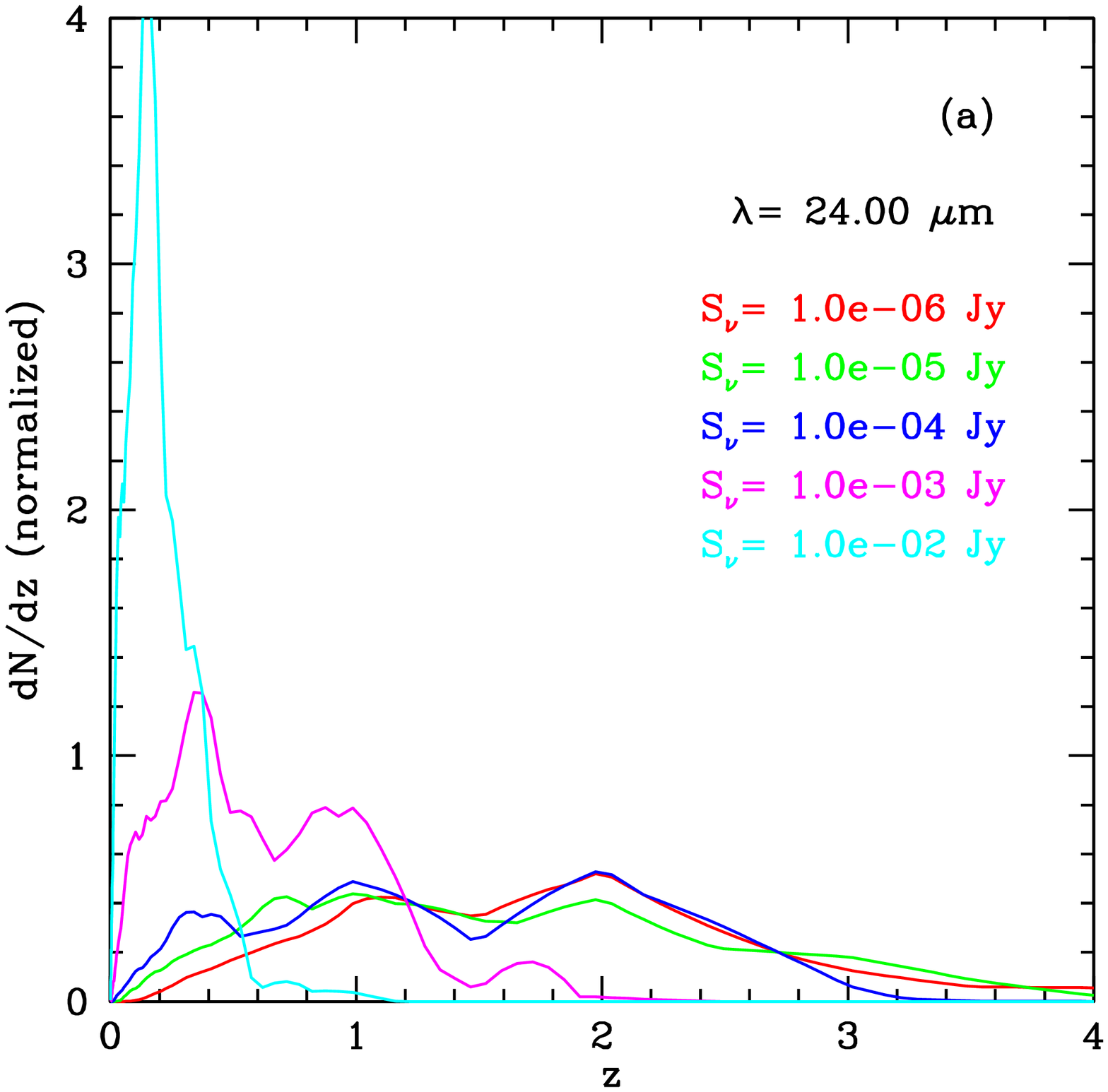}
\includegraphics[width=6.5cm]{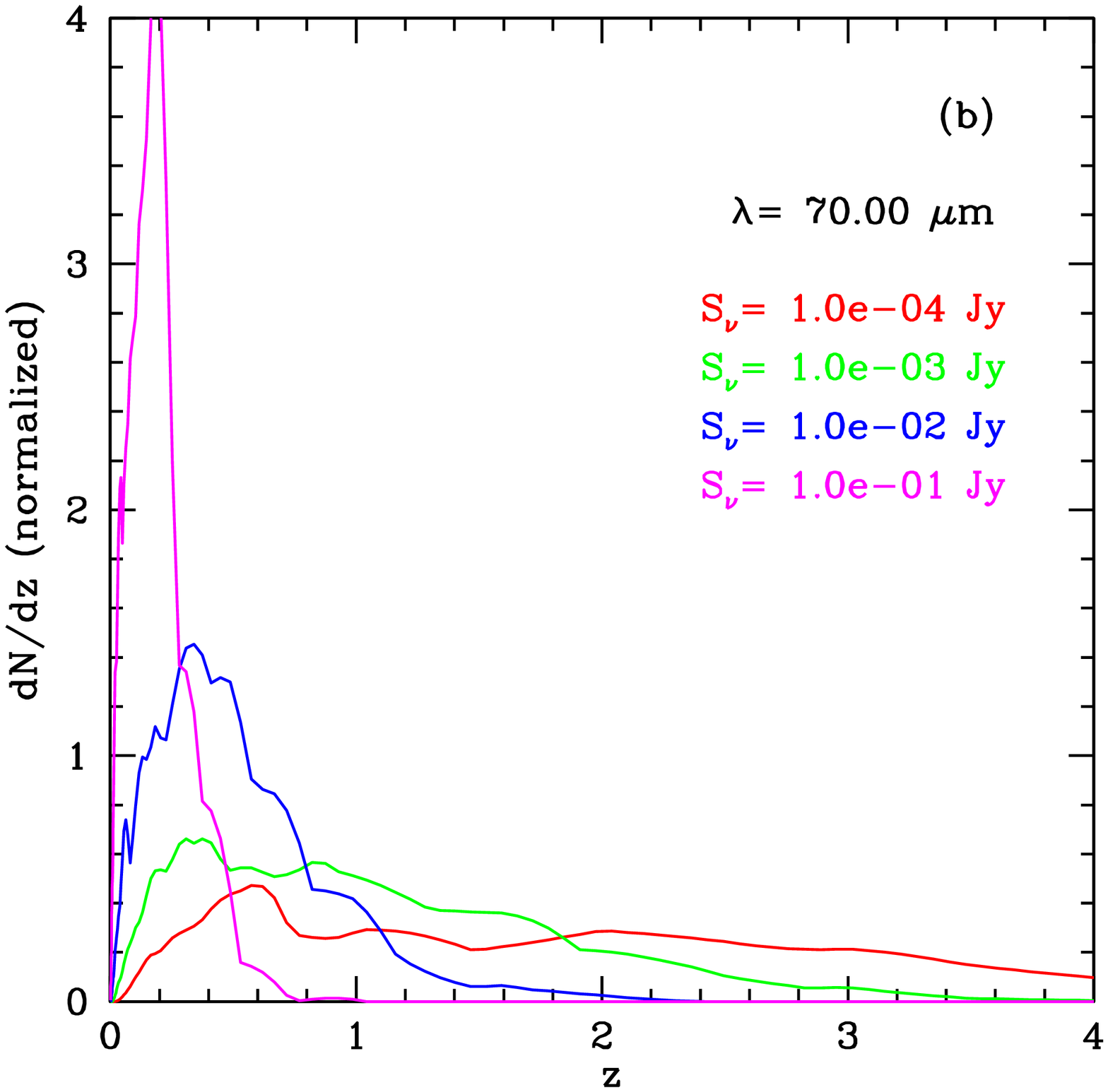}
\includegraphics[width=6.5cm]{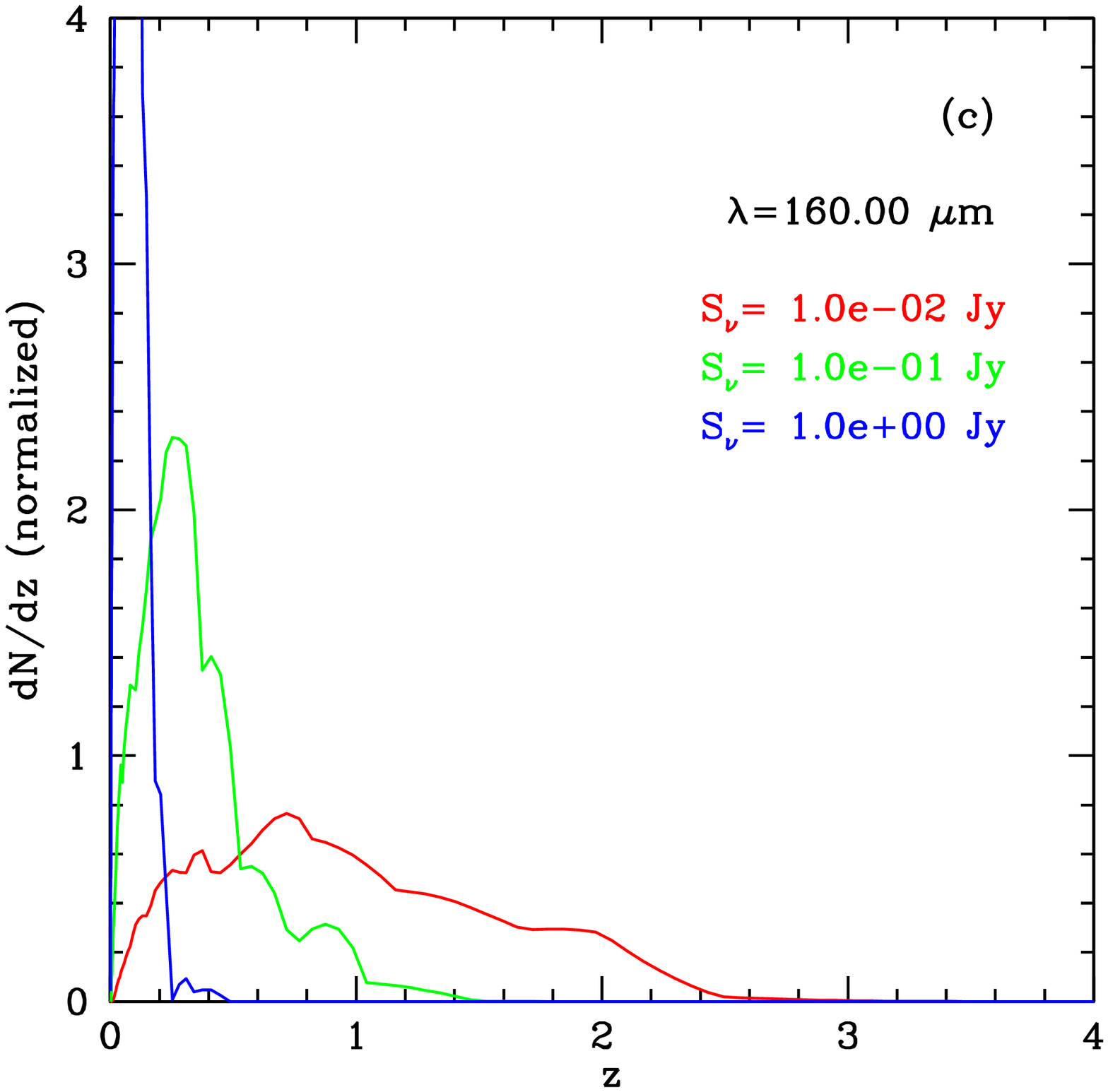}

\end{center}

\caption{Predicted galaxy redshift distributions in the three MIPS
  bands, for different fluxes. (a) 24 $\mum$, (b) 70 $\mum$, and (c)
  160 $\mum$. In this figure, the redshift distributions are for
  galaxies at a particular flux, as shown in the key in each panel. In
  all cases, the model curves are normalized to unit area, and include
  the effects of dust. }

\label{fig:dndzs-MIPS}
\end{figure}

Figs.~\ref{fig:dndz-MIPS} and \ref{fig:dndzs-MIPS} show for the three
MIPS bands the equivalent of Figs.~\ref{fig:dndz-IRAC} and
\ref{fig:dndzs-IRAC} for the IRAC bands. In Fig.~\ref{fig:dndz-MIPS},
we show the predicted redshift distributions for galaxies brighter
than a particular flux, where this flux limit is taken to be 83
$\muJy$ at 24 $\mum$, 10 $\mJy$ at 70 $\mum$ and 100 $\mJy$ at 160
$\mum$. The flux limit at 24 $\mum$ has been chosen to match that used
in the deep observational samples of \citet{LeFloch05},
\citet{Perez05} and \citet{Caputi06}, while the flux limits at 70
$\mum$ and 160 $\mum$ have been chosen to be roughly 3 times brighter
than the source confusion limits in these bands. We see in
Fig.~\ref{fig:dndzs-MIPS} that the redshift distributions at 24 $\mum$
show much more structure than at other wavelengths. This results from
different PAH emission features moving through the 24 $\mum$ band with
changing redshift.

In Fig.~\ref{fig:dndz-MIPS}(a), we compare the predicted redshift
distribution at 24 $\mum$ with observational determinations from
\citet{Perez05} (dashed black line) and \citet{Caputi06} (solid black
line). The observed distributions have been separately normalized to
unit area under the curve, as for the model distribution. Both
observed distributions are based primarily on photometric redshifts,
but the photometric redshifts of \citet{Caputi06} are likely to be
more accurate than those of \citet{Perez05}, since the former are
based on deeper optical and K-band data than the
latter. (\citeauthor{Perez05} found optical counterparts with $B_{AB}
\lsim 24.7$ or $R_{AB} \lsim 23.7$ for $\sim70\%$ of their $S_{\nu}(24
\mum) > 83\muJy $ sources, but relied on IRAC fluxes in deriving
photo-z's for the remaining $\sim 30\%$ of their sample. On the other
hand, \citeauthor{Caputi06} found K-band counterparts with
$K(Vega)<21.5$ for 95\% of their $S_{\nu}(24 \mum) > 80\muJy$ sample,
and derived photo-z's for essentially all of these sources using
optical and K-band data alone).  Both observed distributions are
similar, but the \citeauthor{Caputi06} distribution shows more
structure. This is a combination of the effects of more accurate
photometric redshifts but also a 9 times smaller survey area, which
means that fluctuations due to galaxy clustering are
larger. \citeauthor{Caputi06} argue that the separate peaks at $z\sim
0.7$ and $1.1$ result from large-scale structure, but that the bump at
$z\sim 1.9$ results from PAH emission features entering the observed
24 $\mum$ band. We see that the model also predicts peaks in the
redshift distribution at $z \sim 0.3$, $z \sim 1$ and $z\sim 2$, which
can be explained by different PAH features moving through the 24
$\mum$ band, although the $z \sim 2$ peak is more prominent than is
seen in the observational data. Overall, the model redshift
distribution at this flux limit is too skewed to high redshift
compared to the observations, predicting too few galaxies at $z\sim
0.5-1$, and too many in the peak at $z\sim 2$.

\begin{figure*}
\begin{center}

\begin{minipage}{7cm}
\includegraphics[width=7cm]{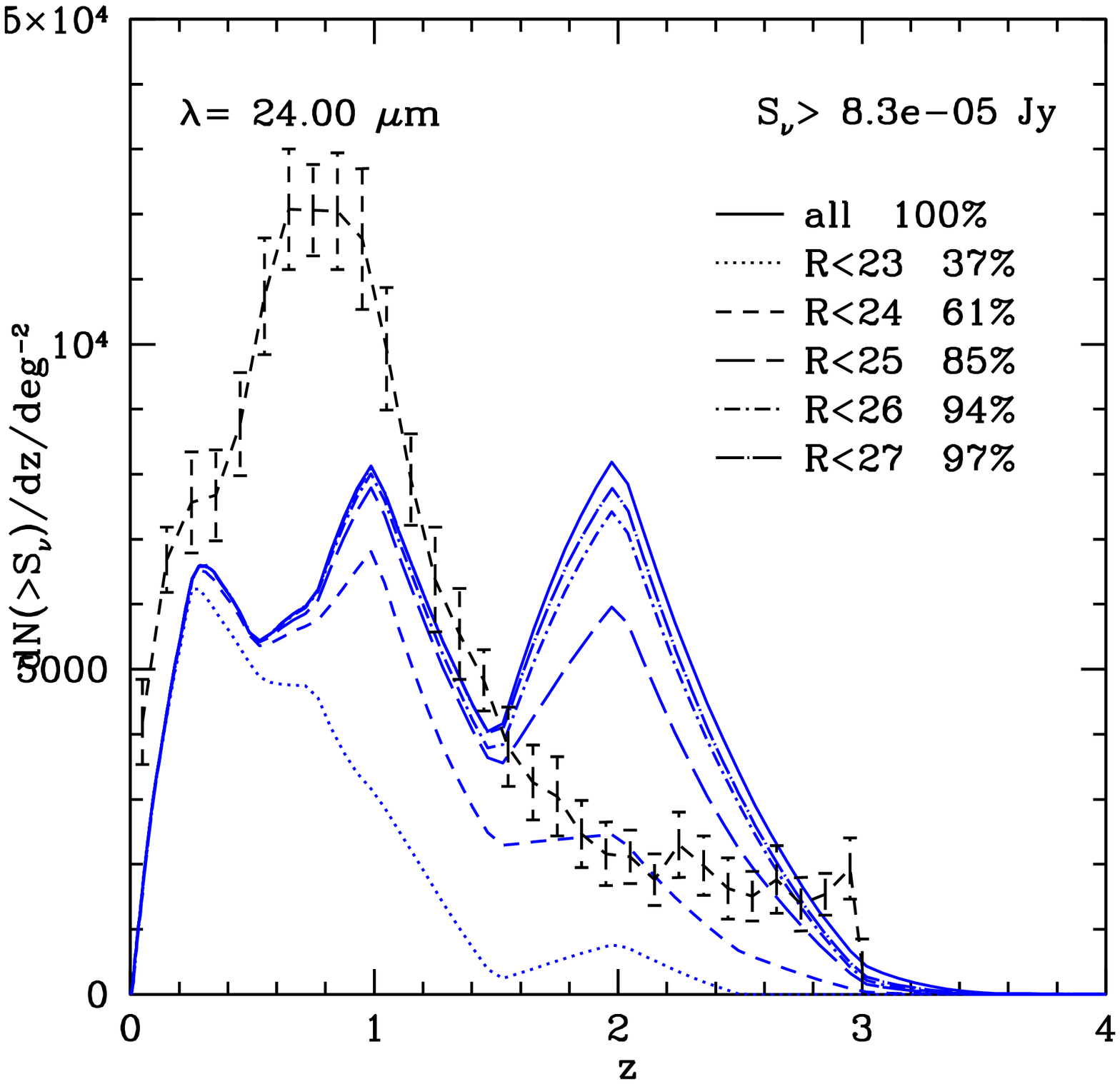}
\end{minipage}
\hspace{1cm}
\begin{minipage}{7cm}
\includegraphics[width=7cm]{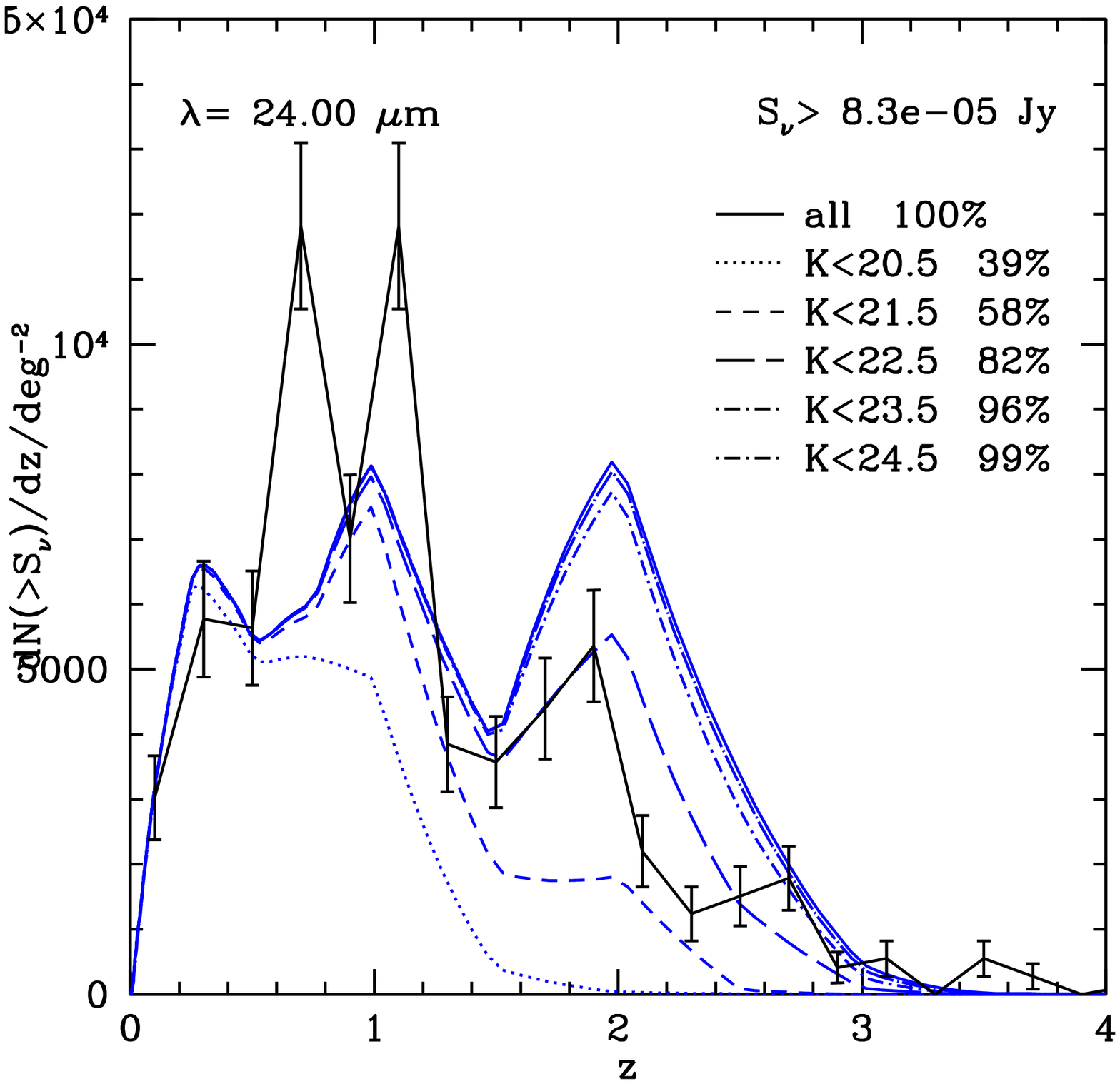}
\end{minipage}

\end{center}
\caption{Predicted redshift distributions at 24$\mum$, showing the
  effects of optical or near-IR magnitude limits. Model galaxies are
  selected with $S_{\nu} > 83\muJy$ together with the optical/NIR
  magnitude limits as shown in the key. The fraction of 24 $\mum$
  sources brighter than each magnitude limit is also given. (a) R-band
  magnitude limit. The observed redshift distribution from
  \citet{Perez05} is overplotted in black. Note \citet{LeFloch05} used
  $R<24$ and obtained 54\% completeness. (b) K-band magnitude
  limit. The observed redshift distribution from \citet{Caputi06}
  (with $K<21.5$) is overplotted. Magnitudes are on the Vega system.}

\label{fig:dndz24-maglim}
\end{figure*}

We investigate further this apparent discrepancy in the 24 $\mum$
redshift distribution in Fig.~\ref{fig:dndz24-maglim}, where we show
the effects of apparent magnitude limits in the R and K-bands on the
predicted redshift distributions for $S_{\nu}(24 \mum) > 83\muJy$. In
this plot, the redshift distributions are plotted as number per solid
angle, without normalizing to unit area under the curve. The left and
right panels respectively have the redshift distributions of
\citeauthor{Perez05} and \citeauthor{Caputi06} overplotted. We
concentrate on the comparison with \citeauthor{Caputi06}, since this
has the simpler sample selection and more accurate redshifts. The
model prediction for $K<21.5$ (which is the magnitude limit used by
\citeauthor{Caputi06}) is shown by the short-dashed blue line, while
the prediction with no limit on the K-magnitude is shown by the solid
blue line. The model $dN/dz$ with no limit on the K magnitude is most
discrepant with the \citeauthor{Caputi06} data at $z\sim 2$, where it
predicts $\sim 2$ times too many galaxies. This is directly related to
the fact that the predicted luminosity function at $z=2$ at rest-frame
wavelength 8 $\mum$ (corresponding to observed wavelength 24 $\mum$)
and luminosity $\sim 10^{11} \Lsol$ is also $\sim 2$ times too high
compared to what \citeauthor{Caputi06} estimate from their data, as
shown in Fig.~\ref{fig:lf8.0-evoln-obs}(c). When the effect of the
$K<21.5$ limit is included, the predicted redshift distribution is
closer to the observational data, but only $58\%$ of the model
galaxies are brighter than this K-band magnitude limit, as against
95\% in the observed sample of \citeauthor{Caputi06}. We conclude that
the main reason for the discrepancy between the predicted and observed
redshift distributions at 24 $\mum$ is that the model predicts a
rest-frame 8 $\mum$ luminosity function at $z\sim 2$ which is somewhat
too high at luminosities $\sim 10^{11} \Lsol$, even though it
reproduces quite well the general features of the evolution of the
mid-IR luminosity function.


\clearpage

\end{document}